\documentclass[10pt,journal,compsoc]{IEEEtran}

\usepackage{cite}

\ifCLASSINFOpdf
  \usepackage[pdftex]{graphicx}
  \DeclareGraphicsExtensions{.pdf,.jpeg,.png}
\else
\fi

\usepackage{amsmath}

\usepackage{multirow}
\usepackage{bm}

\usepackage{hyperref}

\usepackage{nicefrac}
\usepackage{xcolor}

\usepackage[normalem]{ulem} % for striking through text
\usepackage{tcolorbox}

\renewcommand{\vec}[1]{\mathbf{#1}}

\usepackage{soul}
\soulregister\cite7
\soulregister\ref7
\soulregister\pageref7
\soulregister\autoref7

\DeclareRobustCommand{\hsout}[1]{\texorpdfstring{\sout{#1}}{#1}}

\newcommand{\delete}[1]{\textcolor{red}{\hsout{#1}}}
\newcommand{\deletedEqn}[1]{\begin{equation}\cancel{#1}\end{equation}}

\ifdefined\showMarkUps\else

\renewcommand{\delete}[1]{\ignorespaces}

\renewcommand{\deletedEqn}[1]{}
\fi

\newcommand{\deleteF}[1]{\textcolor{red}{\hsout{#1}}}
\ifdefined\showMarkUpsFinal\else

\renewcommand{\deleteF}[1]{\ignorespaces}

\fi

\newcommand{\showDOI}[1]{}
\newcommand{\showURL}[1]{}
   
\graphicspath{{FiguresEPS/}}

\DeclareGraphicsExtensions{.pdf,.jpg,.png,.eps} %high-quality
\newcommand{\figureheight}{\columnwidth}

\usepackage{caption}
\usepackage{xcolor}
\definecolor{darkgrey}{rgb}{0.2, 0.2, 0.2}
\definecolor{amaranth}{rgb}{0.9, 0.17, 0.31}
\definecolor{mygreen}{rgb}{0.1, 0.6, 0.1}

\begin{document}

\newcommand{\mytitle}[1]{Particle Merging-and-Splitting}
\title{\mytitle{\\*}}

\author{Nghia~Truong,
        Cem~Yuksel,
        Chakrit~Watcharopas,
        Joshua~A.~Levine,
        Robert~M.~Kirby
\IEEEcompsocitemizethanks{
\IEEEcompsocthanksitem N. Truong, C. Yuksel, and R. M. Kirby is with University of Utah.
\IEEEcompsocthanksitem C. Watcharopas is with Kasetsart University.
\IEEEcompsocthanksitem J. A. Levine is with University of Arizona.
}
\thanks{Manuscript received November 13, 2020; accepted June 25th, 2021.}}
% \thanks{Manuscript received November 13, 2020; revised April 27, 2020; accepted June 25th, 2021.}}

\markboth{Journal of \LaTeX\ Class Files,~Vol.~14, No.~8, August~2015}%
{Shell \MakeLowercase{\textit{et al.}}: Bare Advanced Demo of IEEEtran.cls for IEEE Computer Society Journals}

\IEEEtitleabstractindextext{%
\begin{abstract}
Robustly handling collisions between individual particles in a large particle-based simulation has been a challenging problem.
We introduce particle \emph{merging-and-splitting}, a simple scheme for robustly handling collisions between particles that prevents inter-penetrations of separate objects without introducing numerical instabilities. This scheme merges colliding particles at the beginning of the time-step and then splits them at the end of the time-step. Thus, collisions last for the duration of a time-step, allowing neighboring particles of the colliding particles to influence each other. We show that our merging-and-splitting method is effective in robustly handling collisions and avoiding penetrations in particle-based simulations. We also show how our merging-and-splitting approach can be used for coupling different simulation systems using different and otherwise incompatible integrators. We present simulation tests involving complex solid-fluid interactions, including solid fractures generated by fluid interactions.
\end{abstract}

}

\maketitle

\ifCLASSOPTIONpeerreview
\begin{center} \bfseries EDICS Category: 3-BBND \end{center}
\fi
\IEEEpeerreviewmaketitle

\section{Introduction}

Particle-based simulations are commonplace in computer graphics, used for
simulating a wide variety of physical phenomena for different material types and phases.
In particular, high-resolution simulations involving a large number of particles can deliver complex animations with rich visual detail.
Yet, handling interactions between individual particles, especially collisions, has been a challenging problem. This is not only because of the sheer number of interactions that can occur but also due to the difficulty of robustly enforcing collision constraints in the presence of other interactions between particles.

Not all particle-based simulations must explicitly consider pair-wise particle collisions. For example, Eulerian fluid simulation handle interactions among particles differently, without directly considering collisions. In such cases, explicit collisions are needed for handling boundaries and coupling with other simulation systems.

However, particle-level collisions cannot always be safely ignored. For example, particle-based simulations involving (non-granular) solids must properly incorporate particle collisions. Unfortunately, common collision handling techniques based on force or impulse formulations can be unstable or fail to resolve collisions, as we show in this paper. Poorly handled particle collisions can lead to catastrophic problems, such as inter-penetrations of separate objects and instabilities that cause numerical failure or unnatural material behavior.
In particular, fracture simulations are highly sensitive to such instabilities, as the velocity spike of a single particle can cause an entire object to instantly crumble.

In this paper, we introduce \emph{merging-and-splitting}, a simple method for robustly handling collisions between individual particles in particle-based simulations. When two (or more) particles come into contact, we first \emph{merge} them into a larger \emph{meta-particle} based on an inelastic collision formulation. A meta-particle behaves as a single particle during numerical integration. This treatment allows us to compute the collective momentum of the colliding particles, considering the influence of the surrounding particles. After numerical integration, we \emph{split} the colliding particles, following the principles of elastic collision with momentum conservation and energy preservation. This approach facilitates information exchange between the colliding particles for the duration of the time-step. Our tests show that our merging-and-splitting scheme completely prevents inter-penetrations without introducing instabilities.

Merging-and-splitting also allows coupling different particle-based simulations using different and otherwise incompatible numerical systems. 
Our tests with coupling different particle-based simulations include mass-spring systems for deformable objects, peridynamics \cite{Levine:2014} for brittle solids, and Smoothed Particle Hydrodynamics (SPH)~\cite{Desbrun:1996,Muller:2003} and Fluid-Implicit Particle (FLIP)~\cite{Zhu:2005} for fluids. 
In our tests, we handle solids using implicit integration, SPH using explicit integration, and FLIP using a semi-implicit scheme (explicit advection and implicit pressure projection).
We use our merging-and-splitting technique for handling collisions between particles of the same simulation system for solids and particles simulated using different integration schemes. 
In fact, the coupling between different simulation systems in our tests is handled entirely via merging-and-splitting.
Our results clearly show that our merging-and-splitting scheme is effective in coupling different particle-based formulations with different types of numerical integrators. 
Using our scheme, we also demonstrate unprecedented simulation scenarios, like solid fracture due to fluid interaction.

\begin{figure*}[tb]\centering
    \includegraphics[width=\linewidth]{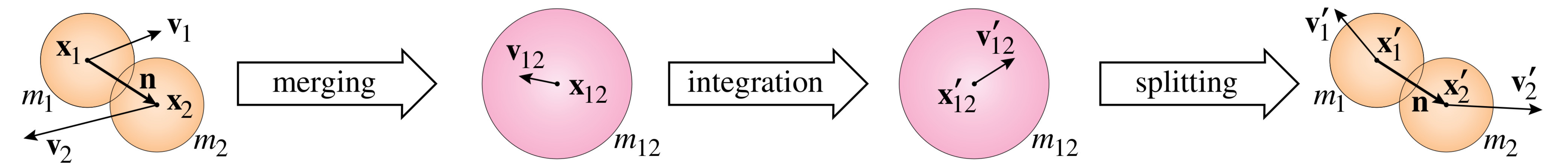}\vspace{-0.3em}\\
    (a)\hspace{0.25\linewidth}%
    (b)\hspace{0.25\linewidth}%
    (c)\hspace{0.27\linewidth}%
    (d)\\
    \caption{\textbf{The overview of merging-and-splitting approach:} (a)~colliding particles are merged into (b)~a meta-particle using the principles of inelastic collision. This meta-particle is used during time integration to compute (c)~its position and velocity at the end of the time-step. Finally, the meta-particle is split into (d)~the two original particles with updated velocities, using the principles of elastic collision with momentum and energy conservation.}
    \label{fig:mering-and-splitting}
\end{figure*}

\begin{comment}
\begin{figure*}[tb]
\centering
\renewcommand{\figureheight}{0.275\textwidth}
\includegraphics[trim=650 200 900 350, clip, height=\figureheight]{BallShootsWall/ball-shoots-f55}\hfill%
\includegraphics[trim=380 0 400 230, clip, height=\figureheight]{bunny_drop_cloth_f100}\hfill%
\includegraphics[trim=430 0 480 430, clip, height=\figureheight]{dambreak-f90}% %\hfill%
\caption{Example frames from simulations of solids and fluids, where interactions between particles of different objects and materials are handled using our merging-and-splitting method.}%\vspace{1em}
\label{fig:teaser}
\end{figure*}
\end{comment}

\section{Related Work}

There is a large body of work on collision detection and handling in computer graphics. In this section we briefly review the most related methods on collision handling.

\textbf{Force-based Collision Response.}
Arguably the most popular technique for handling collisions is by applying penalty forces between colliding objects \cite{Terzopoulos:1987, Terzopoulos:1989, Moore:1988}. The magnitude of a penalty force is determined based on the amount of intersection between the two objects, so typical penalty force formulations do not completely eliminate penetration. More importantly, the results are sensitive to the stiffness of the penalty force, which is often controlled by a user-defined parameter. Penalty forces can be distance-based
\cite{Terzopoulos:1989, Tonnesen:1991} or volume-based \cite{Hasegawa:2004,Faure:2008}. Their stability can be improved using multiple contact points \cite{Drumwright:2008} or a global contact treatment \cite{Heidelberger:2008}, and a friction term can be included \cite{Mishra:2003}. Smooth collision responses can be achieved using continuous penalty forces \cite{Tang:2012}.
Harmon et al.~\cite{Harmon:2009, Harmon:2011} present a method using asynchronous variational integrators and a contact barrier potential to avoid penetrations and conserve momentum and energy, but simulations involving a large number of collisions require numerous iterations, making this approach inefficient for particle-based simulations. Penalty forces are also commonplace in cloth simulations~\cite{Baraff:1998,Provot:1997,Bridson:2002,Selle:2009}.
Notably, \emph{rigid impact zones} were introduced \cite{Provot:1997,Bridson:2002} to handle collisions by rigidifying the entire collision region, which have similarities with our approach.

Similar force-based formulations are also used in solid-fluid coupling \cite{Koschier:2019} by directly applying forces between nearby particles \cite{Becker:2009b,Shao:2014} or introducing additional pressure force \cite{Akinci:2012,Akinci:2013}. 
The simplicity of the force formulation allows coupling Eulerian fluid simulations with particle-based solid simulations \cite{Oliver:2003} and particle-based fluid simulations with deformable solids \cite{Muller:2004,Yang:2012,Shao:2015} and, using continuous collision detection, sheet-based cloth simulations \cite{Du:2012, Huber:2015}. 
Recent work on handling collisions between fluid particles and rigid boundaries include
density maps~\cite{Koschier:2017}, volume maps~\cite{Bender:2019}, liquid boundaries~\cite{Cornelis:2015} and pressure boundaries~\cite{Band:2018a, Band:2018b}.
Density maps can also be augmented to handle frictional contact~\cite{Bender:2020}, and strong fluid-rigid body coupling simulation is achieved by interlinked pressure solvers~\cite{Gissler:2019}.

\textbf{Impulse-based Collisions.}
Another popular alternative for handling collisions is impulse-based formulations, which are common in rigid-body dynamics \cite{Hahn:1988, Mirtich:1995, Guendelman:2003, Weinstein:2006, Vouga:2017}. The momentum and energy transfer between two colliding objects are handled instantly by directly modifying the velocities of colliding objects. Rest-in-contact situations can be handled by careful treatments~\cite{Mirtich:1995,Bender:06}.
Notably, iteratively solving for the impulses needed has been widely used for achieving complex scenarios, such as stacking and static friction \cite{Bender:06,Weinstein:2006,Macklin:2014}.

\textbf{Constrained Dynamics.}
Collision response can be formulated as constraints that prevent penetration~\cite{Witkin:1997}. It is often implemented using Linear Complementary Programming (LCP) for searching the reaction space for the feasible contact behavior~\cite{Erleben:2013} and it can be used for modeling interactions of rigid bodies and deformable bodies \cite{Lotstedt:1984, Stewart:1996, Anitescu:1997, Duriez:2006, Otaduy:2009}, granular materials \cite{Alduan:2011}, fluids \cite{Batty:2007, Chentanez:2011}, and quasi-rigid bodies~\cite{Pauly:2004}.
Formulated as an optimization problem, LCP is suitable for solving contact problems for which the optimal solutions are rest positions with minimum energy exchange \cite{Baraff:1994,Trinkle:1995,Stewart:1996,Kry:2003}.
Robust contact handling which facilitates smooth rolling and sliding, stacking and impact handling was achieved by a formulation based on implicit complimentary constraints and Lagrange multipliers \cite{Otaduy:2009}.

\textbf{Grid-based Solutions.} 
Hybrid Eulerian-Lagrangian simulation methods can avoid explicitly handling collisions by solving the aggregate collision behavior on a grid. 
For example, fluid simulations using FLIP have been coupled with hair~\cite{Fei:2017} and cloth~\cite{Fei:2018} simulations.
The Material Point Method~\cite{Stomakhin:2013} is another good example, providing simulations of various material types, such as snow~\cite{Stomakhin:2013},
multi-species~\cite{Stomakhin:2014, Tampubolon:2017} with phase transition~\cite{Stomakhin:2014}, sand~\cite{Daviet:2016, Klar:2016}, elastoplastic solids, viscoelastic fluids, foams and sponges~\cite{Ram:2015,Yue:2015, Fang:2019}, anisotropic elastoplastic materials~\cite{Jiang:2017, Wolper:2020}, 
fluid-sediment mixture~\cite{Gao:2018}.
MPM can also achieve solid-fluid coupling simulation~\cite{Tampubolon:2017, Gao:2018,Fang:2020}, dynamic fracture~\cite{Wolper:2020}, ductile fracture~\cite{Wang:2019} and frictional contact~\cite{Han:2019}.

\section{Merging-and-Splitting}

We assume that each particle has a non-zero mass and a spherical shape with non-zero radius. This is not necessarily the case for all particle-based simulations. Some of them use massless particles and some treat them as point samples with no size. Therefore, we begin with assigning a mass and a radius to each particle for collision purposes.

Our approach for robustly handling particle-level interactions is a scheme
that we call \emph{merging-and-splitting}. This scheme first \emph{merges}
colliding particles (\autoref{fig:mering-and-splitting}a) into a larger \emph{meta-particle} (\autoref{fig:mering-and-splitting}b). 
We treat this meta-particle as a regular particle, rather than a compound object made up of multiple particles. This crucial simplification allows seamlessly including the meta-particle in arbitrary particle-based simulation systems without modification. 
The meta-particle is integrated
along with the other particles in the system for computing its updated position and velocity at the end of the
time-step (\autoref{fig:mering-and-splitting}c). 
Finally, we \emph{split} the meta-particle into the original colliding particles (\autoref{fig:mering-and-splitting}d).

Consider two particles with masses $m_1$ and $m_2$, positions $\vec{x}_1$ and $\vec{x}_2$, and velocities $\vec{v}_1$ and $\vec{v}_2$
colliding with each other. We merge the two particles based on an inelastic collision
formulation, such that the total mass, position, and velocity of the meta-particle become
\begin{align}
m_{12} &= m_1 + m_2 \;, \label{eq:mass1} \\
\vec{x}_{12} &= \left( m_1 \vec{x}_1 + m_2 \vec{x}_2 \right)/m_{12} \;, \label{eq:position1} \\
\vec{v}_{12} &= \left( m_1 \vec{v}_1 + m_2 \vec{v}_2 \right)/m_{12} \;. \label{eq:momentum1}
\end{align}
While this merging operation conserves momentum, it does not conserve kinetic energy, because it is based on inelastic collision principles. In our interaction scheme we compute the change in kinetic energy, ${\Delta E}$, and store it as a potential energy in a \emph{virtual
bond} between the two colliding particles, where
\begin{equation}
\Delta E = \frac{m_1 m_2}{2 m_{12}} \left(\vec{v}_1 - \vec{v}_2\right)^2 \;.
\label{eq:energy_loss2}
\end{equation}
Note that ${\Delta E}$ cannot be negative, so we always store a non-negative potential
energy in the virtual bond between the two particles.

The meta-particles are used during the time-step integration instead of the original particles. However, while computing the forces between the meta-particles and the surrounding particles, we use the relative positions of the merged particles that make up the meta-particles.
Let $\vec{n}$ be the vector connecting the centers of the two colliding
particles in the beginning of the time-step, such that
\begin{equation}
\vec{n} = \vec{x}_2 - \vec{x}_1 \;.
\end{equation}

We assume that the meta-particle preserves the relative orientation of the merged particles throughout the time-step integration, so $\vec{n}$ remains constant. Therefore, we can find the positions of the merged particles using the updated position of the meta-particle $\vec{x}_{12}^\prime$, such that
\begin{align}
\vec{x}_1^\prime &= \vec{x}_{12}^\prime - \left(m_2/m_{12}\right)\vec{n} \;, \text{ and}\\
\vec{x}_2^\prime &= \vec{x}_{12}^\prime + \left(m_1/m_{12}\right)\vec{n} \;.
\end{align}
Thus, ${\vec{x}_1^\prime=\vec{x}_1+\Delta\vec{x}}$ and ${\vec{x}_2^\prime=\vec{x}_2+\Delta\vec{x}}$ for ${\Delta\vec{x}=\vec{x}_{12}^\prime - \vec{x}_{12}}$.
Note that these simple equations can be used to calculate the intermediate positions of the merged particles at any time within the time-step for computing forces between the meta-particles and the rest of the simulation system. 
Since meta-particles are treated as regular particles, forces acting on the meta-particles are applied at the center of mass of the meta-particles.

During splitting, a portion of the potential energy $\Delta E$ stored in the virtual
bond converts back into kinetic energy, while the rest dissipates.
The amount of energy restoration is controlled by a user-defined coefficient ${\alpha
\in [0,1]}$ that serves as a restitution parameter (we use $\alpha=1$ in our tests, unless otherwise specified, so we fully conserve energy for collision handling without any dissipation). 
Let $\vec{v}_{12}^\prime$ be the velocity of the meta-particle at the end of the time-step integration.
The final velocities of the particles $\vec{v}_1^\prime$ and $\vec{v}_2^\prime$ satisfy momentum and energy conservation equations
\begin{align}
m_{12} \vec{v}_{12}^\prime &= m_1 \vec{v}_1^\prime + m_2 \vec{v}_2^\prime\label{eq:momentum}\\
\alpha \Delta E + \frac{1}{2} m_{12} {\vec{v}_{12}^\prime}^2 &=
\frac{1}{2} m_1 {\vec{v}_1^\prime}^2 + \frac{1}{2} m_2 {\vec{v}_2^\prime}^2 \;.
\label{eq:energy_conservation}
\end{align}
Using these two equations we can write
\begin{equation}
\left(\vec{v}_{12}^\prime - \vec{v}_1^\prime\right)^2 = s^2 \label{eq:quadratic} \;,
\end{equation}
where
\begin{equation}
s^2 = \frac{2\alpha \Delta E}{m_{12}\left(m_1/m_2\right)} \;.
\label{eq:s}
\end{equation}
Here the only unknown is $\vec{v}_1^\prime$, and once it is solved, $\vec{v}_2^\prime$ can be
calculated using momentum conservation (\autoref{eq:momentum}).

\autoref{eq:quadratic} describes the general energy and momentum conservation
laws for splitting a meta-particle into any two particles (with masses $m_1$ and $m_2$) without considering the initial configuration prior to merging. Therefore, \autoref{eq:quadratic} has infinitely many solutions. We consider the initial conditions of the collision event to narrow down the solution space.

We can split the change in velocity $\vec{v}_1^\prime-\vec{v}_1$ into two components: one along the collision direction ${\vec{\hat{n}}=\vec{n}/\left|\vec{n}\right|}$; and one along an orthogonal direction, using
\begin{align}
\vec{v}_1^\prime = \vec{v}_1 + \mu \vec{\hat{n}} + \bm{\epsilon} \;,
\label{eq:ray}
\end{align}
where $\mu$ is some scalar value and $\bm{\epsilon}$ the orthogonal velocity change, such that $\bm{\epsilon}\cdot\vec{\hat{n}}=0$. 
Substituting this into \autoref{eq:quadratic} yields the quadratic equation
\begin{equation}
\mu^2 - \left(2\vec{\hat{n}}\cdot(\vec{v}_{12}^\prime-\vec{v}_1 -\bm{\epsilon})\right)\mu + (\vec{v}_{12}^\prime-\vec{v}_1)^2 - s^2 = 0 \;,
\label{eq:quadratic2}
\end{equation}
which has a closed form solution to $\mu$ for a given $\bm{\epsilon}$.

For simplicity, we favor solutions with minimal orthogonal momentum exchanges between colliding particles and take $\bm{\epsilon}=\vec{0}$, resulting in momentum exchange along $\vec{\hat{n}}$ only.
If the quadratic equation has two real roots for $\mu$, we take the smaller one, since it ensures that the velocities $\vec{v}_1^\prime$ and $\vec{v}_2^\prime$ separate the particles, such that
\begin{equation}
\left( \vec v_2^\prime - \vec v_1^\prime \right)\cdot \vec{\hat{n}} \geq 0 \;.
\label{eq:velocity_nice_property}
\end{equation}
The other (larger) root for $\mu$ would lead to final velocities pointing towards each other, which would be an unacceptable solution for two colliding particles.

Yet, since we impose no restrictions on the time-step integration, $\vec{v}_{12}^\prime$ can theoretically be any arbitrary value and the resulting \autoref{eq:quadratic2} may have no real solution for $\mu$ with $\bm{\epsilon}=\vec{0}$.
In this case, we consider the solution that minimizes the magnitude of the orthogonal term
$\left|\bm{\epsilon}\right|$.

\begin{figure}[tb]
\centering
\begin{tabular}{cc}
\includegraphics[width=0.35\linewidth]{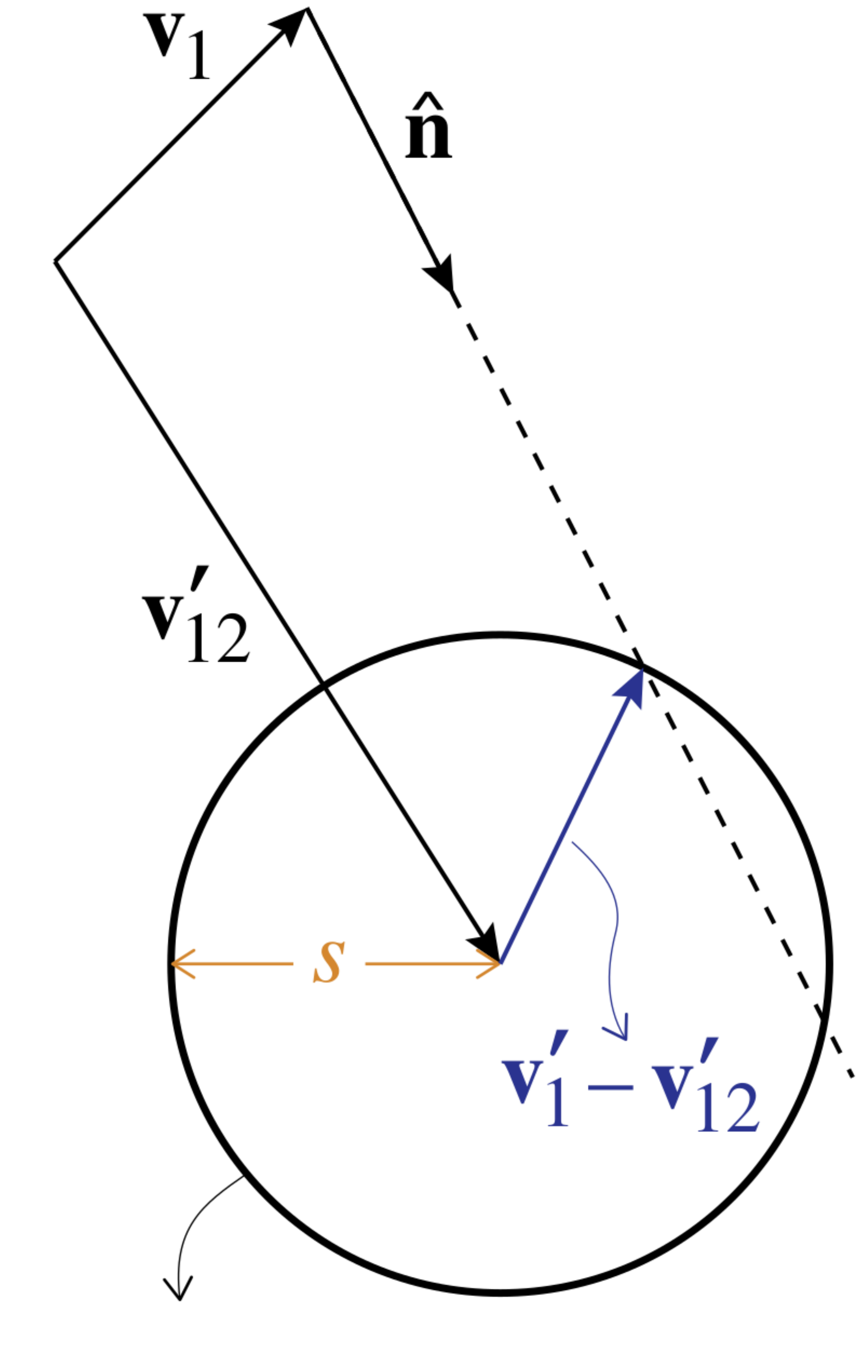}&%
\hspace{1em}%
\includegraphics[width=0.35\linewidth]{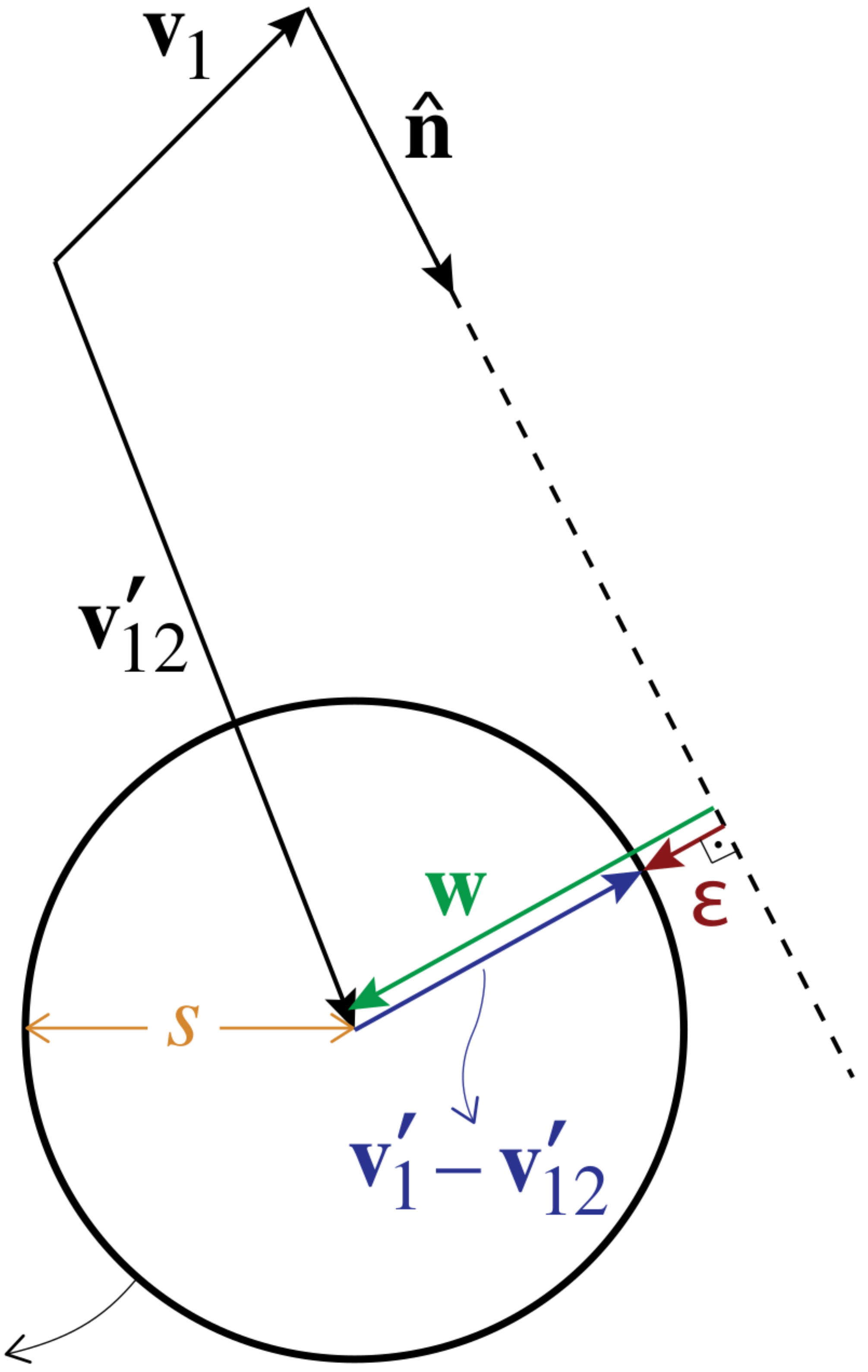}\\
(a)&(b)\\
\end{tabular}\vspace{-2.3em}\\
\hspace*{-9em}\small \emph{the sphere of valid solutions}%
\vspace{1.5em}\\
\caption{Geometric solutions to the quadratic equation in \autoref{eq:quadratic2} showing the sphere of valid solutions centered around $\vec{v}_{12}^\prime$ that conserve momentum and energy: (a)~quadratic equation has roots, and (b)~quadratic equation has no real roots.}
\label{fig:geometric_solution}
%\vspace{-2.5em}
\end{figure}

We derive a closed form solution to this minimization geometrically, as shown in \autoref{fig:geometric_solution}. Notice that 
\autoref{eq:quadratic} defines a sphere of valid solutions for $\vec{v}_1^\prime$ that conserve energy centered
at $\vec{v}_{12}^\prime$ with radius $s$. If the ray $\vec{v}_1+\mu\vec{\hat{n}}$
intersects with this sphere, we get up to two solutions for $\mu$, corresponding to the two intersection points of the ray with the sphere of valid solutions (\autoref{fig:geometric_solution}a). If there is no real root for \autoref{eq:quadratic2}, the ray does not
intersect with this sphere (\autoref{fig:geometric_solution}b). In that case, the point along the ray that minimizes
$\left|\bm{\epsilon}\right|$ is the closest point between the line $\vec{v}_1+\mu\vec{\hat{n}}$ and the sphere in \autoref{eq:quadratic}. Using this
property and solving for $\mu$ we can write
\begin{align}
\mu &= \vec{\hat{n}} \cdot \left(\vec{v}_{12}^\prime - \vec{v}_1\right) \label{eq:solution_mu}\\
\bm{\epsilon} &= \left( \left|\vec{w}\right| - s \right) \frac{\vec{w}}{\left|\vec{w}\right|}
\label{eq:solution_epsilon}
\end{align}
where $\vec{w} = \vec{v}_{12}^\prime - \vec{v}_1 - \mu \vec{\hat{n}}$.
Notice that in this solution $\bm{\epsilon}$ is perpendicular to $\vec{\hat{n}}$.

An important property of our formulation is that, in any case, the resulting velocities are guaranteed to separate the particles, as in \autoref{eq:velocity_nice_property}.
The solution to $\vec{v}_2^\prime$ can be computed from momentum conservation, such that
\begin{equation}
\vec{v}_2^\prime = \frac{m_{12}}{m_2} \vec{v}_{12}^\prime - \frac{m_1}{m_2} \vec{v}_1^\prime \;.  \nonumber
\end{equation}
Therefore, we can write
\begin{equation}
\left(\vec{v}_2^\prime - \vec{v}_1^\prime\right) \cdot \vec{\hat{n}} =
- \frac{m_{12}}{m_2} \left(\vec{v}_1^\prime - \vec{v}_{12}^\prime \right) \cdot \vec{\hat{n}} \;.  \nonumber
\end{equation}
If Equation~\ref{eq:quadratic2} has no root, ${\left(\vec{v}_1^\prime - \vec{v}_{12}^\prime\right) \cdot \vec{\hat{n}}=0}$, since ${\left(\vec{v}_1^\prime - \vec{v}_{12}^\prime\right)}$ is perpendicular to $\vec{\hat{n}}$ (Fig.~\ref{fig:geometric_solution}b). Otherwise, we pick the smaller root, which makes ${\left(\vec{v}_1^\prime - \vec{v}_{12}^\prime\right) \cdot \vec{\hat{n}}\le0}$ (Fig.~\ref{fig:geometric_solution}a). 
Therefore, Equation~\ref{eq:velocity_nice_property} is always satisfied.

Note that any solution that satisfies \autoref{eq:quadratic} conserves both
momentum and energy. The procedure described above picks a valid solution that conforms to the initial collision configuration of the merged particles. Therefore, whether \autoref{eq:quadratic2} has a real root and whether some minimal orthogonal momentum exchange between particles must be assumed has no practical consequence, since both momentum and energy are always conserved. Furthermore, we impose no restrictions on the time integration, so we have no control over the resulting $\vec{v}_{12}^\prime$ value, which may require a non-zero orthogonal term to preserve the initial particle configuration. 
Thus, it would be unreasonable to always expect a solution with ${\left|\bm{\epsilon}\right|=0}$.
Nonetheless, in any case we pick a consistent solution that minimizes the orthogonal term.

In our tests, we observed that when colliding particles are free (i.e. not connected to other particles) or have weak connections to other particles, we often find a solution for $\mu$ with ${\left|\bm{\epsilon}\right|=0}$. On the other
hand, if the motion of one particle is restricted (such as when the object is pushed
against the simulation boundary or particles are attached to immobile points), which in turn restricts the motion of the meta-particle, a high percentage of
the solutions contain a non-zero orthogonal term (i.e. ${\left|\bm{\epsilon}\right|>0}$).

If more than two particles collide within the same time-step, we recursively merge these particles into larger meta-particles.
After computing the updated momentum for the final meta-particle, we recursively split them in the inverse order of merging.
We have empirically verified that the order of merging does not affect the outcome of the final particle velocities after splitting.
Thus, recursive merging can begin with any pair of particles.

Note that by using merging-and-splitting, we assume that collisions between the particles are not instantaneous, but instead they can take as long as one time-step. 
This allows the neighboring particles to interact with the colliding particles and influence the outcome of the collision event. 
Collisions that require even longer interactions than one time-step, such as rest-in-contact situations, are handled in the subsequent time-steps.

\section{Coupling Different Integrators}

Our merging-and-splitting approach can also be used for coupling different particle-based simulation systems using different integrators. We achieve this by introducing collision-based interactions between particles of different simulation systems.

When the particles of the two systems come into contact, we merge them into meta-particles, as explained in the previous section. We include these meta-particles in \emph{both} simulation systems. However, since we impose no restrictions on the integrators used for the two systems, the two integrators are likely to produce two different results for each meta-particle. Since we cannot allow two different solutions for one meta-particle, we must combine the two results into one solution. We do so by producing a \emph{synchronized velocity} for each meta-particle using the two solutions. The meta-particle positions are updated using the synchronized velocities to ensure that the two systems produce consistent results.
Combining the solutions of two different integrators is not a new concept \cite{Farhat:1990,Olson:2014,Newmark:1959}. What is different about our approach is that we formulate our synchronization similar to the merging operations we use for generating meta-particles.

Let $\vec{v}_{12}^{A}$ and $\vec{v}_{12}^{B}$ be the velocities of a meta-particle generated by the two systems at the end of the time-step integration. We calculate the synchronized velocity $\vec{v}_{12}^\prime$ using a weighted average of the two solutions. Consider that before merging the particles with masses $m_1$ and $m_2$ originally belong to the first and the second simulation systems, respectively. We compute the synchronized velocity $\vec{v}_{12}^\prime$ using
\begin{equation}
\vec{v}_{12}^\prime = \left( m_1 \vec{v}_{12}^{A} + m_2 \vec{v}_{12}^{B} \right) \Big/ m_{12} \;.
\label{eq:meta_merge}
\end{equation}
This provides a weighted average of the meta-particle momenta generated by the two integrators, using the mass percentages of the two particle types in the meta-particle as weights. Yet, similar to merging, this operation leads to energy dissipation. To avoid this, we also consider the weighted average of the kinetic energy
\begin{equation}
\bar{E}_k = \frac{1}{2} m_{1} \left(\vec{v}_{12}^{A}\right)^2 + \frac{1}{2} m_{2} \left(\vec{v}_{12}^{B}\right)^2 \;.
\end{equation}
Since \autoref{eq:meta_merge} cannot preserve all of this kinetic energy, we add ${\Delta E_k}$, the energy lost in \autoref{eq:meta_merge}, to the virtual bond of the meta-particle, using
\begin{equation}
\Delta \bar{E}_k = \frac{m_1 m_2}{2 m_{12}} \left( \vec{v}_{12}^{A} - \vec{v}_{12}^{B} \right)^2 \;.
\label{eq:meta_energy_loss}
\end{equation}
During splitting, we multiply this energy with another user-defined parameter, ${\beta \in [0,1]}$, which determines the percentage of this energy that should be preserved, and add it to the left side of \autoref{eq:energy_conservation}. This replaces \autoref{eq:s} with
\begin{equation}
s^2 = \frac{2\left(\alpha \Delta E + \beta \Delta \bar{E}_k\right)}{m_{12}\left(m_1/m_2\right)} \;.
\label{eq:s2}
\end{equation}

The synchronized velocities produced by these operations ensure that we have a single solution for each meta-particle.
Coupling different simulation systems using merging-and-splitting has the obvious advantage that there is no need to track the interface between the two systems. Furthermore, it allows coupling particle-based simulation systems that are designed for different material types and material behavior for handling interesting simulation scenarios that are beyond the capabilities of current unified simulation systems.

\begin{figure*}[tb]
  % For box color around images:
  \fboxsep=0pt%padding thickness
  \fboxrule=1pt%border thickness
  \newcommand{\figurewidth}{0.13\linewidth}
  \setlength{\tabcolsep}{0.1em}
  \renewcommand{\arraystretch}{1.5}
  \centering
\hspace*{-0.013\linewidth}%
\resizebox{1.013\linewidth}{!}{%
\begin{tabular}{@{}cccccccc@{}}
      \multicolumn{3}{c}{Penalty Force}                                                           &
      \multicolumn{3}{c}{SPH-based Force}                                                         &
      Impulse-based                                                                               & \textbf{Merging-and-}                                                                            \\[-1.5em]
      \multicolumn{3}{c}{\rule{0.36\linewidth}{0.4pt}}                                            &                                                                                                  %
      \multicolumn{3}{c}{\rule{0.36\linewidth}{0.4pt}}                                                                                                                                               \\[-1.0em]
      weak                                                                                        & medium                & strong & weak & medium & strong & Collisions & \textbf{Splitting (Ours)} \\
        \includegraphics[trim=240 200 240 0,clip,width=\figurewidth]{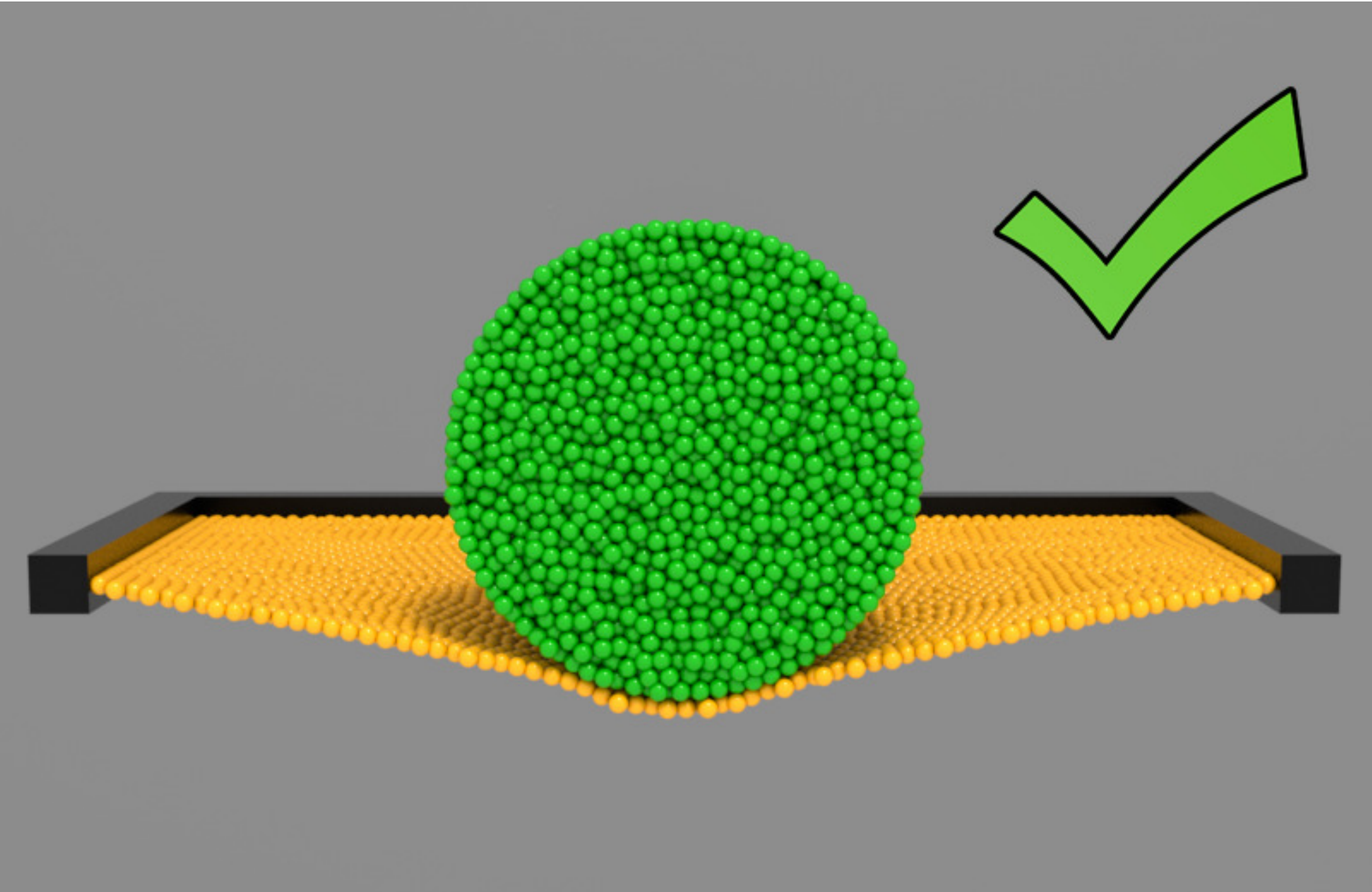} &
      \includegraphics[trim=240 200 240 0,clip,width=\figurewidth]{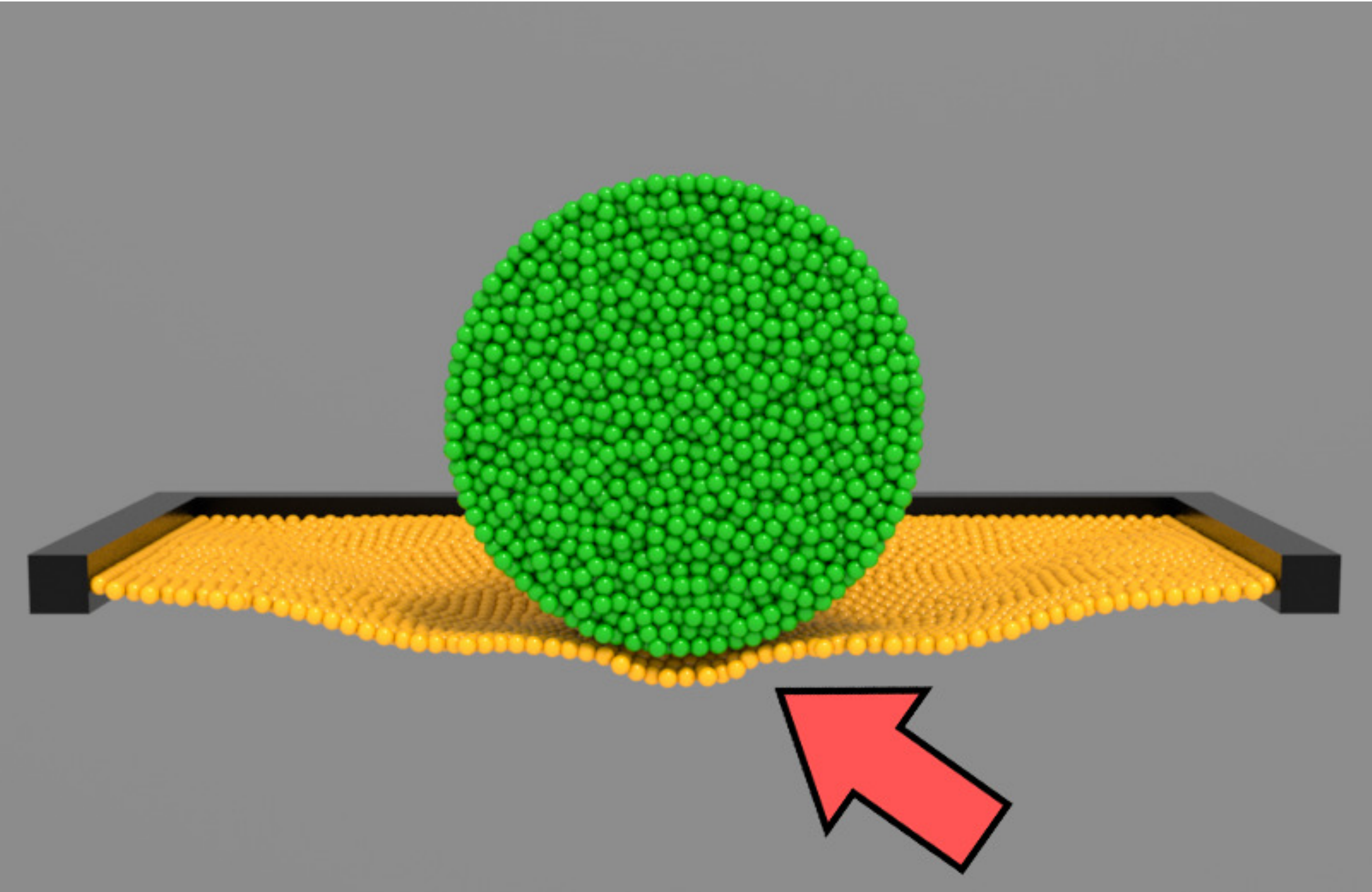}   &
      \includegraphics[trim=240 200 240 0,clip,width=\figurewidth]{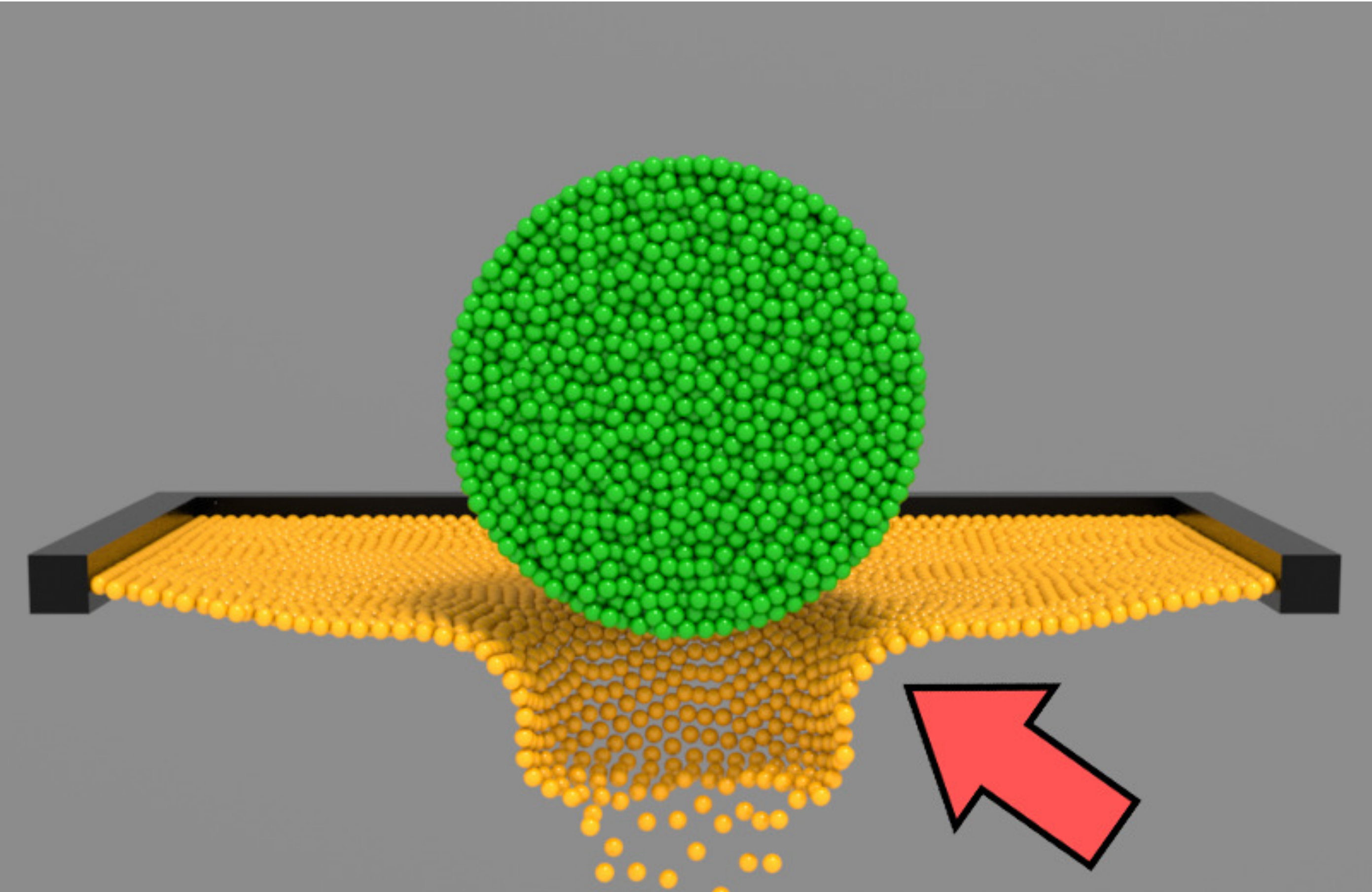}  &
      \includegraphics[trim=240 200 240 0,clip,width=\figurewidth]{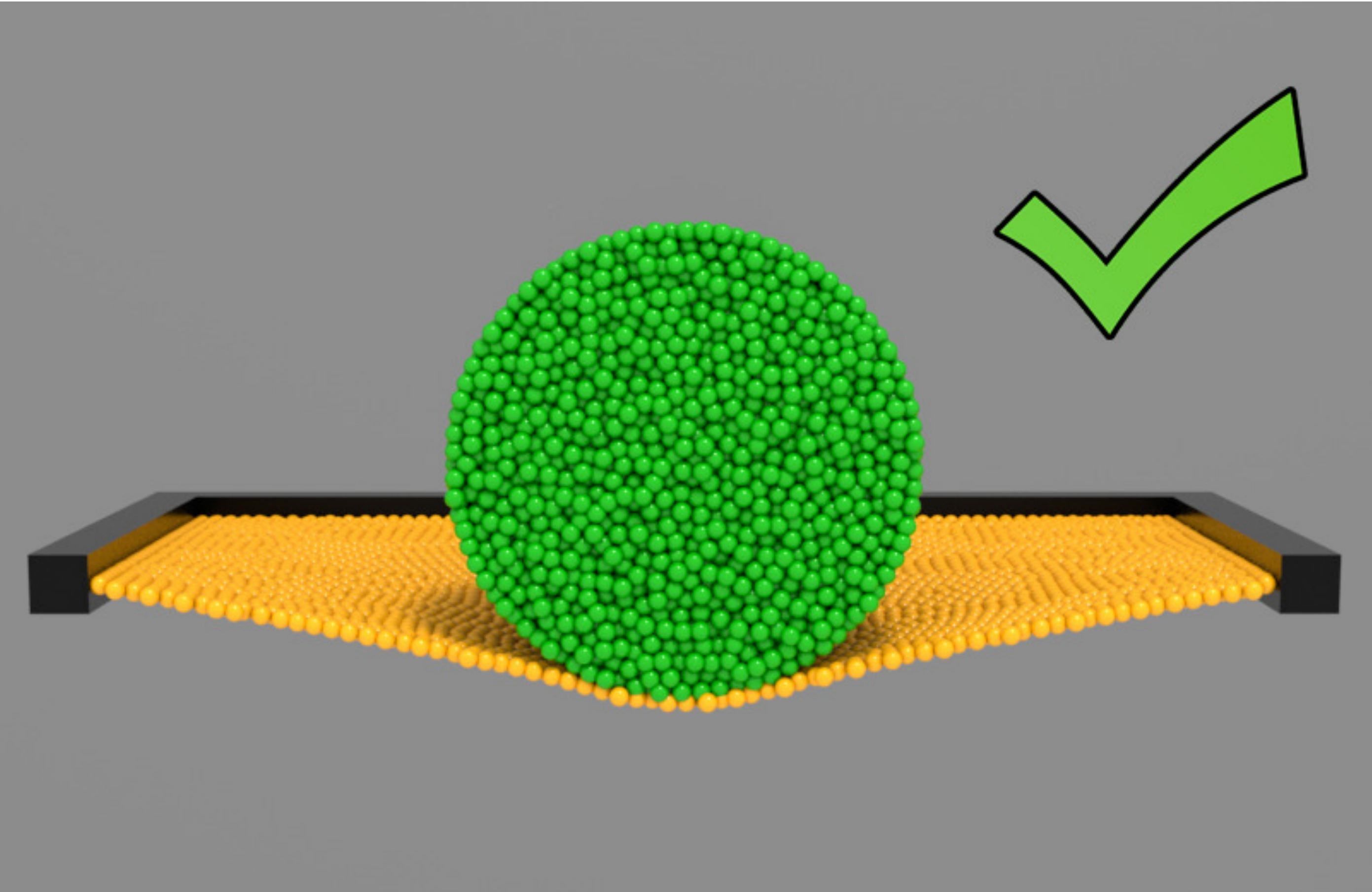}  &
      \includegraphics[trim=240 200 240 0,clip,width=\figurewidth]{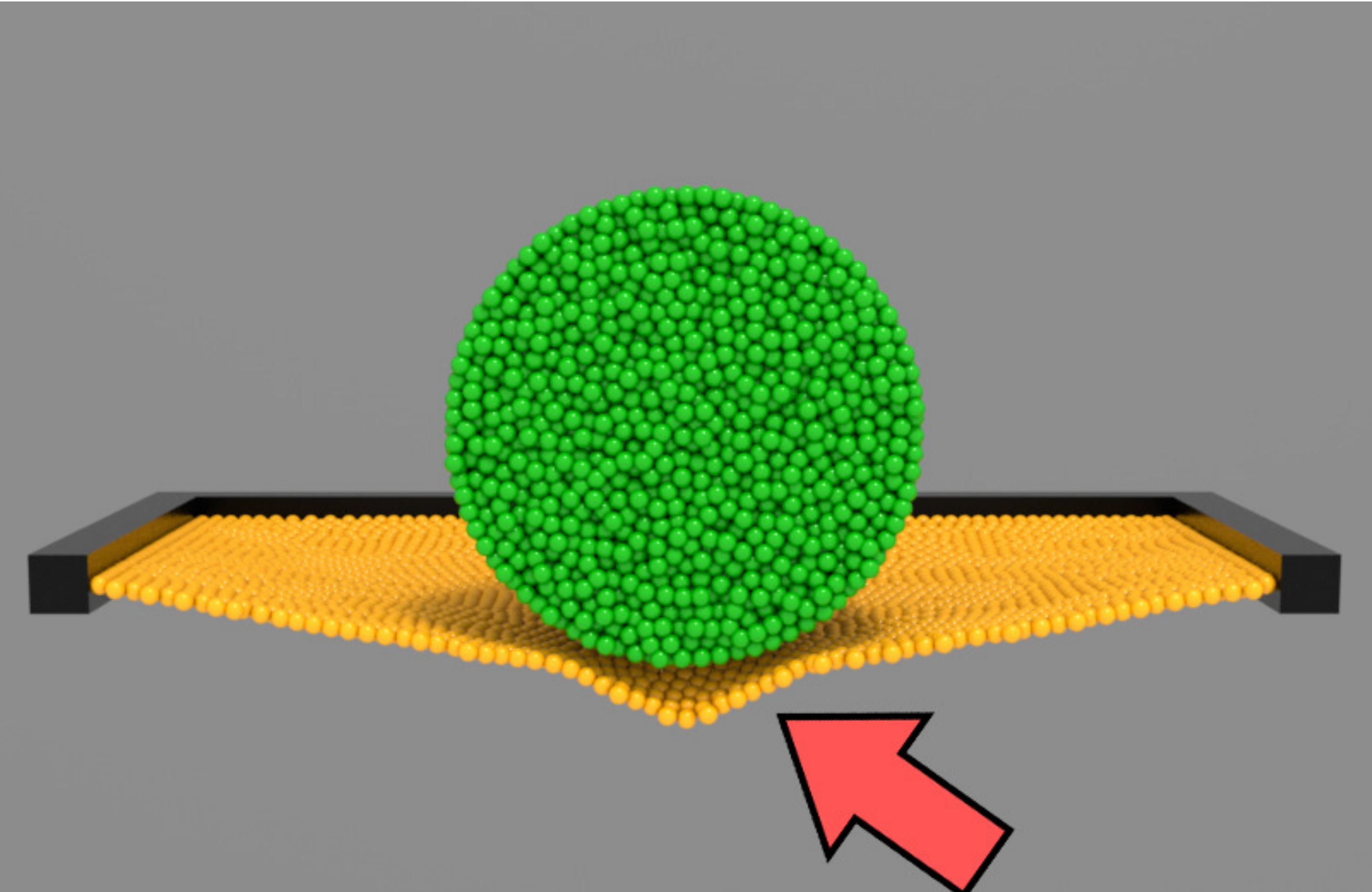}  &
      \includegraphics[trim=240 200 240 0,clip,width=\figurewidth]{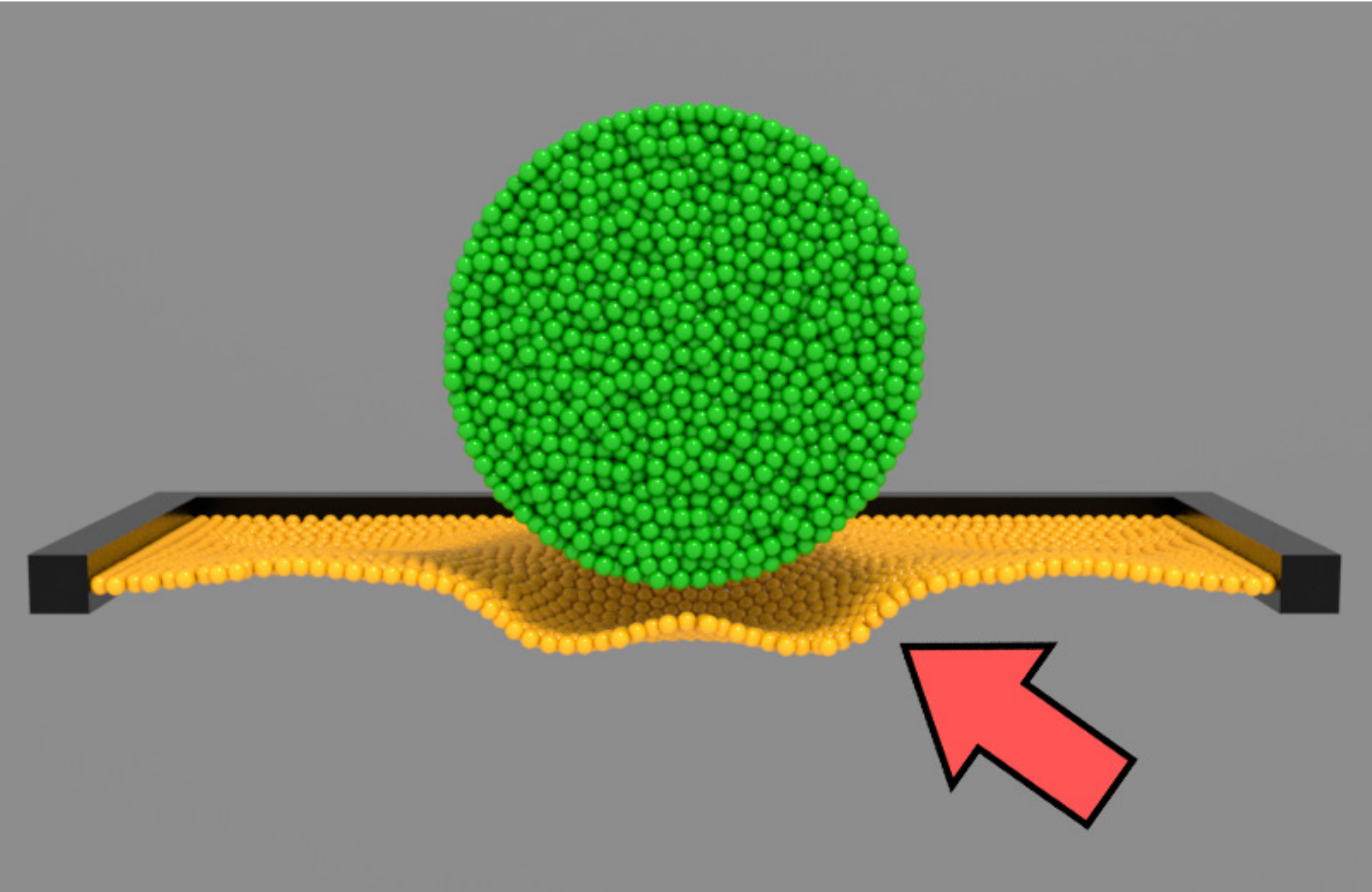}  &
      \includegraphics[trim=240 200 240 0,clip,width=\figurewidth]{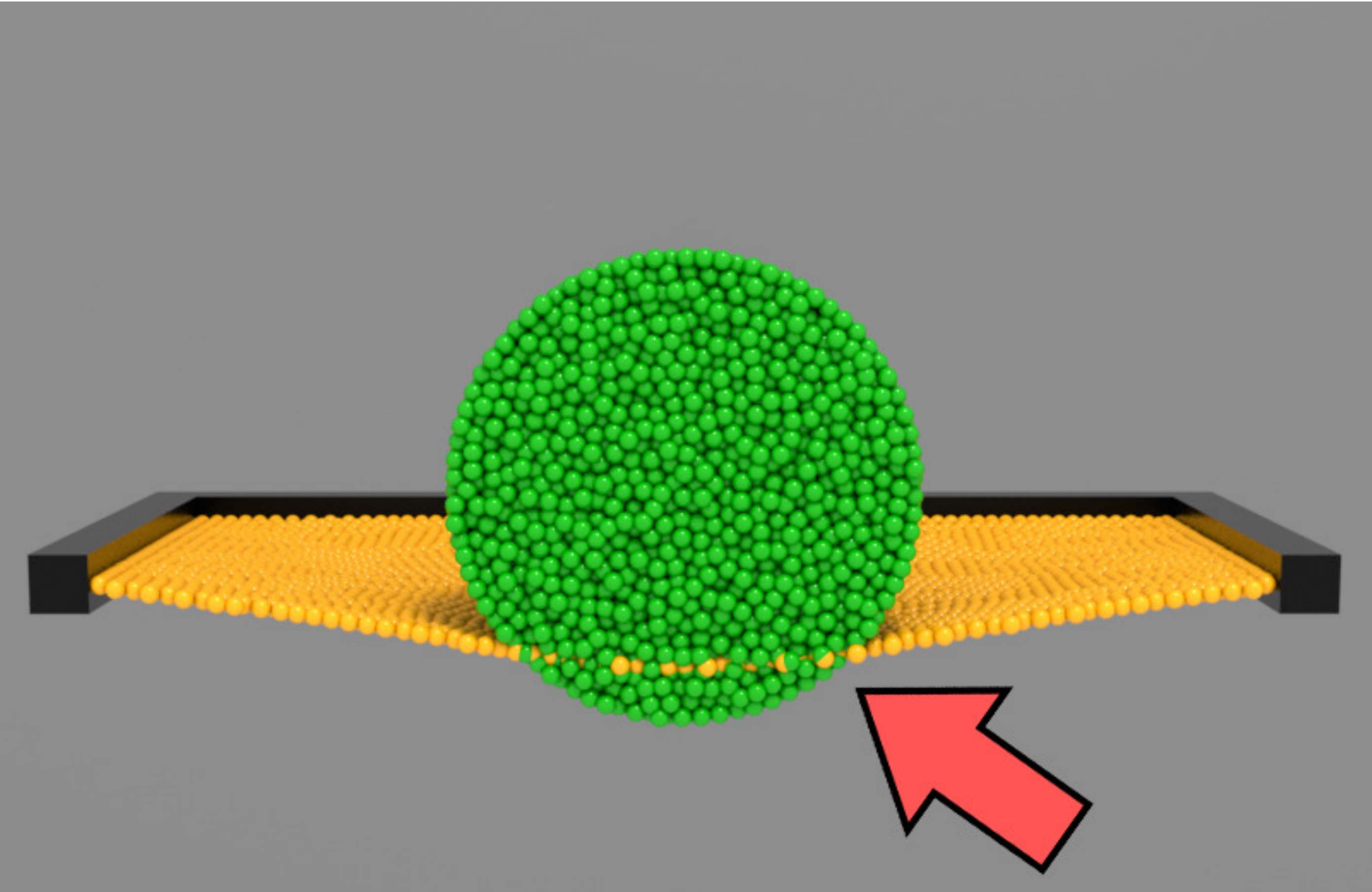}       &
      \includegraphics[trim=240 200 240 0,clip,width=\figurewidth]{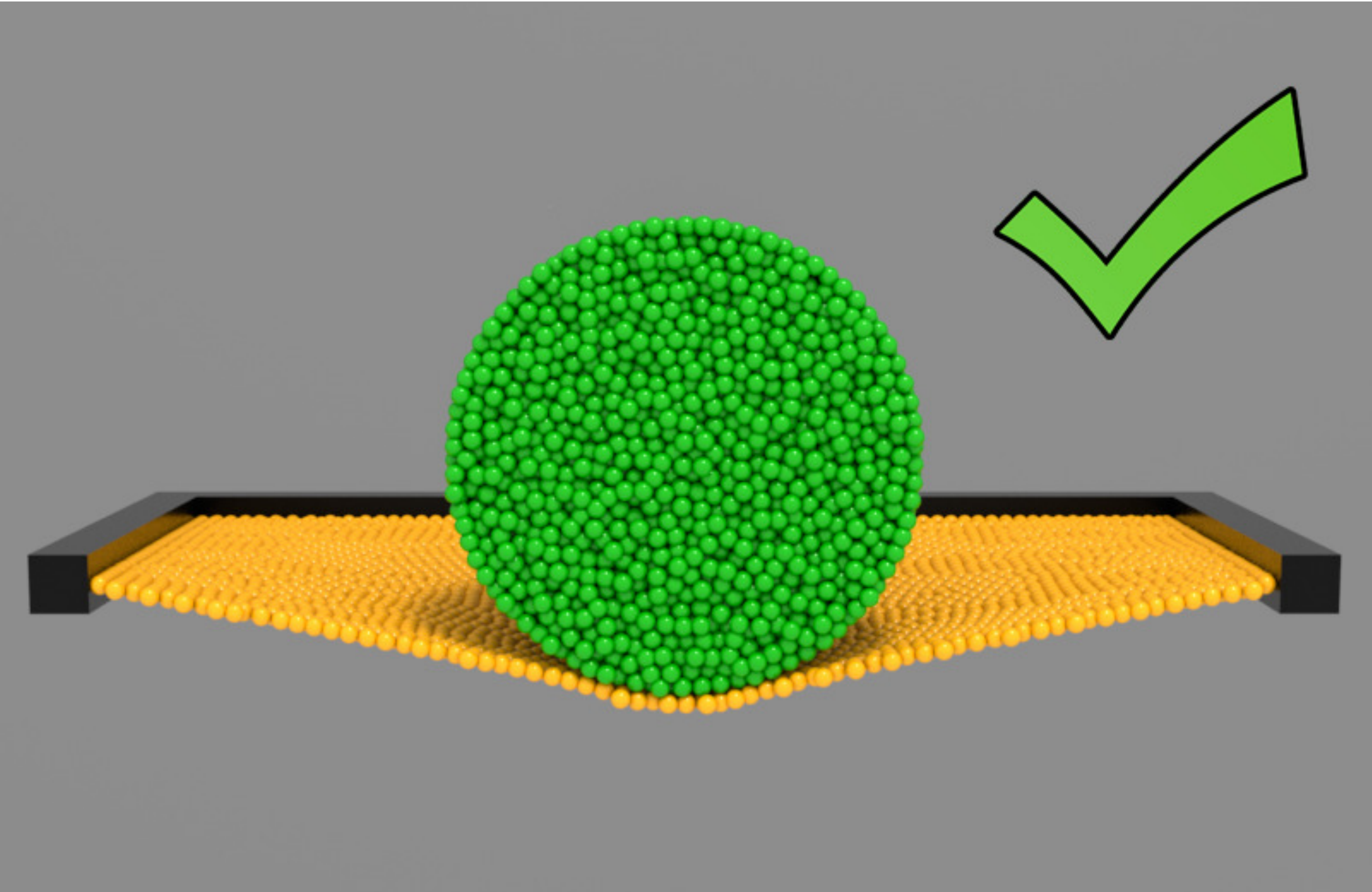}
      \vspace{-0.2em}                                                                                                                                                                                \\
      \includegraphics[trim=240 0 240 200,clip,width=\figurewidth]{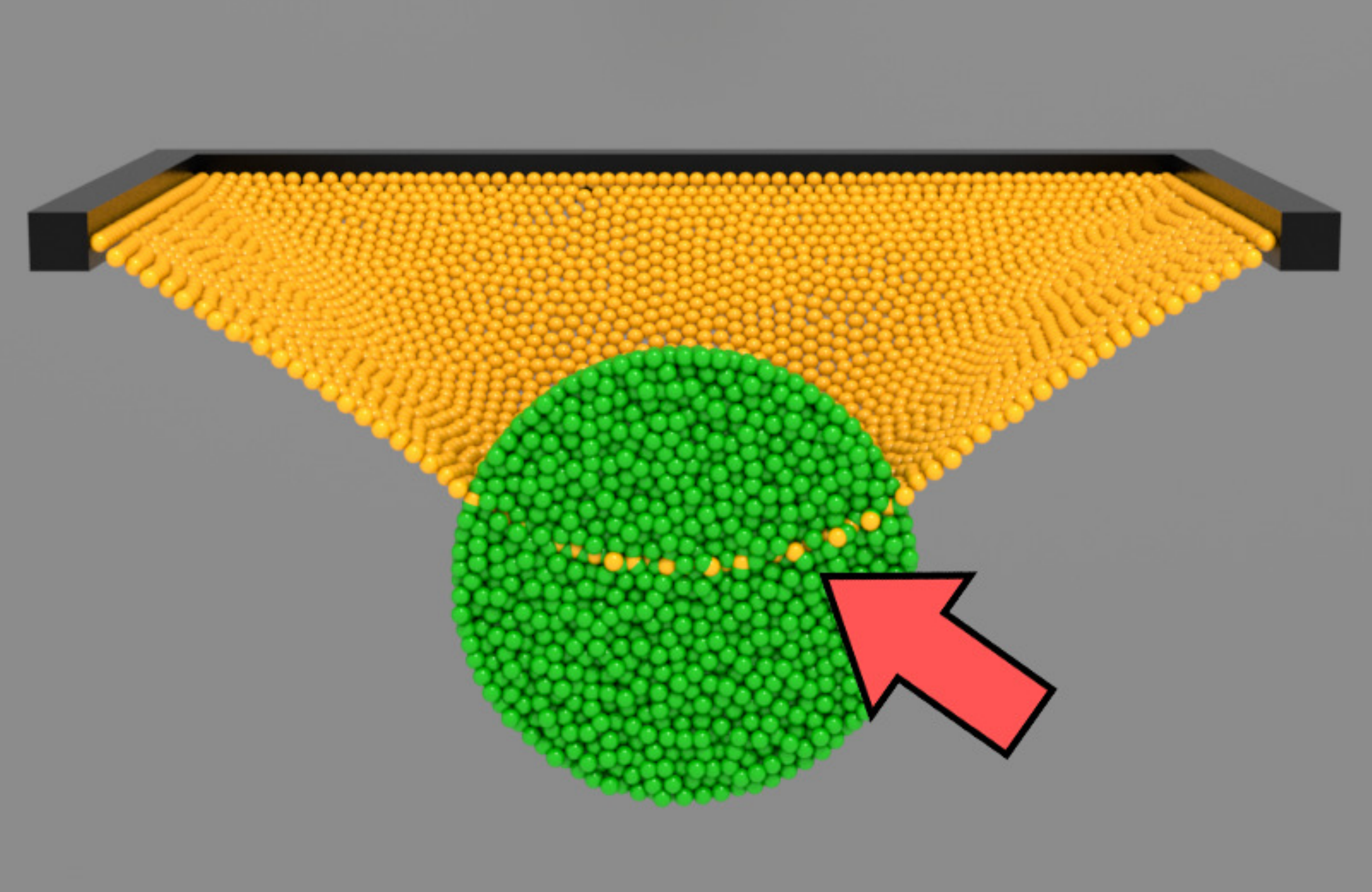}  &
      \includegraphics[trim=240 0 240 200,clip,width=\figurewidth]{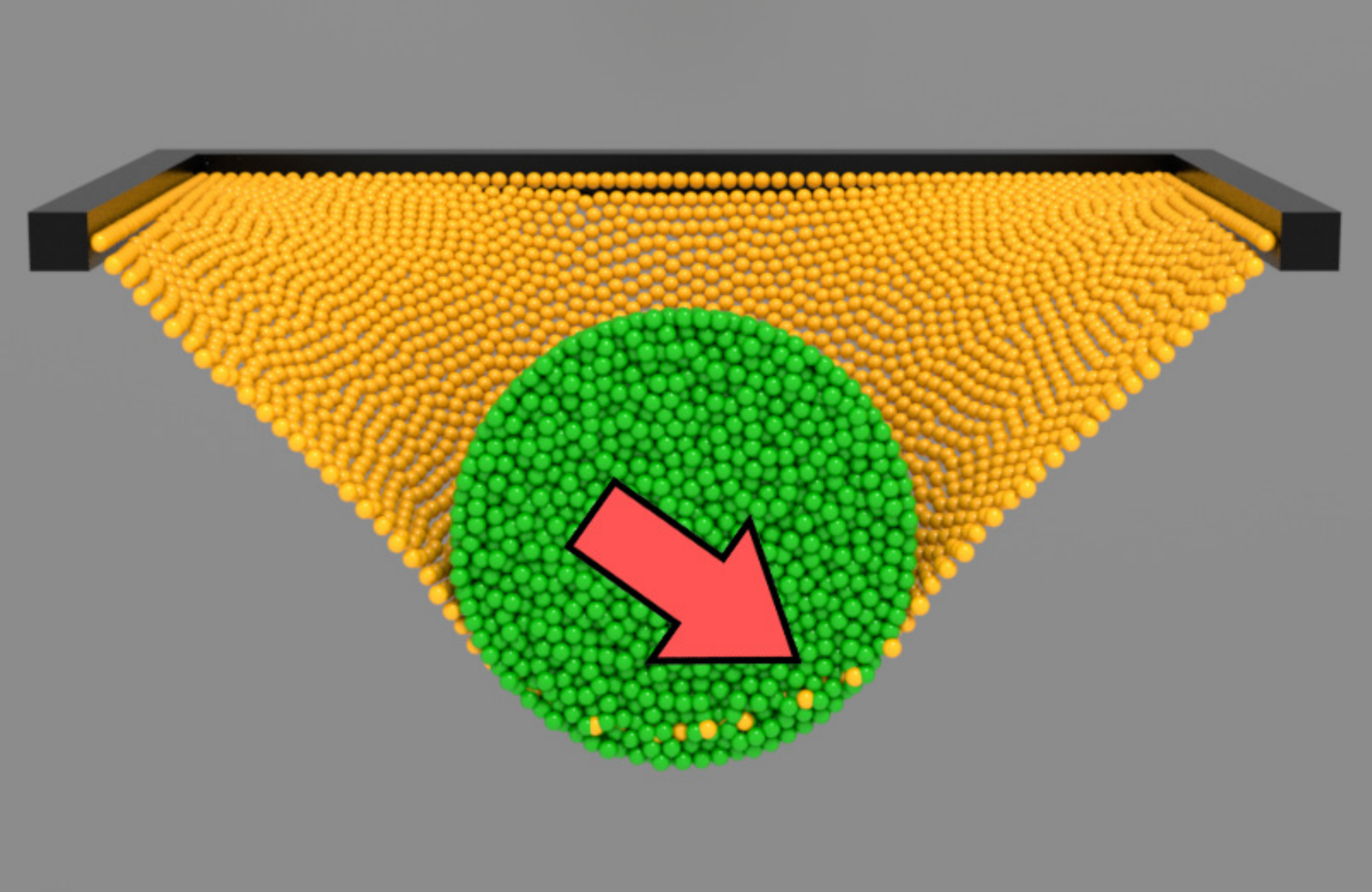}  &
      \includegraphics[trim=240 0 240 200,clip,width=\figurewidth]{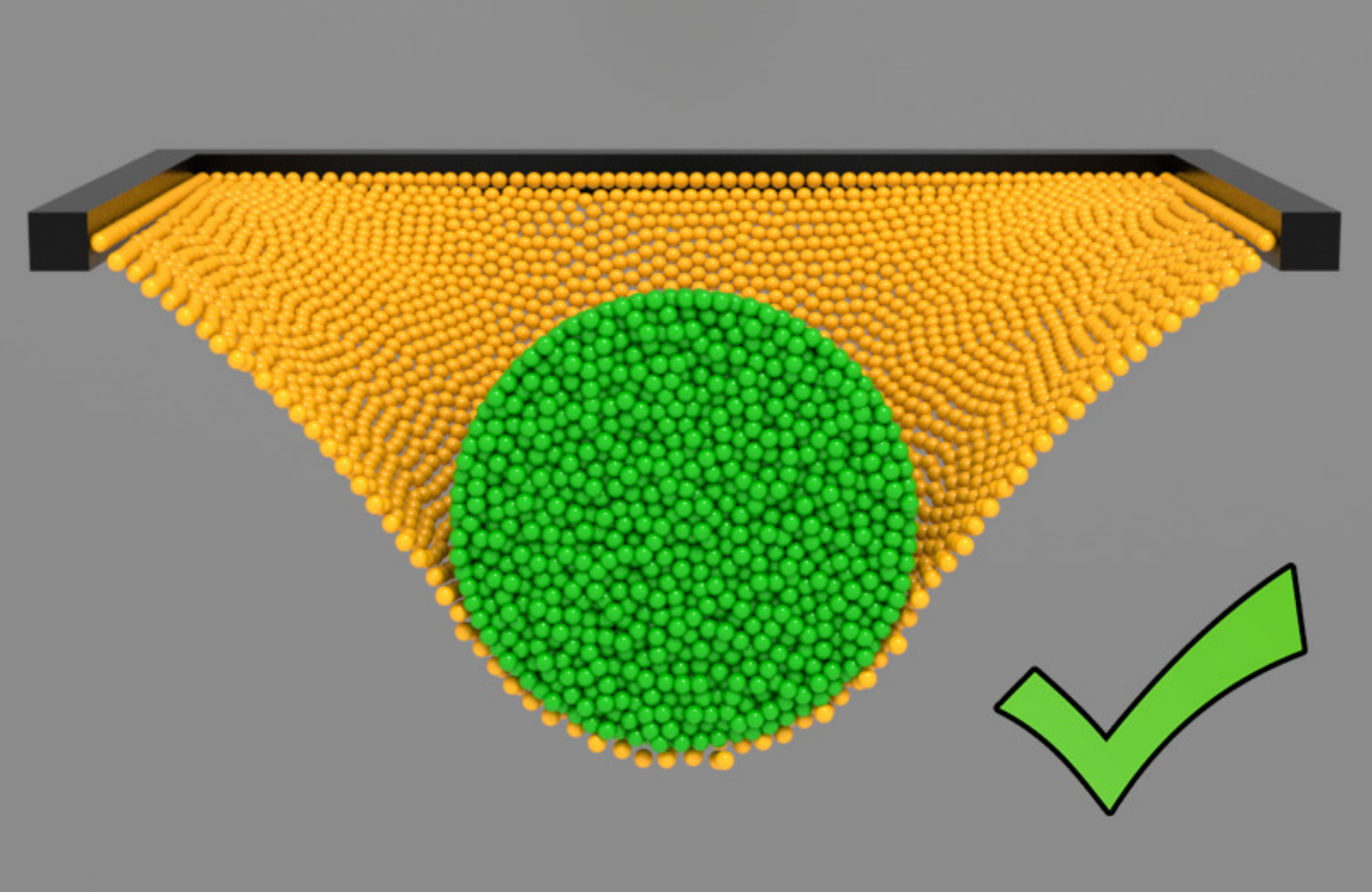} &
      \includegraphics[trim=240 0 240 200,clip,width=\figurewidth]{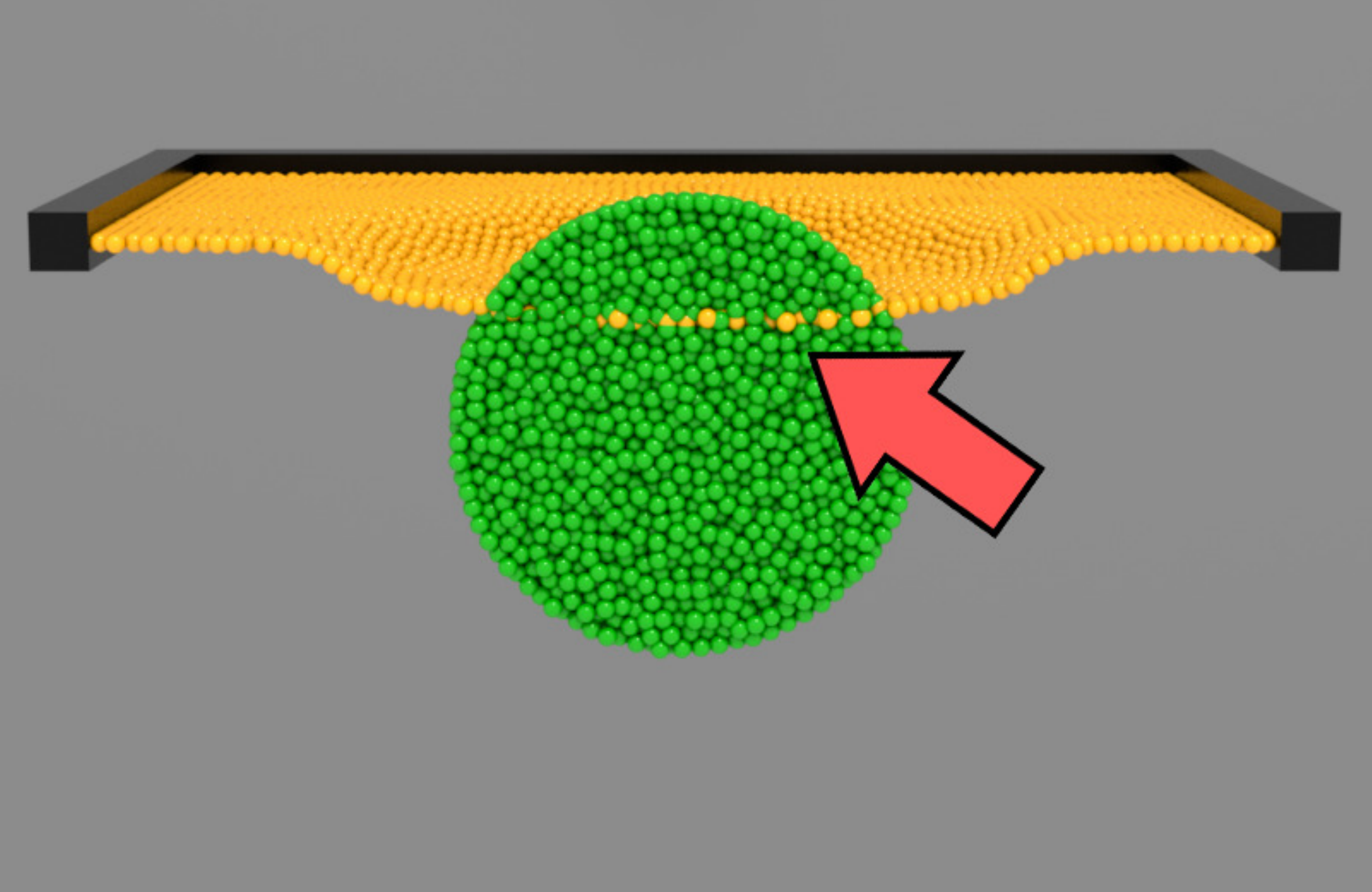} &
      \includegraphics[trim=240 0 240 200,clip,width=\figurewidth]{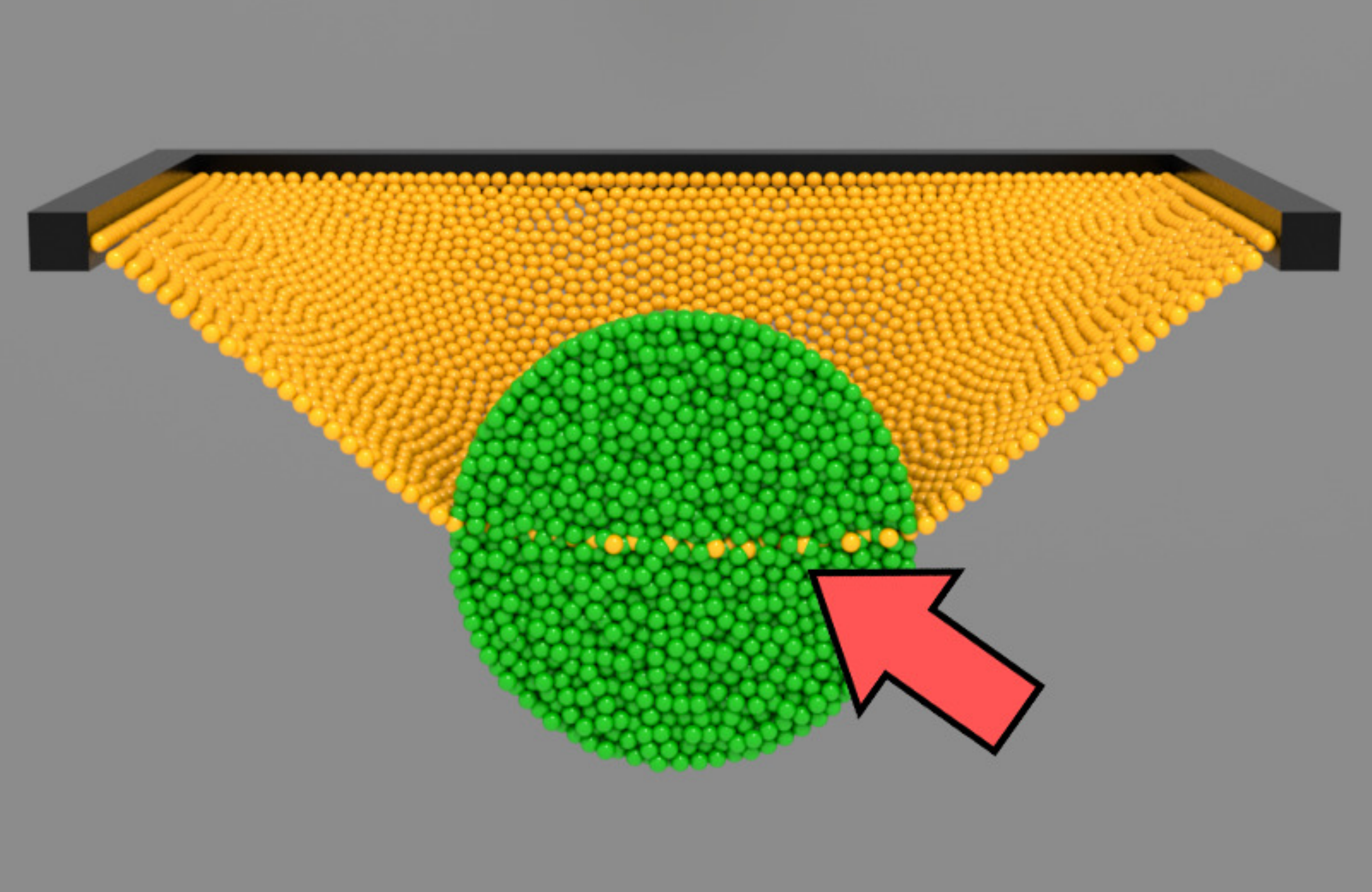} &
      \includegraphics[trim=240 0 240 200,clip,width=\figurewidth]{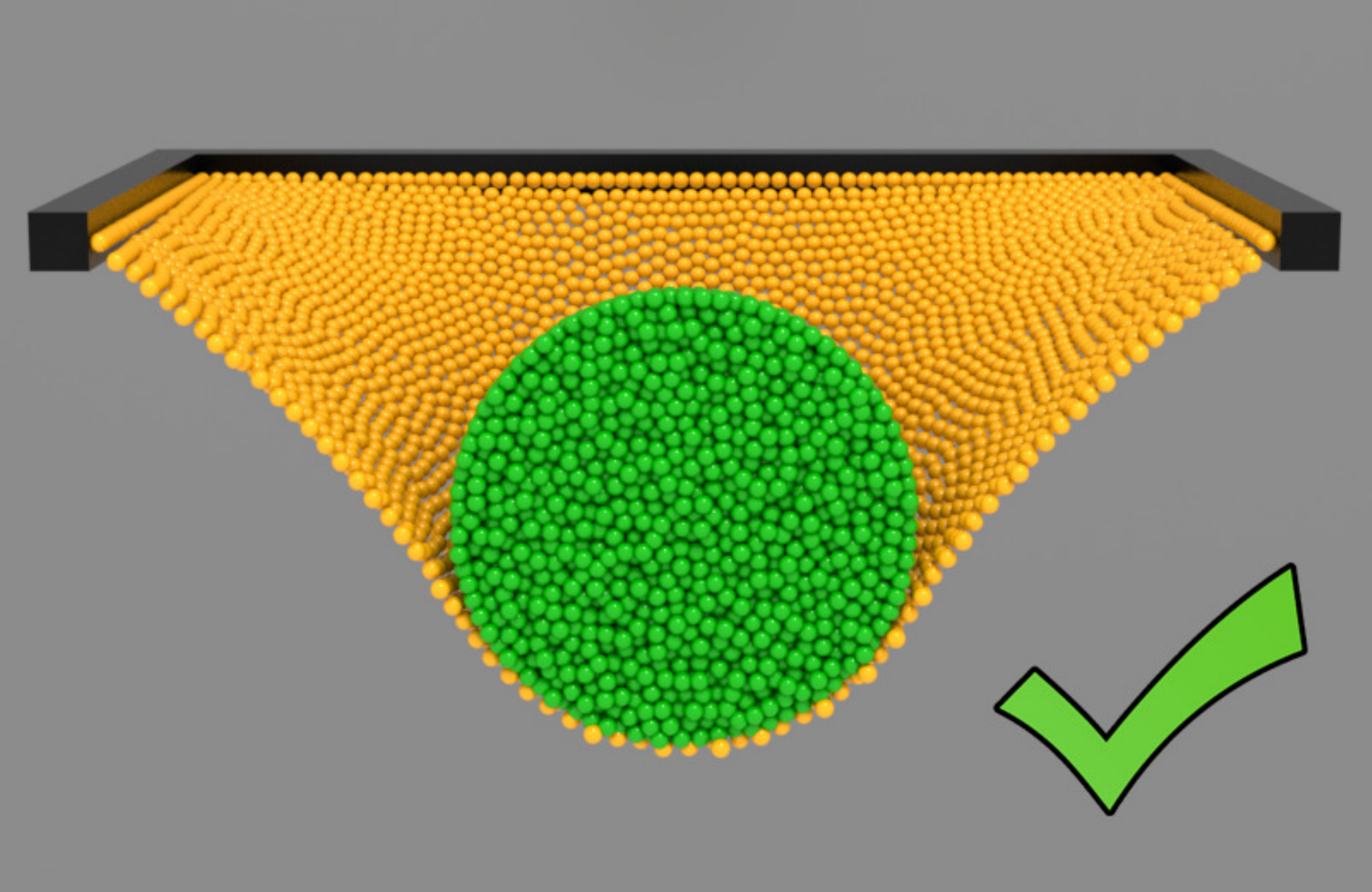} &
      \includegraphics[trim=240 0 240 200,clip,width=\figurewidth]{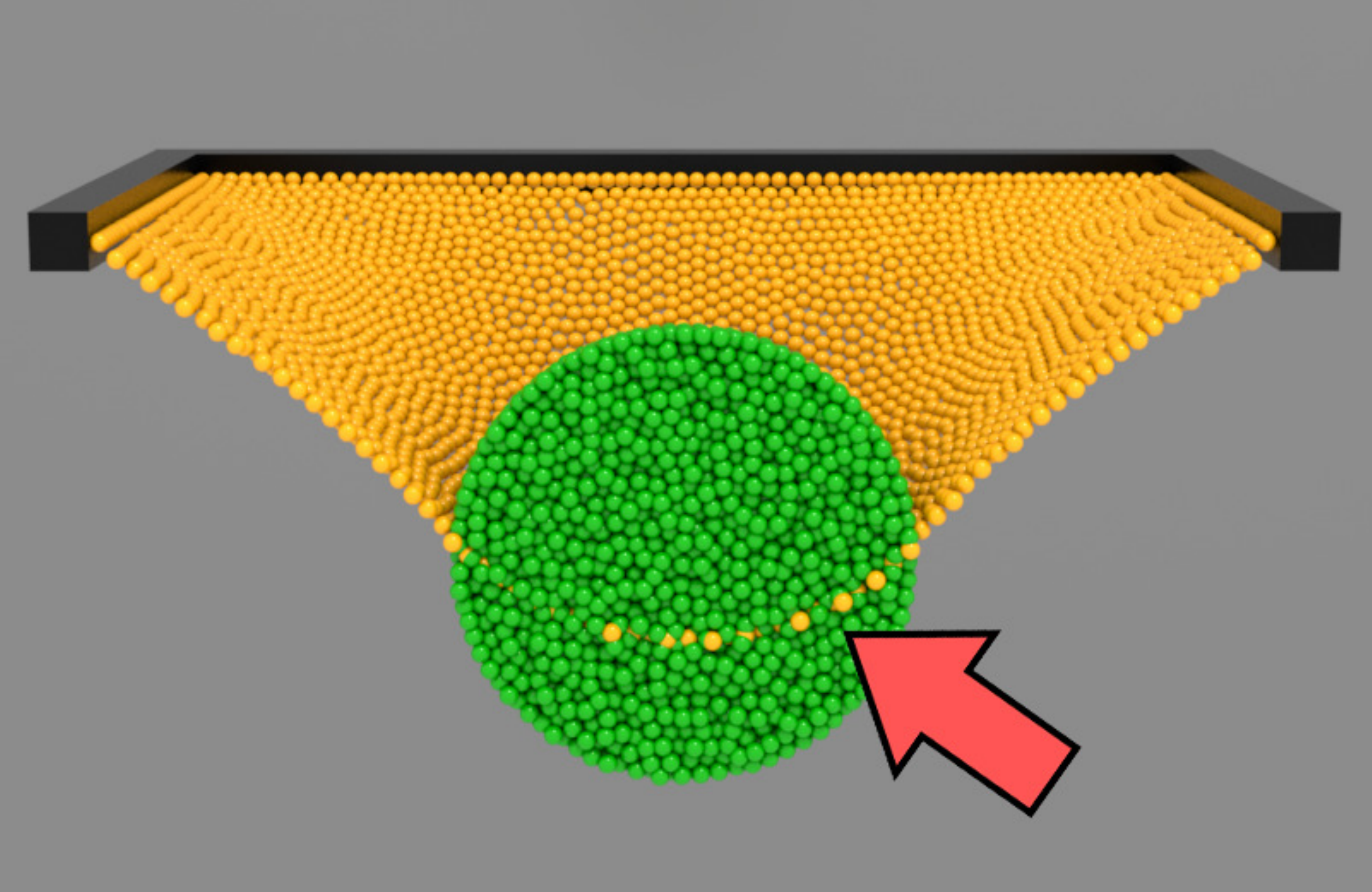}      &
      \includegraphics[trim=240 0 240 200,clip,width=\figurewidth]{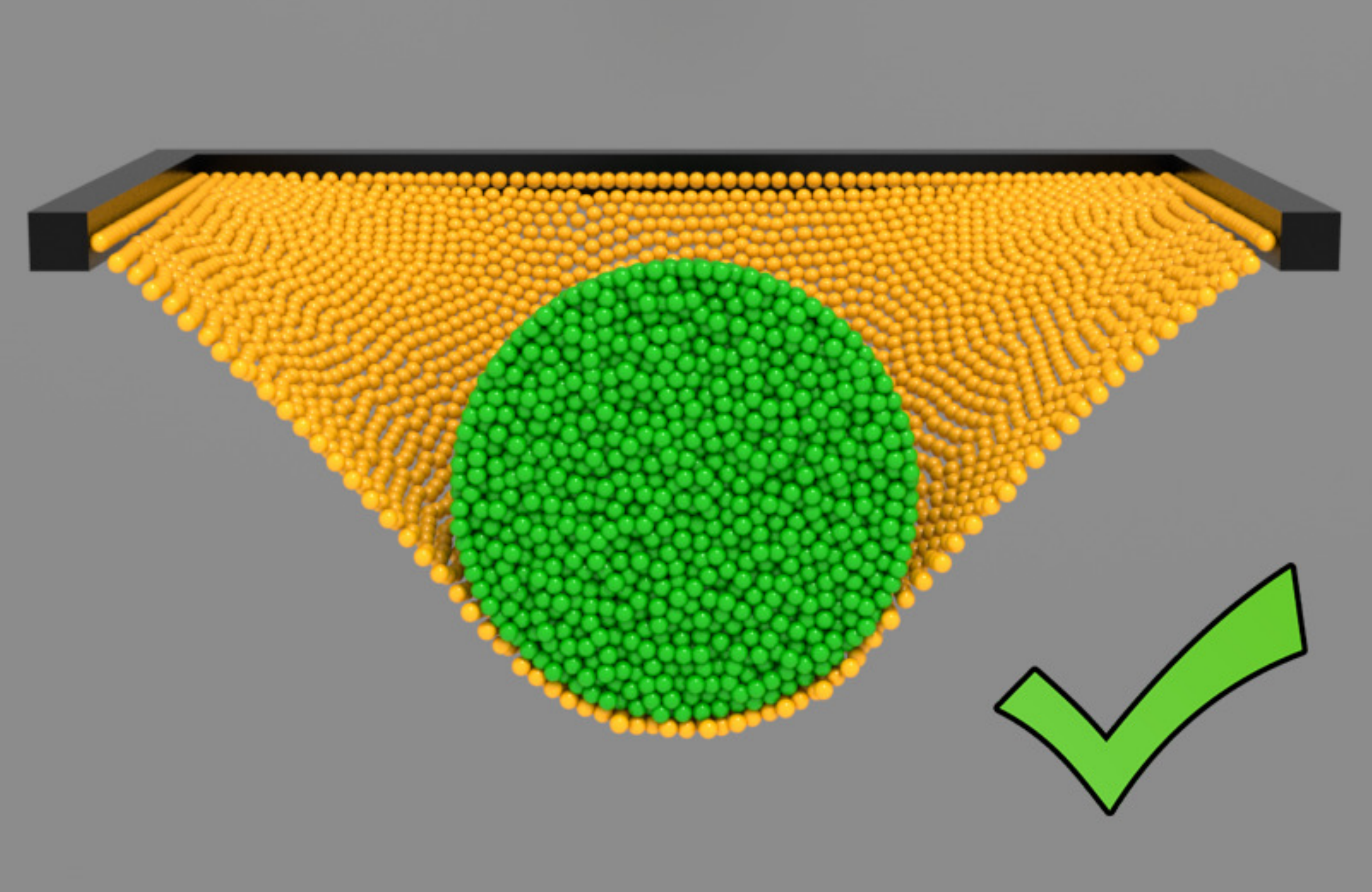}\vspace{-0.2em}
      \\
    \end{tabular}}
  \caption{\textbf{Comparison of different methods for handling solid-solid collisions:} cross-section views of simulations including a soft elastic cloth hit by a solid ball in free fall. The bottom row shows the same simulation with the solid ball falling from a higher initial position (i.e. faster impact). All particles have identical mass and weak/medium/strong penalty and SPH-based forces differ by an order of magnitude.
  %Notice that 
  Only our merging-and-splitting approach prevents penetrations without introducing instability for both cases.
  }
  \label{fig:comparison_sphere_cloth_wall}
\end{figure*}

\makeatletter
\define@key{Gin}{mycrops}[]{\setkeys{Gin}{trim=390 200 390 200,clip}}
\define@key{Gin}{mycropsNew}[]{\setkeys{Gin}{trim=280 150 280 150,clip}}
\makeatother
\begin{figure*}[tb]
  \newcommand{\figurewidth}{0.164\textwidth}
  \setlength{\tabcolsep}{0.1em}
  \renewcommand{\arraystretch}{1.5}
  \centering
  \begin{tabular}{cccccc}
    \multicolumn{2}{c}{Penalty Force}                                                                   &
    \multicolumn{2}{c}{SPH-based Force}                                                                 &
    Impulse-based                                                                                       & \textbf{Merging-and-}                                                          \\[-1.5em]
    \multicolumn{2}{c}{\rule{0.3\linewidth}{0.4pt}}                                                     &                                                                                %
    \multicolumn{2}{c}{\rule{0.3\linewidth}{0.4pt}}                                                                                                                                      \\[-1.0em]
    weak                                                                                                & strong                & weak & strong & Collisions & \textbf{Splitting (Ours)} \\
%            \\
    \includegraphics[mycropsNew,width=\figurewidth]{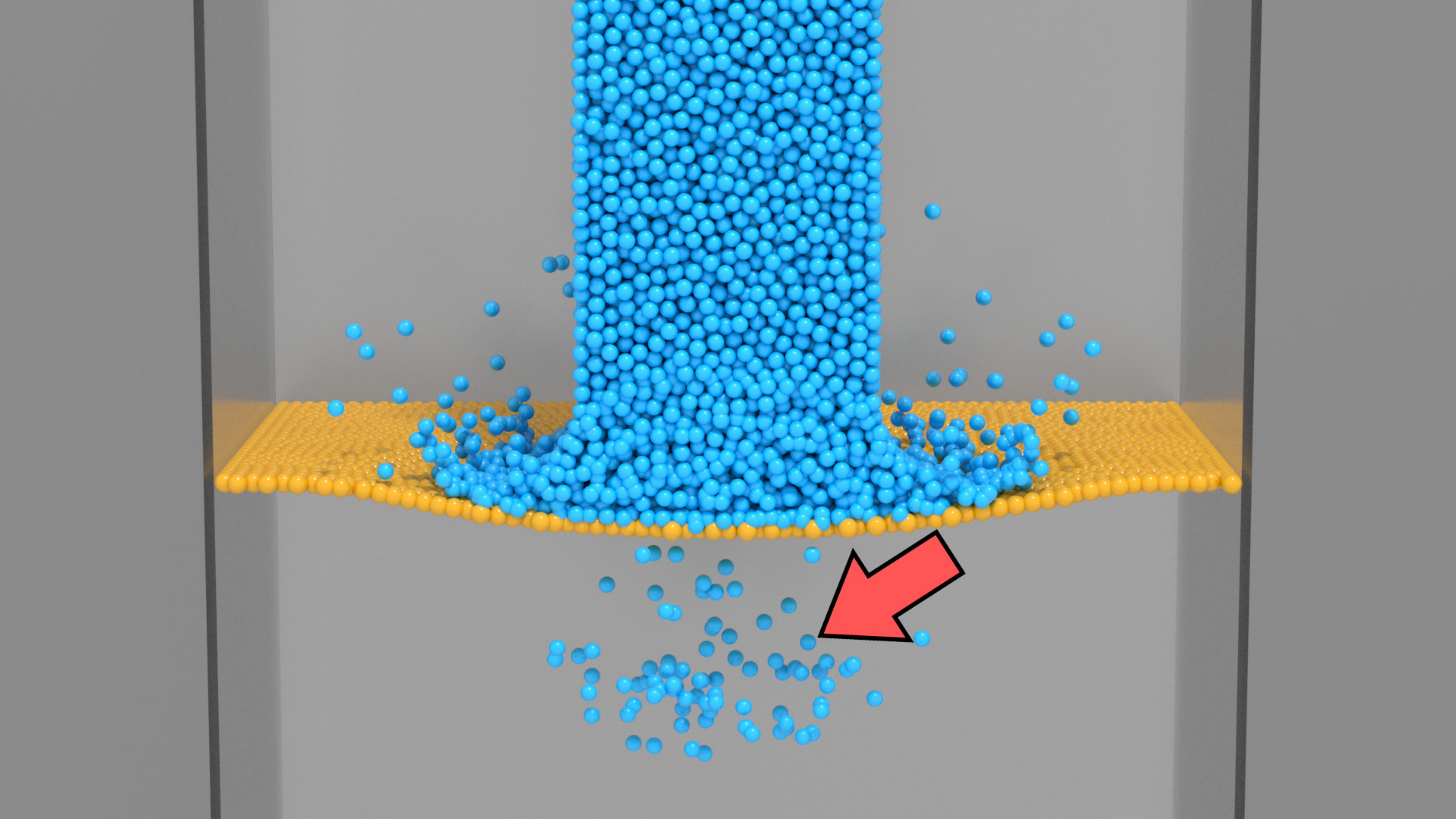}    &
    \includegraphics[mycropsNew,width=\figurewidth]{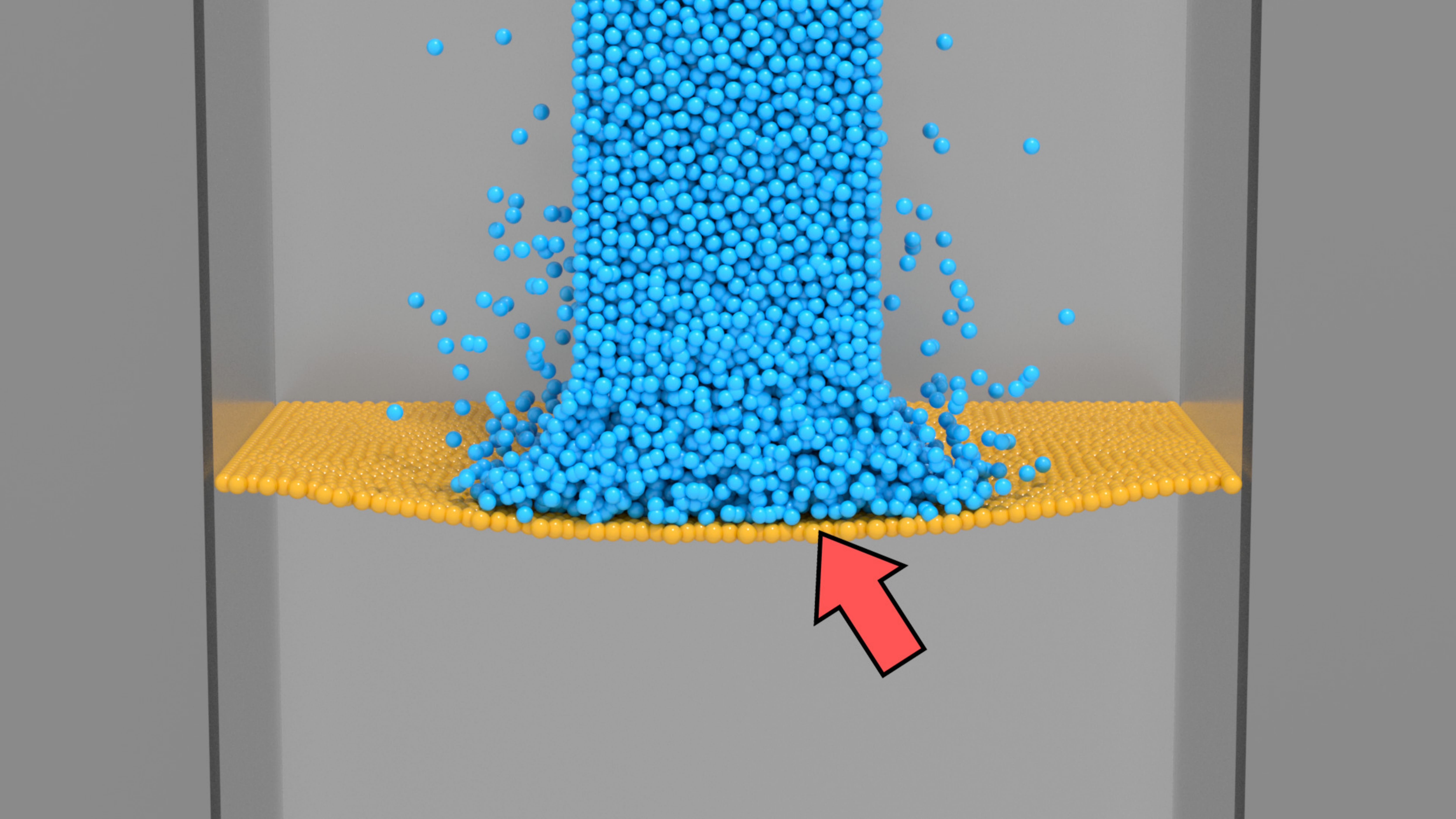}  &
    \includegraphics[mycropsNew,width=\figurewidth]{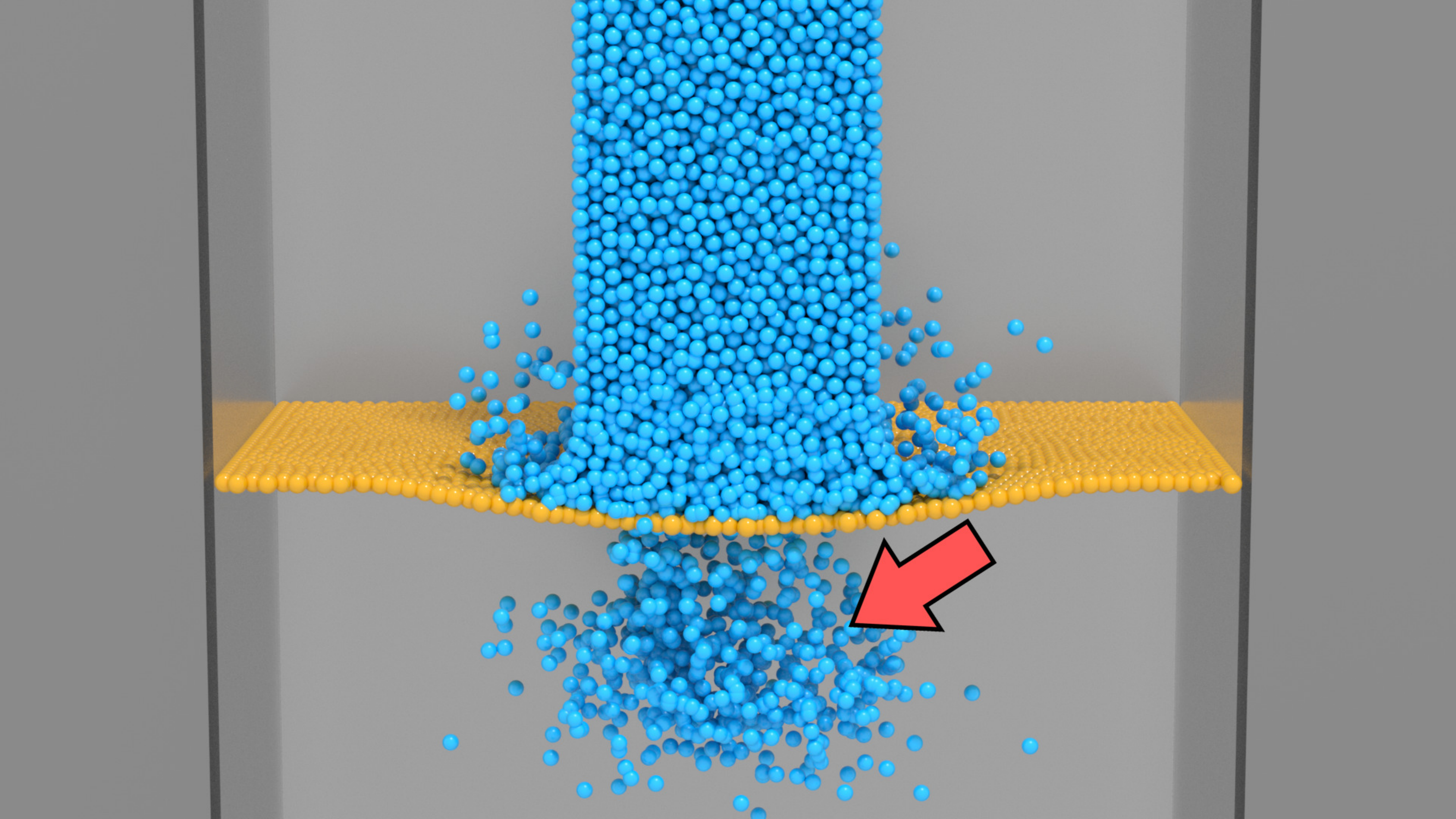}   &
    \includegraphics[mycropsNew,width=\figurewidth]{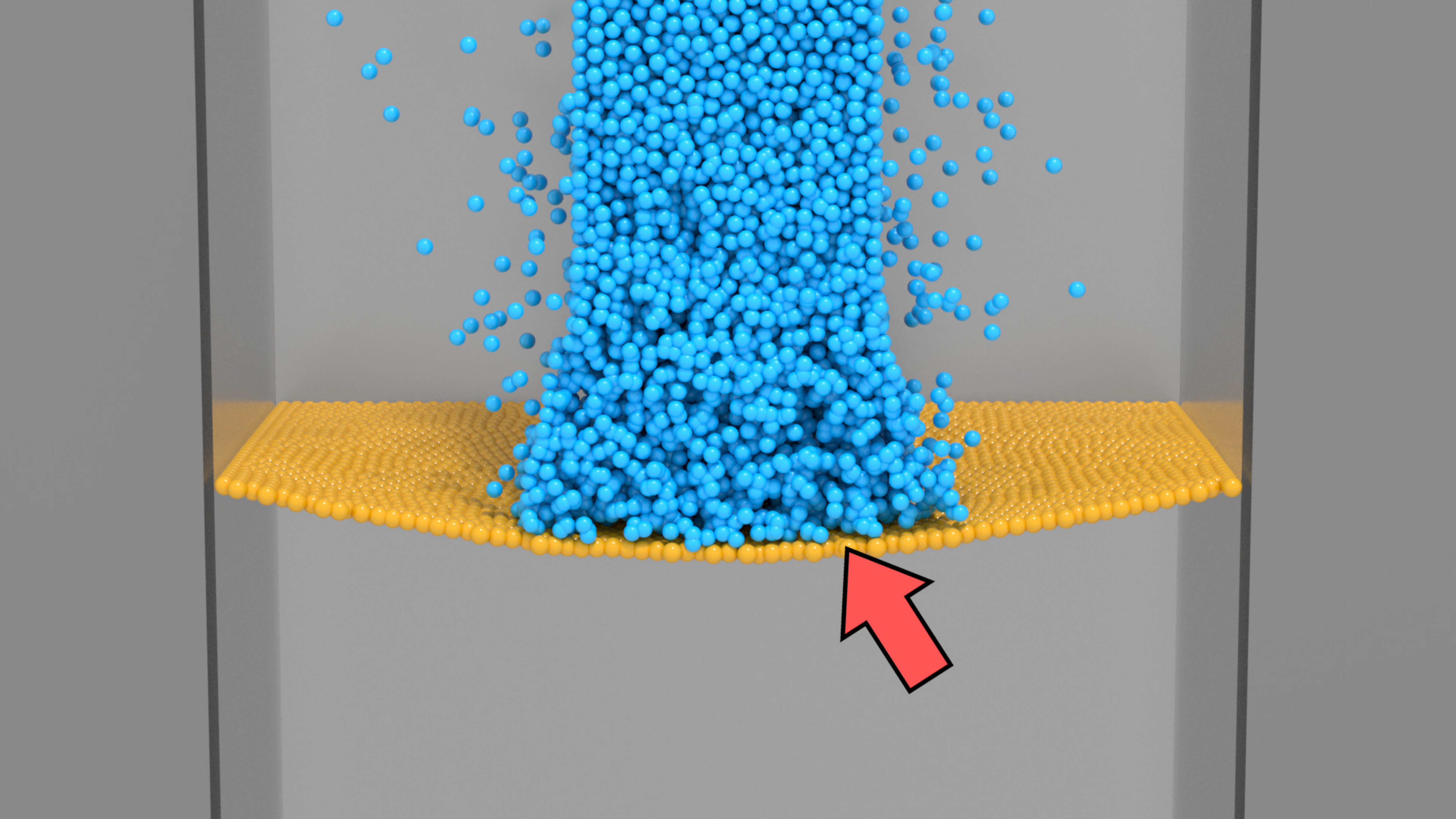} &
    \includegraphics[mycropsNew,width=\figurewidth]{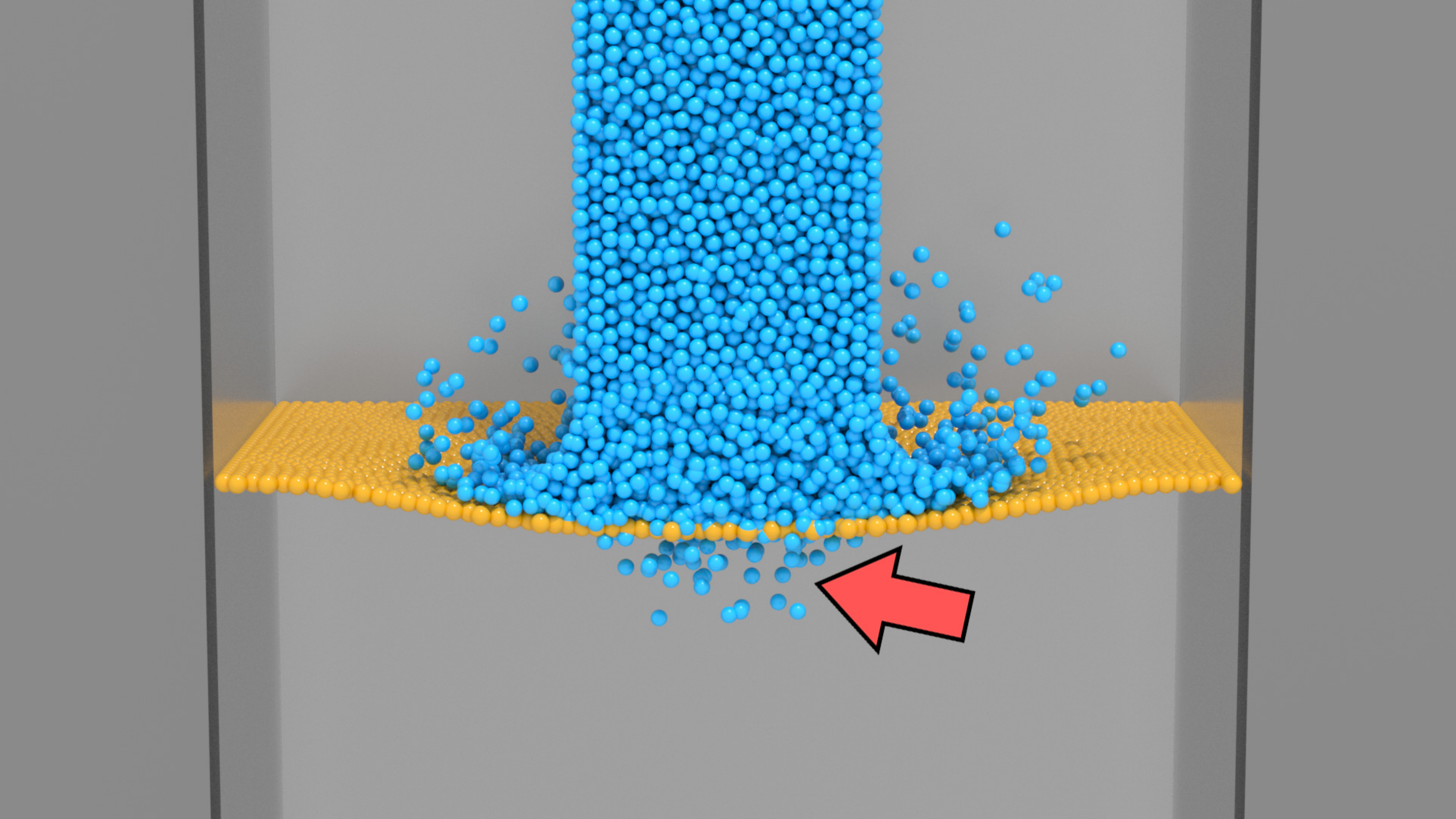}         &
    \includegraphics[mycropsNew,width=\figurewidth]{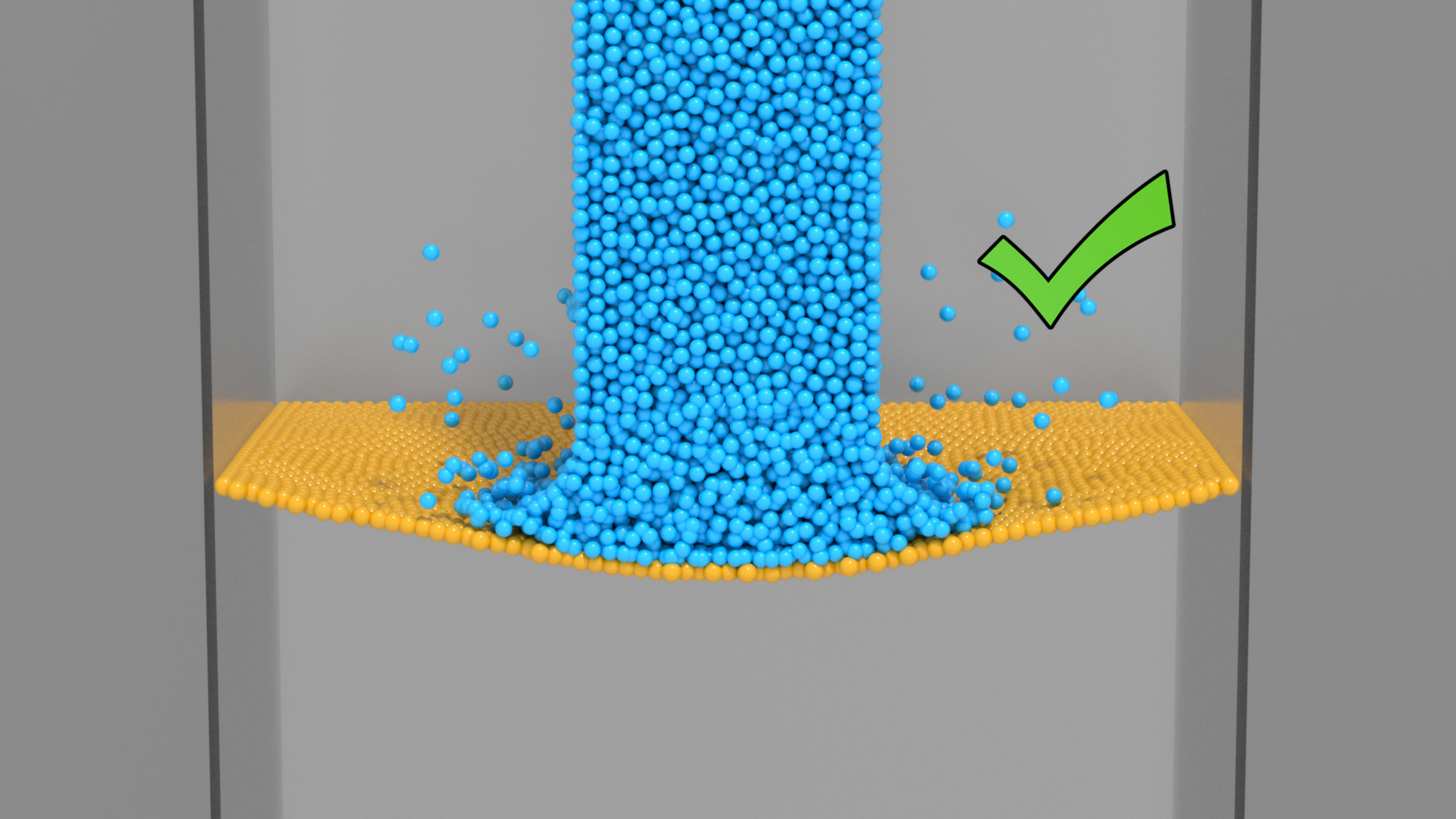}
            \\[-0.5em]
    \includegraphics[mycropsNew,width=\figurewidth]{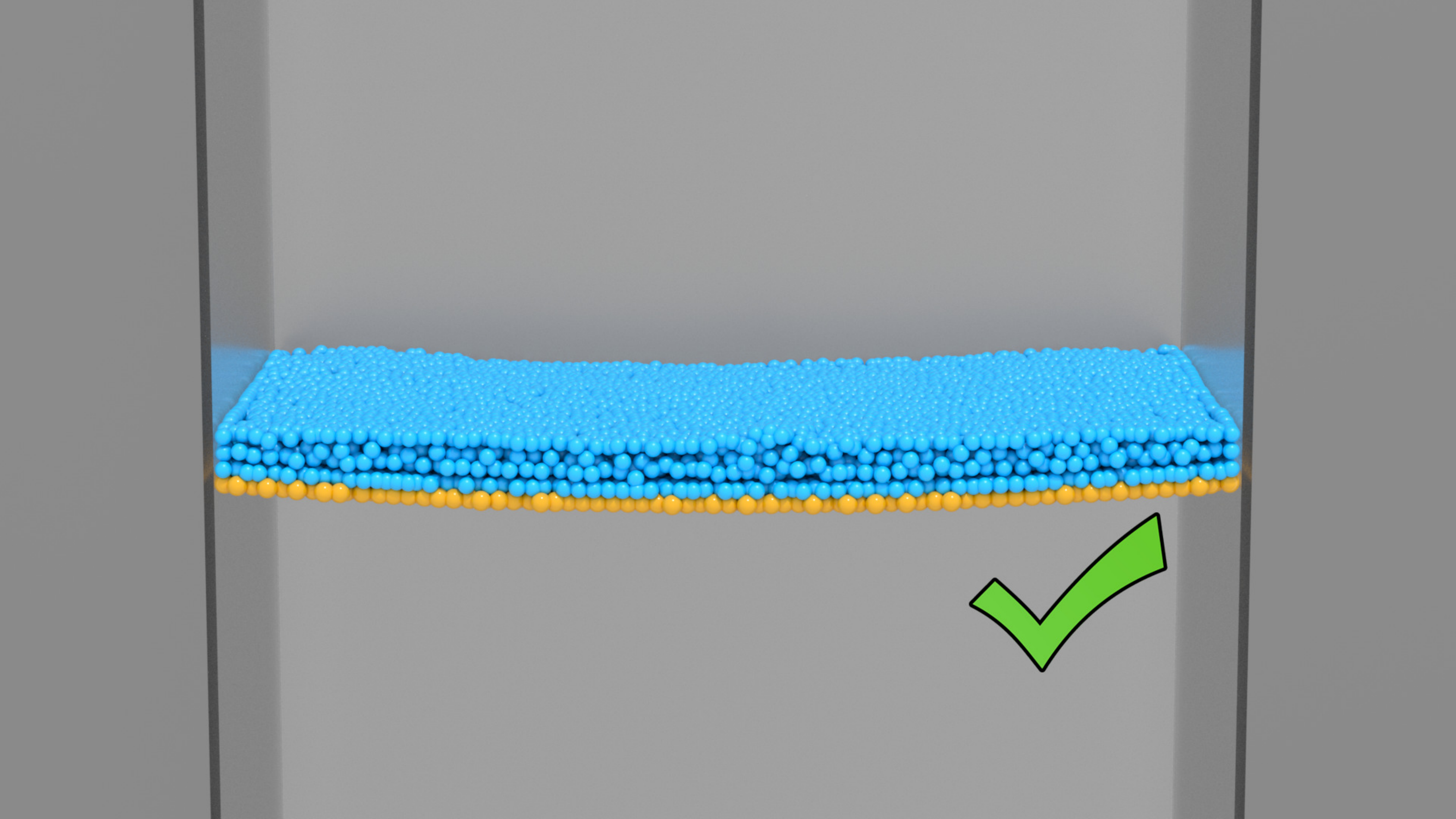}    &
    \includegraphics[mycropsNew,width=\figurewidth]{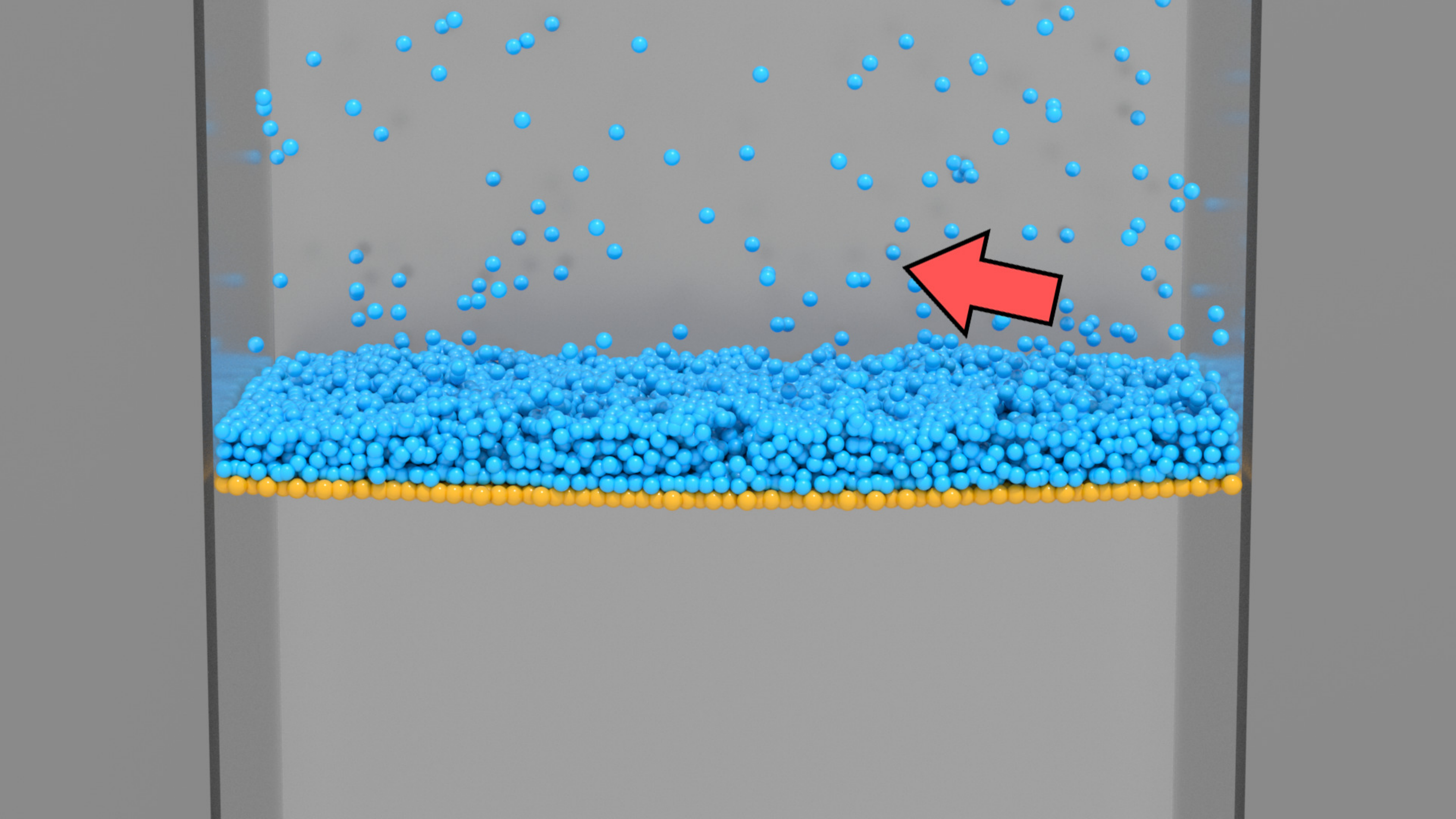}  &
    \includegraphics[mycropsNew,width=\figurewidth]{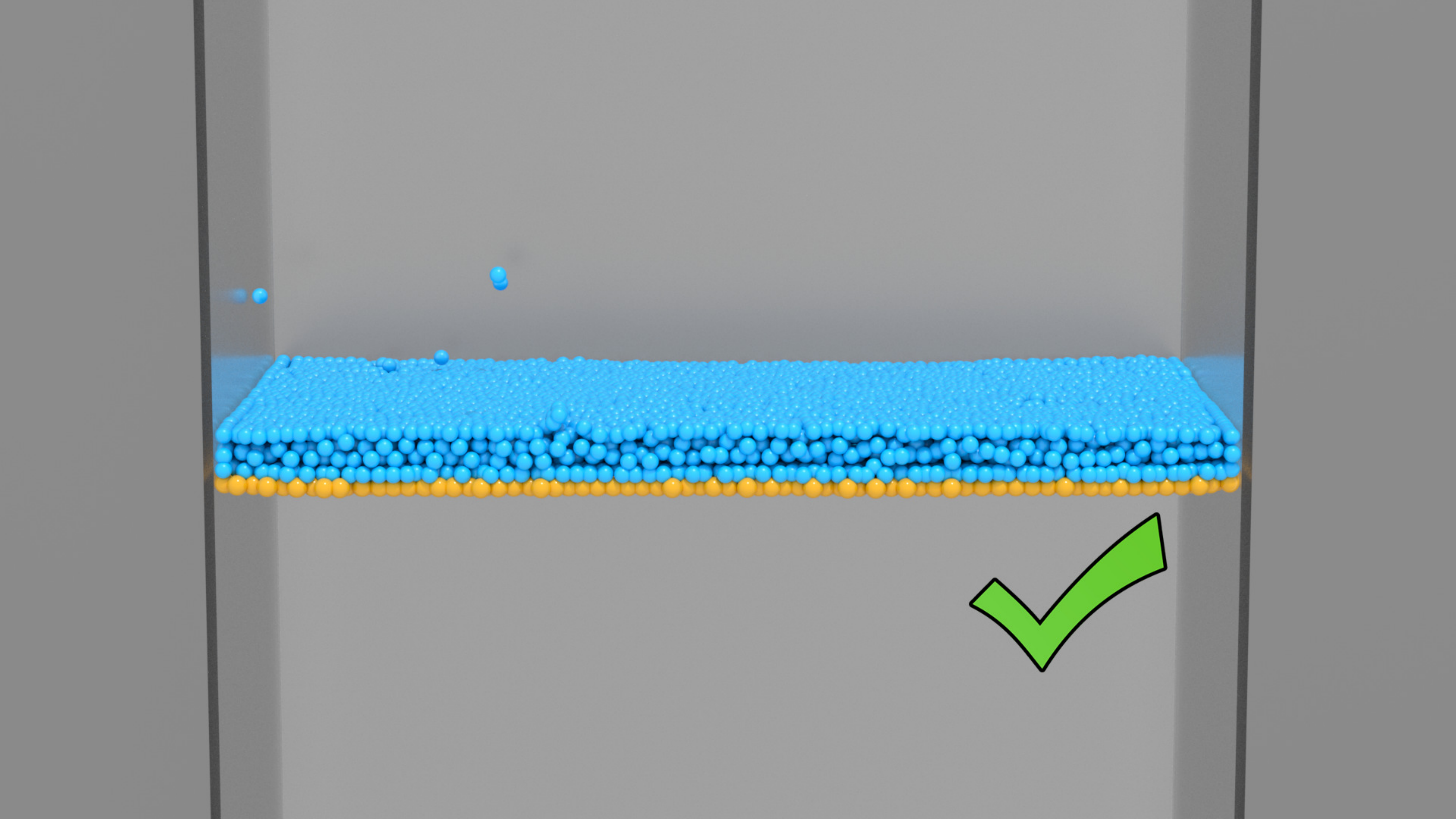}   &
    \includegraphics[mycropsNew,width=\figurewidth]{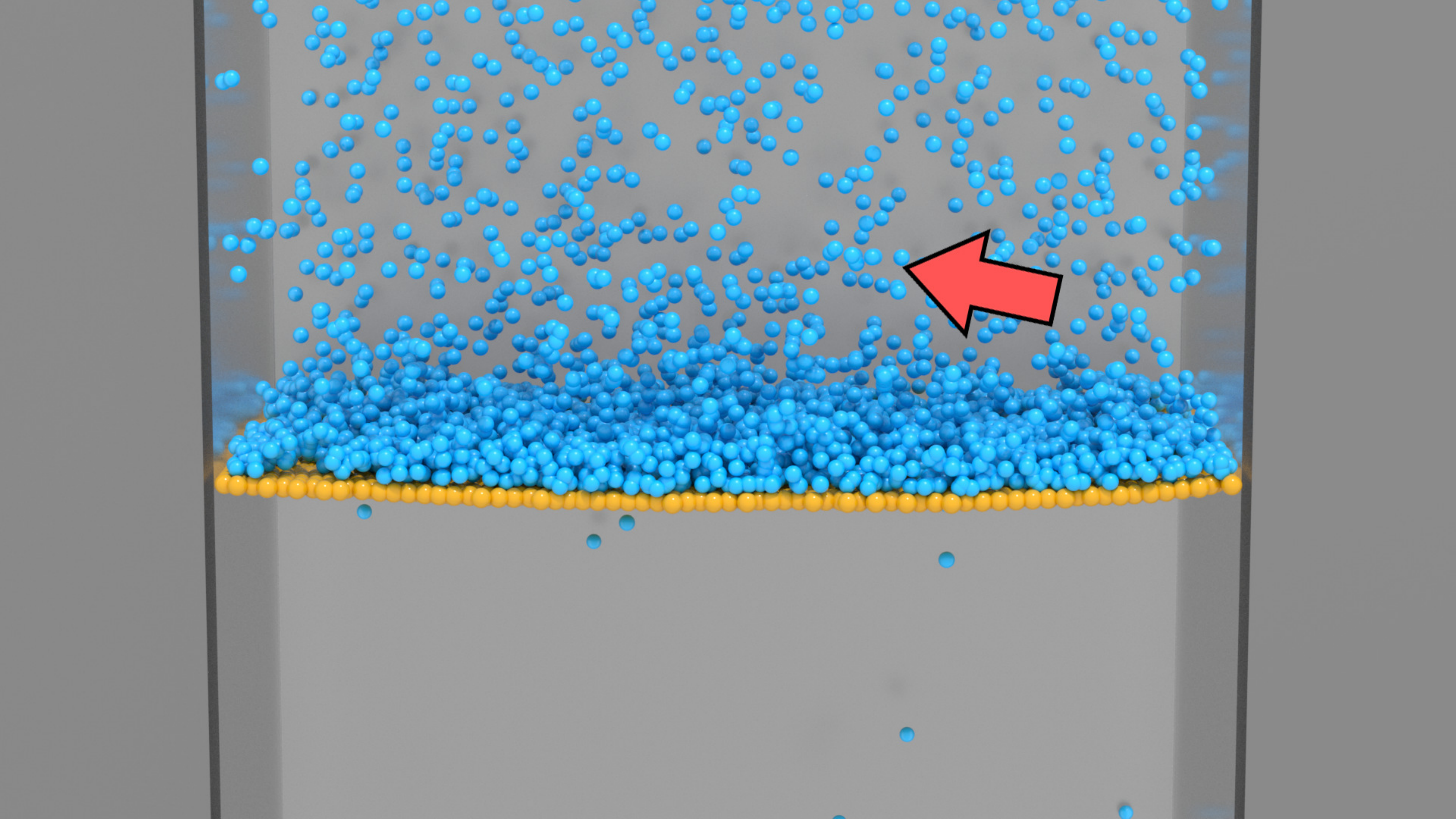} &
    \includegraphics[mycropsNew,width=\figurewidth]{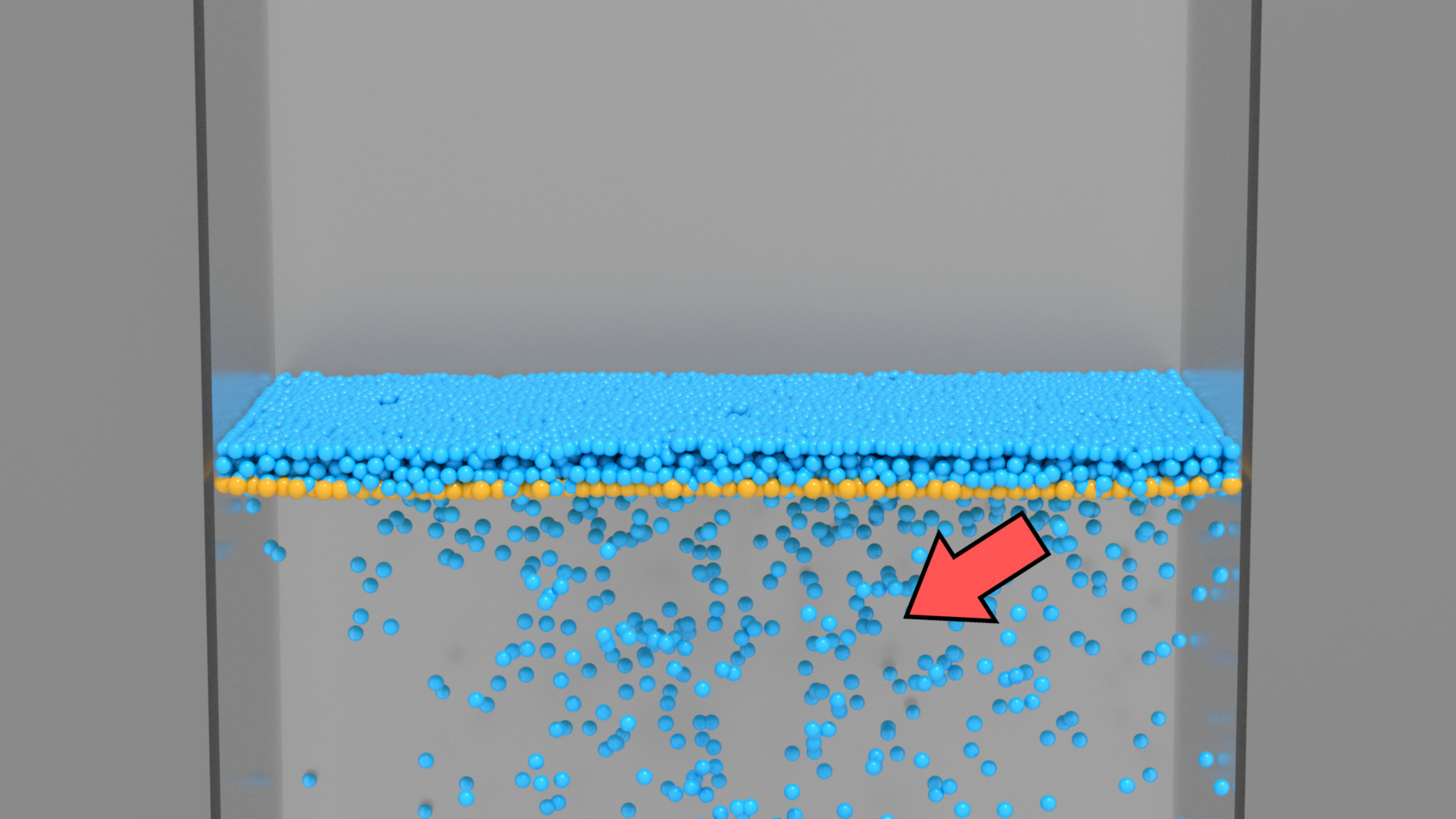}         &
    \includegraphics[mycropsNew,width=\figurewidth]{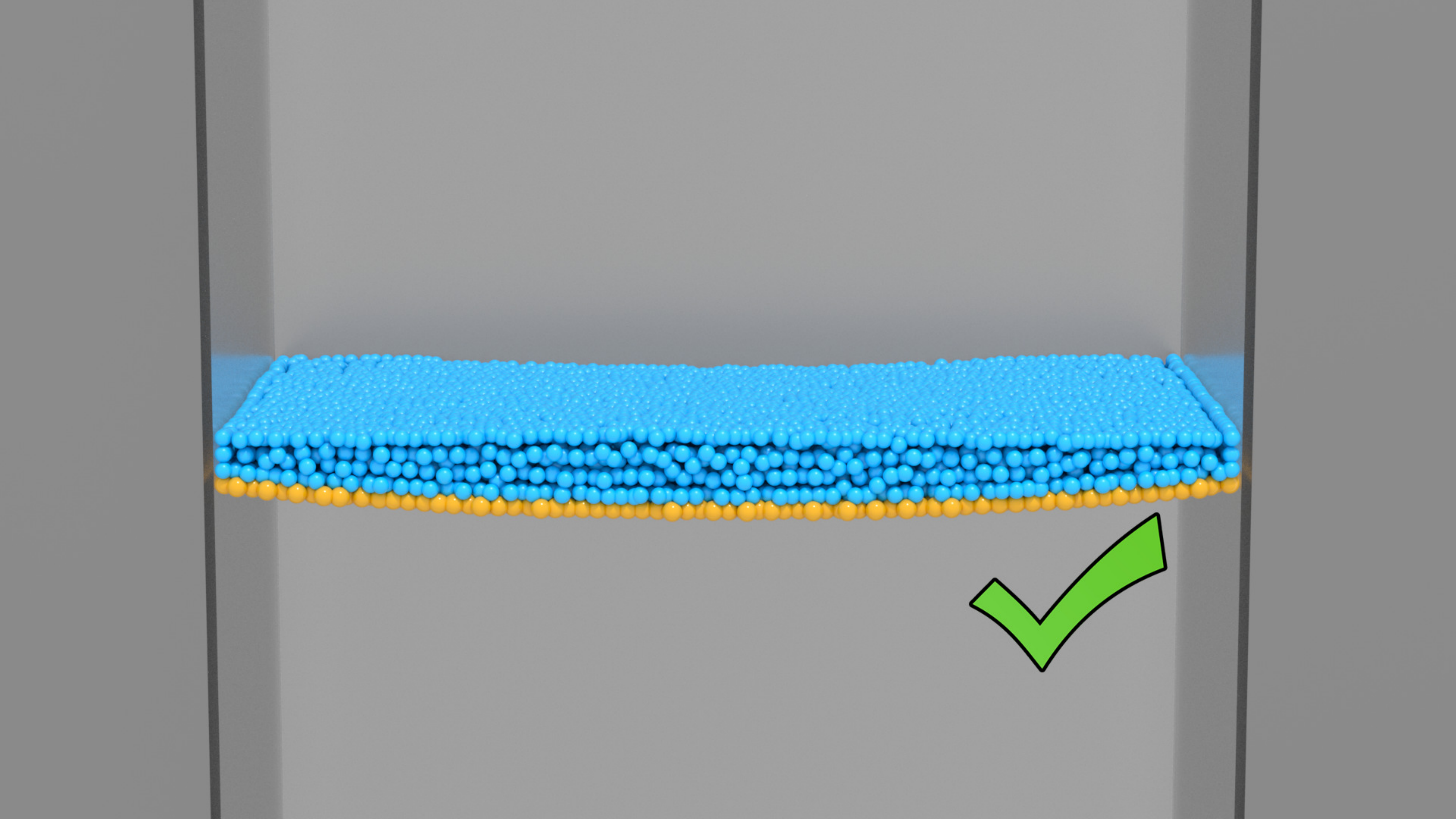}
            \\
    %                                 \\
    %
  \end{tabular}
  \caption{\textbf{Comparison of different methods for handling solid-fluid collisions:} cross-section views of simulations including a soft elastic cloth hit by a fluid beam. Each column shows the same simulation at two different time steps. All particles have identical mass and weak/strong penalty and SPH-based force stiffness parameters differ by an order of magnitude. Notice that only our merging-and-splitting approach prevents penetrations without introducing instability.}
  \label{fig:comparison_fluid_pd}
    %\vspace{-1em}
\end{figure*}

\section{Implementation Details}
\label{sec:implementation}

Our merging-and-splitting scheme can be used in various ways for handling collisions within a simulation system or coupling different simulation systems. Yet, there are a number of details specific to merging-and-splitting that need to be considered. In this section we provide the details of our implementation and the reasons behind our implementation decisions. While most of these details are specific to our implementation, they represent a list of potential issues that one needs to consider for any implementation of merging-and-splitting.
A significant portion of our implementation decisions are related to the collision detection scheme we use.

\subsection{Collision Detection}

Our implementation uses a simple collision detection scheme that merely checks whether particles intersect at the beginning of each time-step. If two particles intersect, they are merged and integrated as a meta-particle. If the meta-particles are split before the positions update, our method guarantees that the split particles will not move towards each other, so they may no longer be in contact in the beginning of the next time-step. While this may appear like a positive outcome at first glance, it can fail to prevent penetrations for contact situations that require multiple time-steps to resolve (such as rest-in-contact). Therefore, in our implementation we perform splitting after position update. This means that the same two particles will remain in contact in the beginning of the next time-step as well. Thus, if we only rely on the particle positions for collision detection, two particles that come into contact would perpetually remain in contact. We avoid this by introducing a secondary rule for collision detection, such that intersecting particles are merged only if their velocities are towards each other. Let $r_1$ and $r_2$ be the collision radii of the colliding particles. We use the following two simple rules for collision detection:
\begin{align}
   & \text{Collision Rule 1:\hspace*{-3em}} & \left\vert \vec x_2 - \vec x_1 \right\vert      & < r_1 + r_2  \label{eq:collision_penetration} \\
   & \text{Collision Rule 2:\hspace*{-3em}} & \vec{n} \cdot \left( \vec v_2 - \vec v_1\right) & < 0  \label{eq:collision_velocity_conditions}
\end{align}

For handling the first rule, limiting the time-step size is important to ensure that we do not miss collisions. To compute a safe time-step size, we employ a typical CFL condition that limits the motion of particles within a time-step to no more than particle radius.
This ensures that our simple collision detection mechanism does not miss head-on
collisions between particles, but collisions at grazing angles can be missed.
Note that alternative approaches such as using a continuous collision detection scheme \cite{Redon:2002} instead would entirely avoid this limit on the time-step size.

The second rule ensures that the colliding particles do not perpetually remain in contact. Indeed, since our merging-and-splitting scheme produces final velocities for the merged particles that point away from each other, it is guaranteed that the two intersecting particles will not be merged in the beginning of the next time-step. However, if the collision event must last longer than one time-step (such as rest-in-contact situations), the intersecting particles must be merged the next time-step as well. To facilitate this, our implementation uses a two-stage integration scheme. In the first stage, we only merge intersecting particles with velocities pointing towards each other (following the second rule). We mark intersecting particles with velocities pointing away from each other, but we do not merge them. If the marked particles at the end of the time-step integration have velocities pointing towards each other, we recompute the time-step integration in a second stage by merging those marked particles in the beginning of the time-step. Obviously, this two-stage integration scheme nearly doubles the computation time.

In theory, at the end of the second stage, we can detect that other marked particles that were not merged prior to the second stage may require merging. Therefore, handling all collisions with this approach may need more than two stages. Yet, in our implementation we only use two stages and we have not observed any practical consequences of not adding extra stages as needed.

An alternative solution would be to modify the intersecting particle positions at the end of each time-step to ensure that the previously merged particles no longer intersect. However, while this is a commonly-used technique in computer graphics \cite{Shao:2014,Shao:2015}, position corrections often inject extra energy into the system that impacts the stability of the simulations or introduces artificial vibrations. Moreover, this seemingly minor energy injection of position correction can lead to catastrophic events with fracture simulations, instantly shattering all bonds and causing entire objects to explode into individual particles. 
Also, position corrections can cause other particles to intersect. That is why we use the more expensive two-stage integration approach in our implementation.

We use our merging-and-splitting scheme for handling collisions between solid particles integrated using the same system. However, we ignore intersections of neighboring particles that are directly connected via springs. Thus, we can have high-resolution solid simulations with neighboring particles almost intersecting with each other, without introducing undesirable internal collisions between neighboring particles.

\subsection{Solid-Fluid Coupling}

Solid-fluid coupling is an important application in computer graphics \cite{Koschier:2019}.
In our system, after collision detection, we merge colliding particle pairs recursively.
When coupling integrators for solid and fluid simulations, large chains of particle intersections can occur at the solid-fluid boundary. Merging all intersecting particles of a chain into a single meta-particle can rigidify the entire solid-fluid interface and lead to unnatural results. We use a simple fix that merely limits the number of particles in a meta-particle using a threshold $n$.
This breaks large chains of intersecting particles into multiple groups. Our tests show that the simulations are not sensitive to the value of this threshold, unless a very small or a very large number is used. When $n$ is too small, ignored collisions due to this limit can lead to penetrations. On the other hand, when $n$ is too large, the entire solid-fluid interface can get rigidified. We use a number between 8 and 64 in our tests, producing similar results.
Our implementation performs merging in the order of particle indices. To introduce randomness in grouping, before starting to merge each group, we can randomly pick a limit between two user-specified limits $n_{\min}$ and $n_{\max}$.

Limiting the number of particles in a meta-particle has the obvious theoretical drawback that some relatively small fraction of collision events would be ignored at each time step. Yet, as our experiments confirm, this is of little practical concern. This is because, with randomized grouping, we do not persistently ignore specific pairs of colliding particles, i.e. a collision event that is ignored in one time step is likely to be considered in the next time step.

\section{Results}

We evaluate our merging-and-splitting method by comparing it to typical alternative techniques for handling collisions in particle-based simulations and presenting large simulations examples, including solid-fluid coupling tests using different simulation systems. We use $\alpha=1$, $\beta=0$, and $n_{\min}=8, n_{\max}=64$ in all simulations, unless otherwise specified. The rendered surfaces are generated using level-sets~\cite{Bhattacharya:2011} for fluid particles and tetrahedralization of the solid mesh in the beginning of the simulation~\cite{Chakrit:2015} for solid particles.

\begin{figure*}[tb]
\centering
\newcommand{\figurewidth}{0.164\textwidth}
\setlength{\tabcolsep}{0.025in}
\renewcommand{\arraystretch}{1.5}
%\small
\hspace*{-0.013\linewidth}%
\resizebox{1.013\linewidth}{!}{%
\begin{tabular}{cccccc}
\multicolumn{2}{c}{Penalty Force}&
\multicolumn{2}{c}{SPH-based Force}&
Impulse-based&\textbf{Merging-and-}\\[-1.5em]
\multicolumn{2}{c}{\rule{0.3\linewidth}{0.4pt}}&%
\multicolumn{2}{c}{\rule{0.3\linewidth}{0.4pt}}\\[-1.0em]
weak&strong&weak&strong&Collisions&\textbf{Splitting (Ours)}\\
\includegraphics[trim=700 50 350 50, clip, width=\figurewidth]{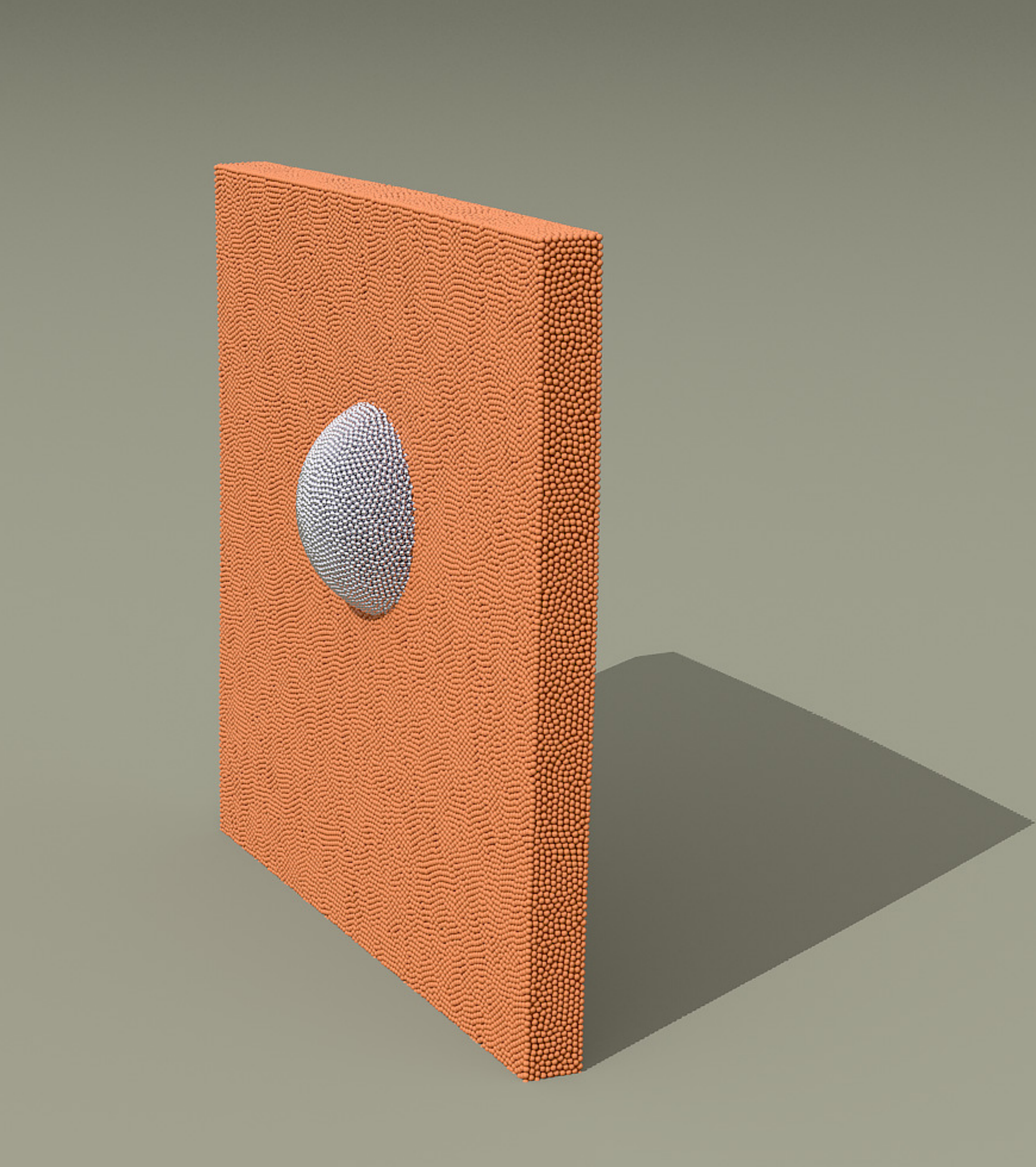}&%
\includegraphics[trim=700 50 350 50, clip, width=\figurewidth]{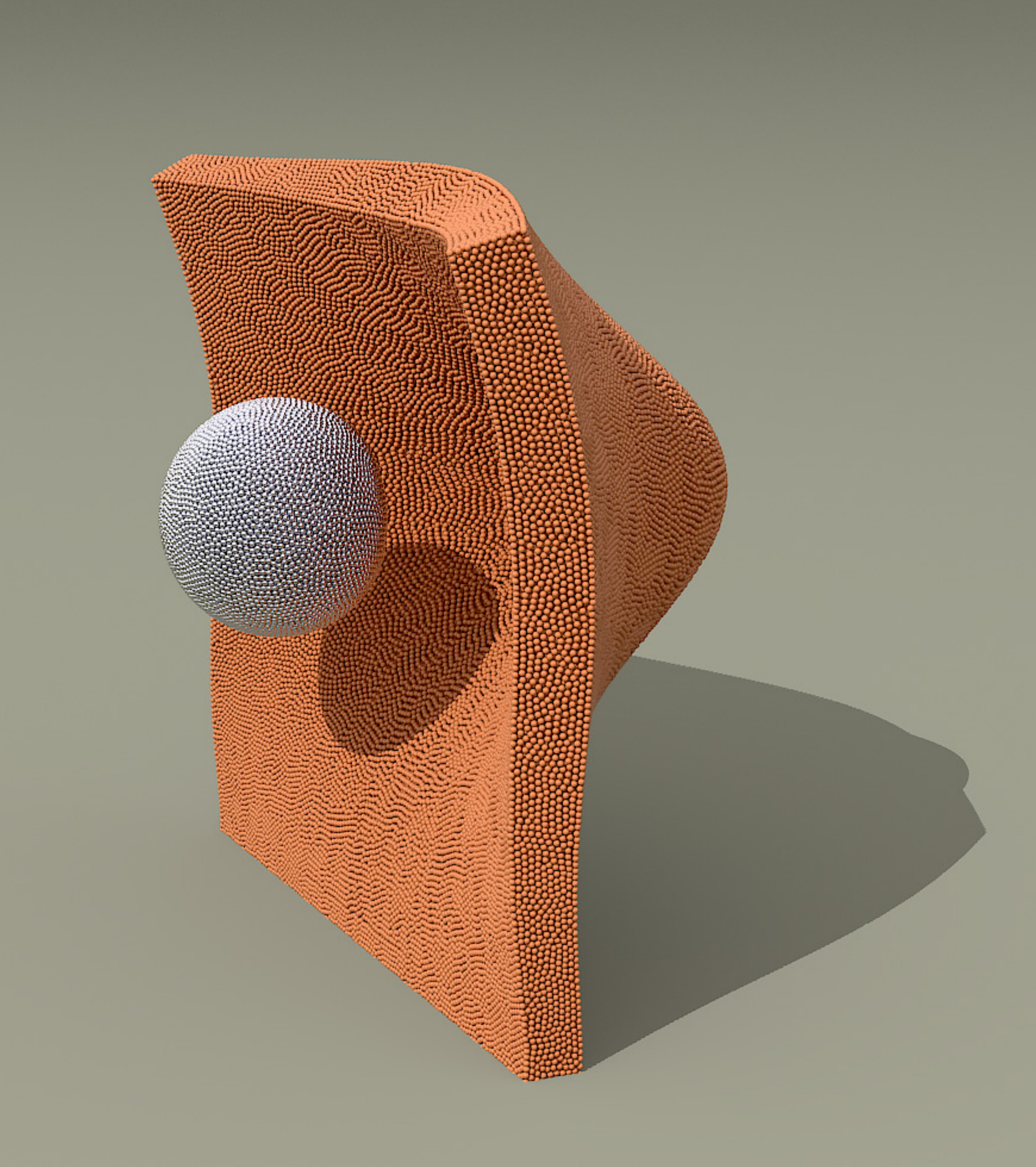}&%
\includegraphics[trim=700 50 350 50, clip, width=\figurewidth]{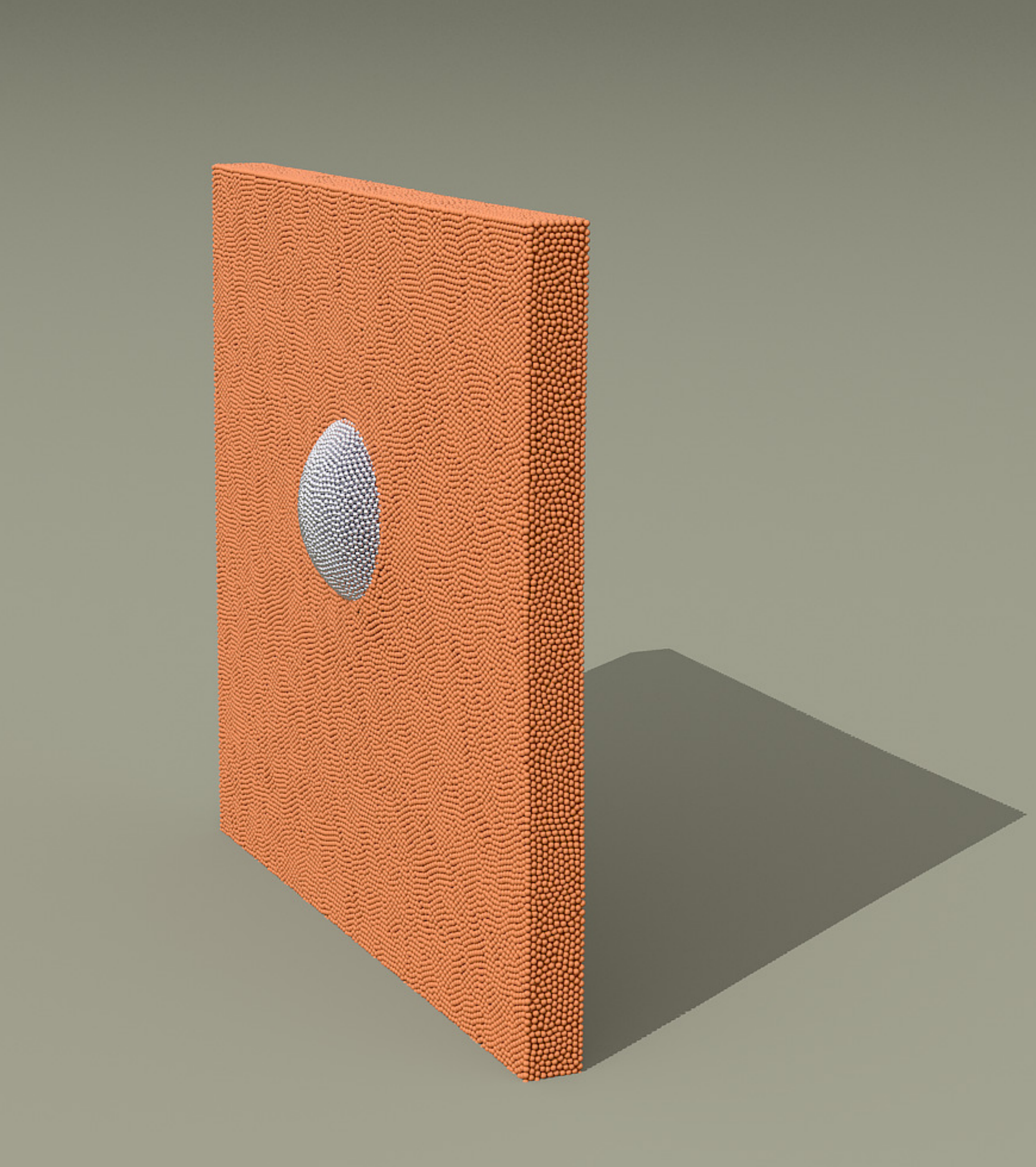}&%
\includegraphics[trim=700 50 350 50, clip, width=\figurewidth]{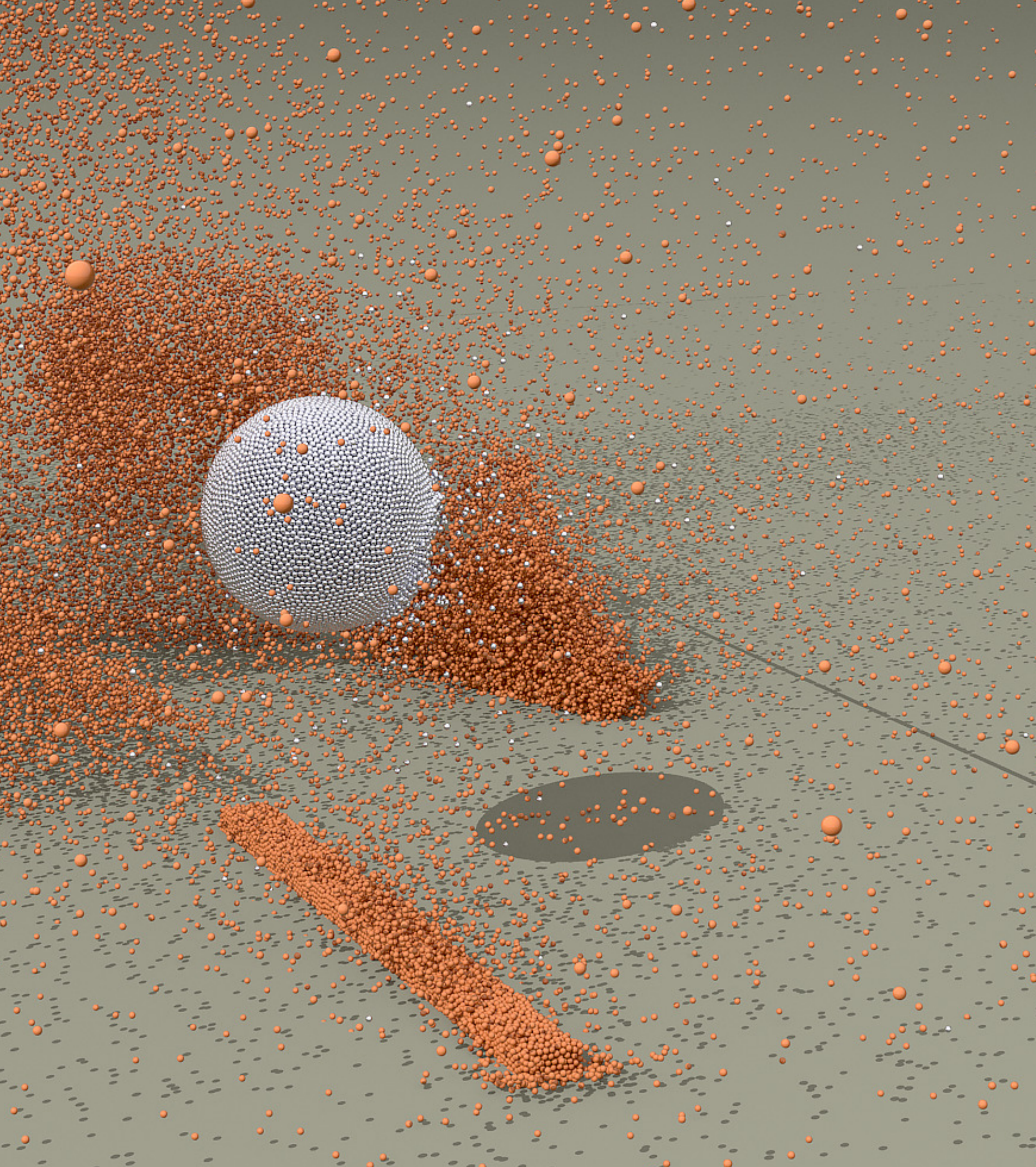}&%
\includegraphics[trim=700 50 350 50, clip, width=\figurewidth]{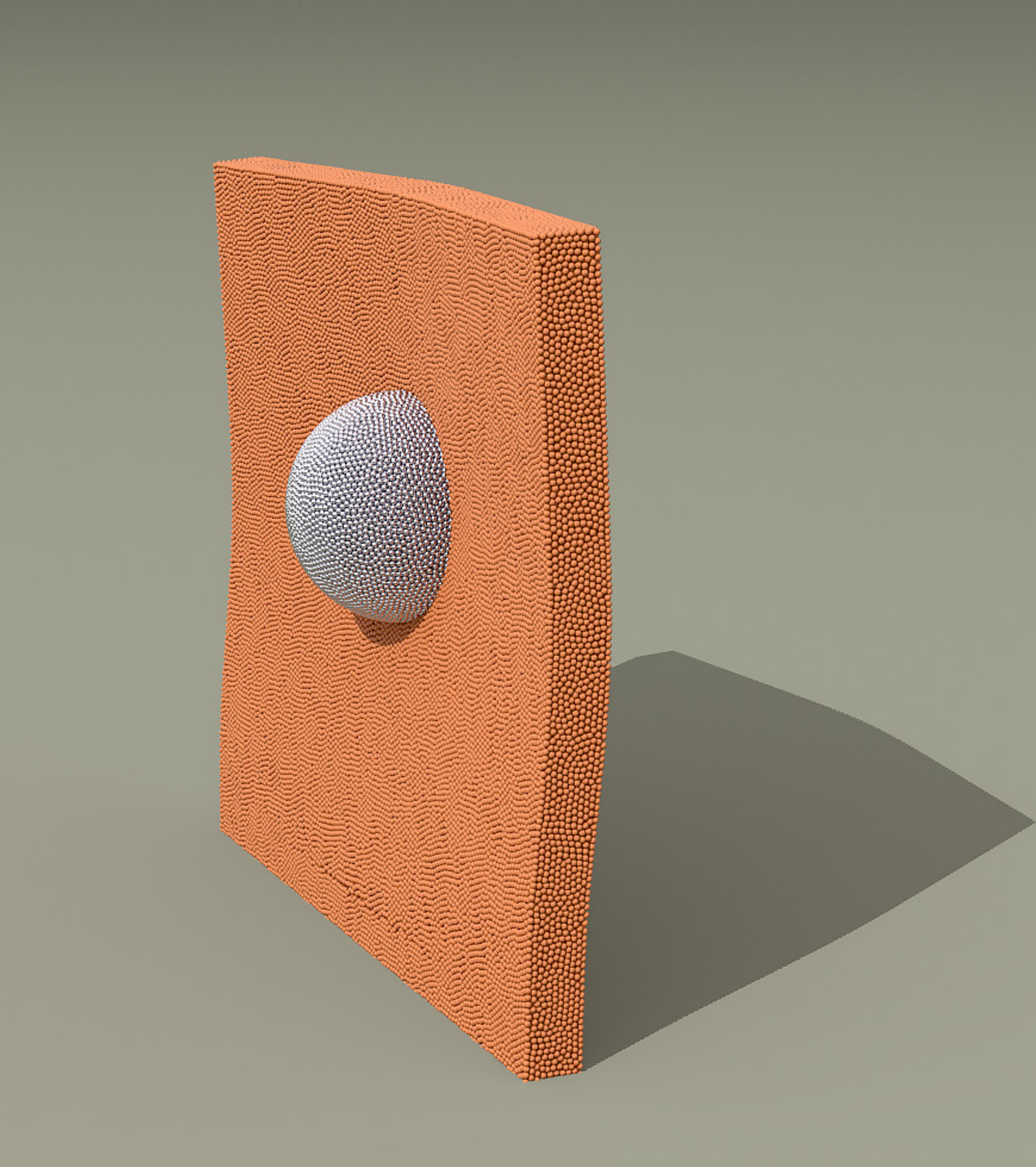}&%
\includegraphics[trim=700 50 350 50, clip, width=\figurewidth]{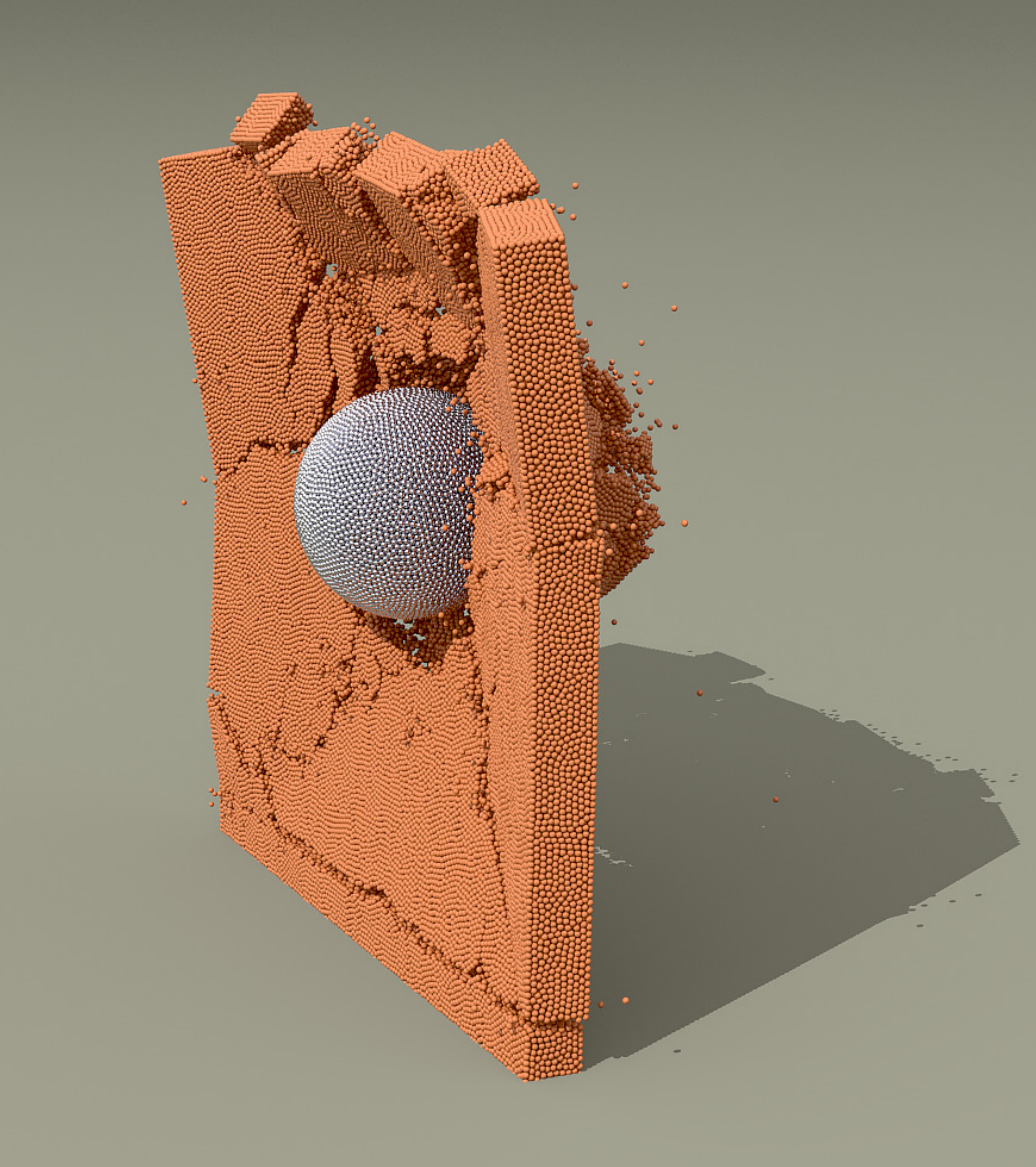}\\
\end{tabular}
}
\caption{\textbf{Comparison of different methods for collision handling in fracture simulation:} a solid brittle wall is hit by a dense ball with $20\times$ heavier particles. The bottom row shows the same simulation from a different view at a later time step. Notice that only our merging-and-splitting approach can produce a stable fracture simulation with pleasing result.}
\label{fig:ball_shoots_wall_comparison}
%\vspace{-1em}
\end{figure*}

\subsection{Comparisons}

We compare our merging-and-splitting method with force-based approaches, using penalty \cite{Terzopoulos:1989, Tonnesen:1991} or SPH-based \cite{Becker:2009b,Shao:2014} force formulations, and impulse-based collision handling methods \cite{Hahn:1988, Mirtich:1995, Guendelman:2003, Weinstein:2006, Vouga:2017}.

\autoref{fig:comparison_sphere_cloth_wall} shows comparisons of different methods for handling collisions within a particle-based solid simulation. The top and bottom rows show a ball falling onto an elastic cloth from two different heights. The ball is simulated using peridynamics~\cite{Levine:2014} with implicit integration and a high threshold that prevents fracturing.
The cloth is simulated using a mass-spring system within the same implicit integration system used for peridynamics.
As can be seen in the figure, using a weak penalty or SPH-based force fails to resolve penetration. Increasing the stiffness of the force formulation helps, but when the force is too strong with either force-based formulation, it can make the simulation system highly unstable. 
In practice, such instabilities can be avoided by carefully tuning the stiffness parameter, but the right stiffness values depend on the collision scenario. Notice that the stiffness that works for the top row does not work for the bottom row showing faster impact and vice-versa.
Impulse-based collisions fail to prevent penetration, since they handle colliding particles in isolation using instant velocity updates. Our merging-and-splitting approach completely prevents penetration and properly resolves the collisions without introducing instabilities for both cases.
Notice that the deformations of the cloth with merging-and-splitting are similar to the deformations achieved with force-based formulations using the right stiffness parameters.

Similar tests involving solid-fluid coupling are shown in \autoref{fig:comparison_fluid_pd}. In these tests, a cylinder-shaped fluid column with high velocity falls onto an elastic cloth. The fluid particles are simulated using SPH~\cite{Muller:2003} with explicit integration and the elastic cloth is simulated using a mass-spring system with implicit integration. The solid-fluid coupling is handled entirely via particle-level collisions. 
As can be seen in the figure, penetrations cannot be avoided with weak penalty or SPH-based forces. Stronger forces lead to fluid particles bouncing back with high velocity. Again, these problems can be avoided by tuning the stiffness parameter accordingly. Impulse-based collisions cannot prevent penetration either and consistent low-velocity impacts without position correction lead to fluid particles slowly passing through the cloth layer. Our merging-and-splitting method produces perfect separation between the fluid and the solid systems and completely prevents penetration without introducing instabilities.

Particle-based fracture simulations using peridynamics are typically handled using penalty forces \cite{Levine:2014}, which require extremely small time step size ($\Delta t \approx 10^{-7}$) and parameter tuning until a desirable animation is produced.

\begin{figure}[tb]
\centering
\renewcommand{\figureheight}{0.372\linewidth}
\includegraphics[trim=650 200 900 350, clip, height=\figureheight]{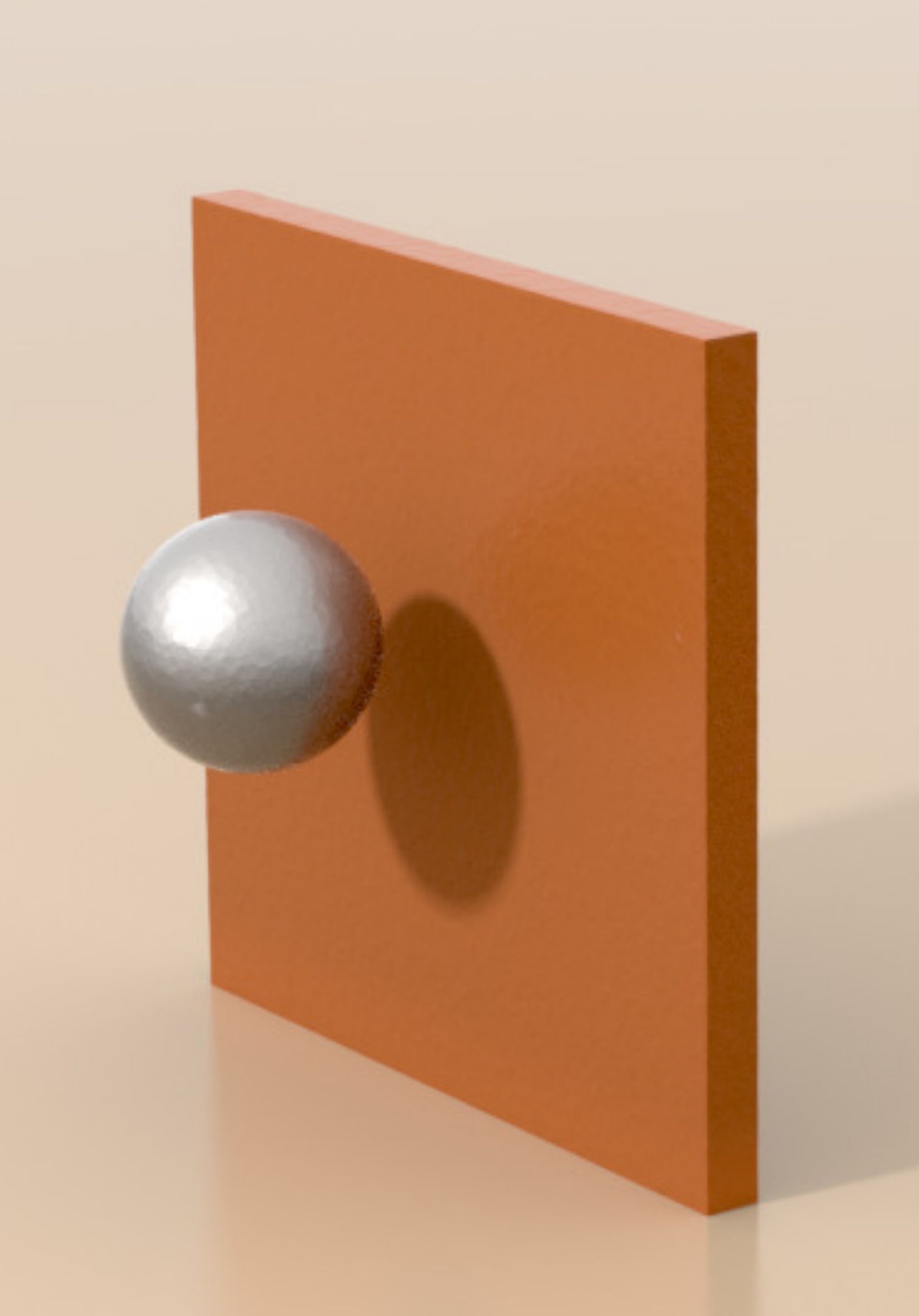}\hfill%
\includegraphics[trim=650 200 900 350, clip, height=\figureheight]{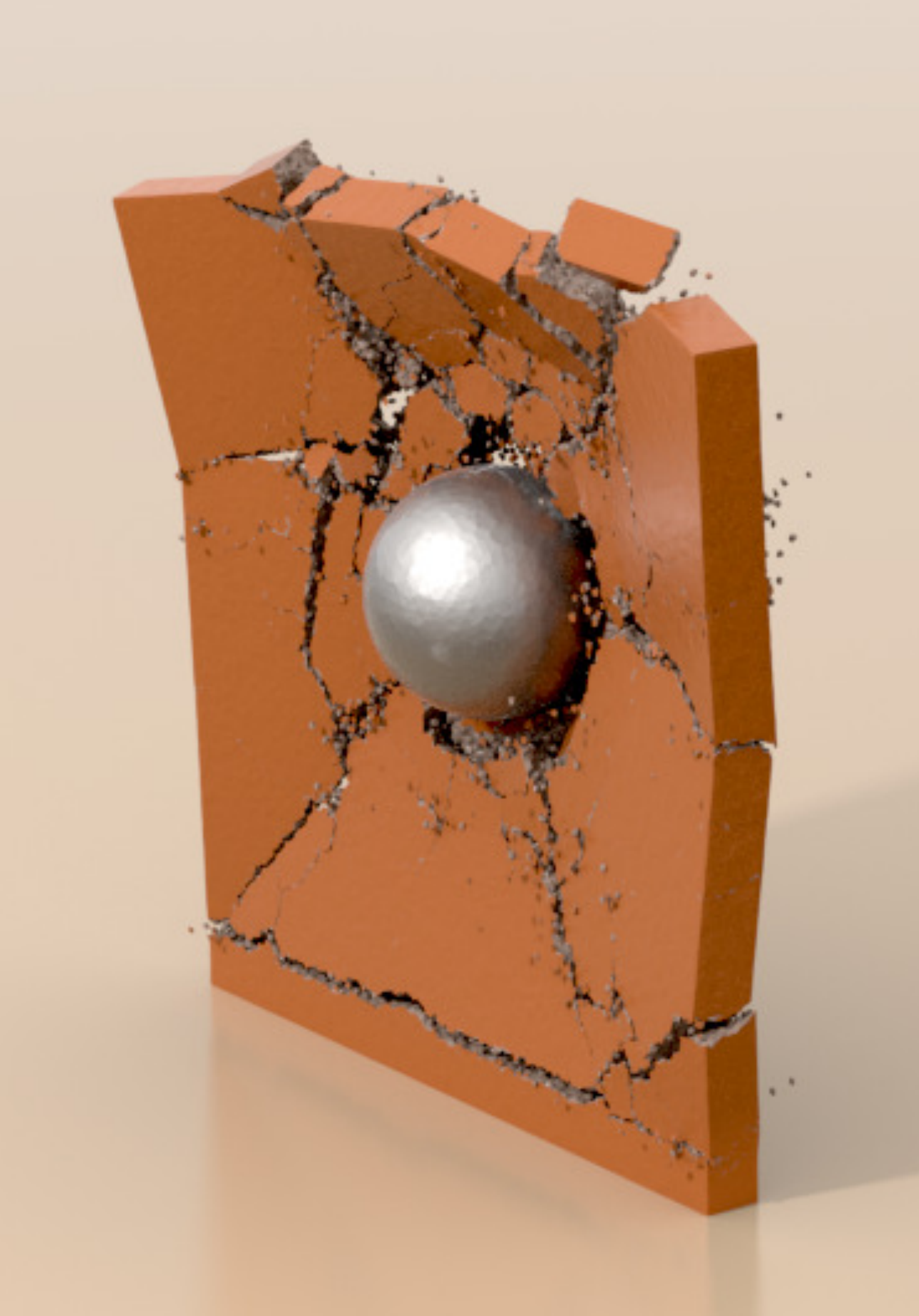}\hfill%
\includegraphics[trim=600 200 450 200, clip, height=\figureheight]{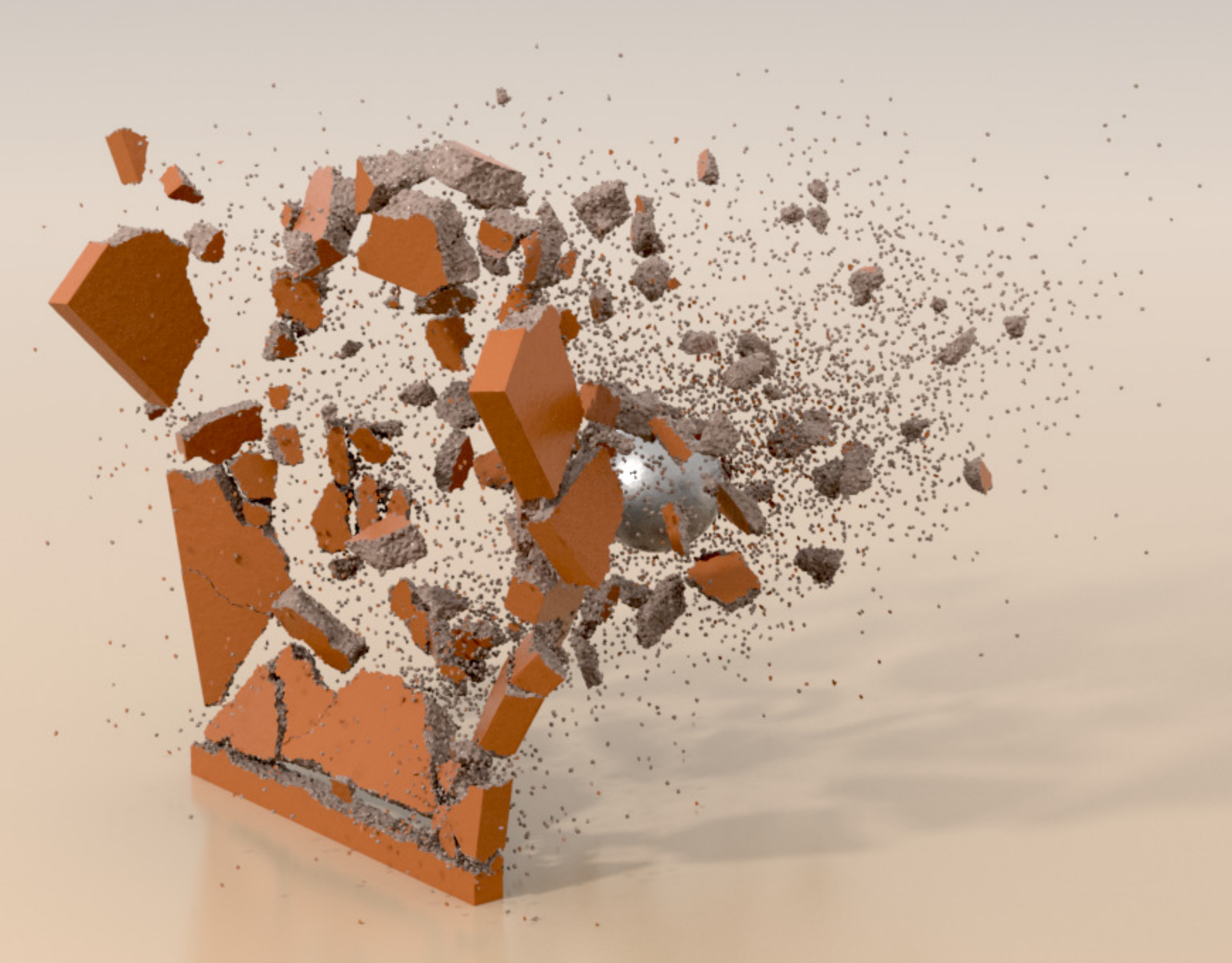}\\
\caption{Frames from our simulation with high impact collision between a ball and a brittle 
wall using peridynamics with merging-and-splitting.}
\label{fig:ball_shoots_wall}
\end{figure}

This is presented in \autoref{fig:ball_shoots_wall_comparison} with a brittle wall hit by a high-velocity ball with $20\times$ heavier particles, both simulated using peridynamics. 
When the collision forces are too strong, the wall crumbles into tiny pieces. Weak forces or impulse-based collisions lead to penetrations before fracture.
Eventually, all collision handling methods lead to fracture, but they differ by the amount of inter-penetration occurring prior to fracture and the visual quality of the fracture. Our merging-and-splitting scheme, in comparison, produces stable and expected results without any parameter tuning (\autoref{fig:ball_shoots_wall}).

\begin{figure}[tb]
\newcommand{\figurewidth}{0.33\linewidth} % for 4 columns
\setlength{\tabcolsep}{0.1em}
\renewcommand{\arraystretch}{1.5}
\centering
\begin{tabular}{ccc}
\includegraphics[trim=390 0 390 150,clip,width=\figurewidth]{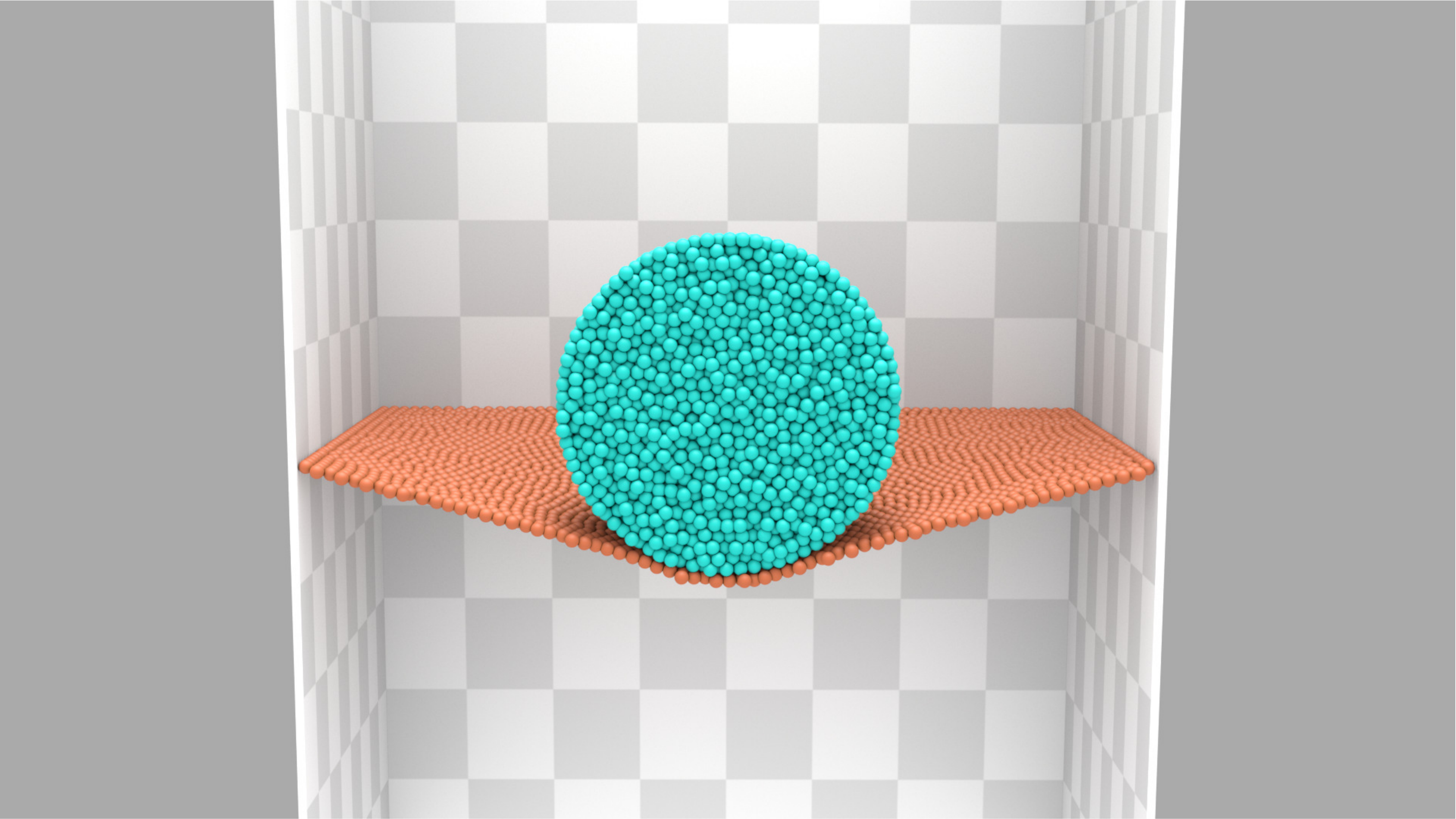}&%
\includegraphics[trim=390 0 390 150,clip,width=\figurewidth]{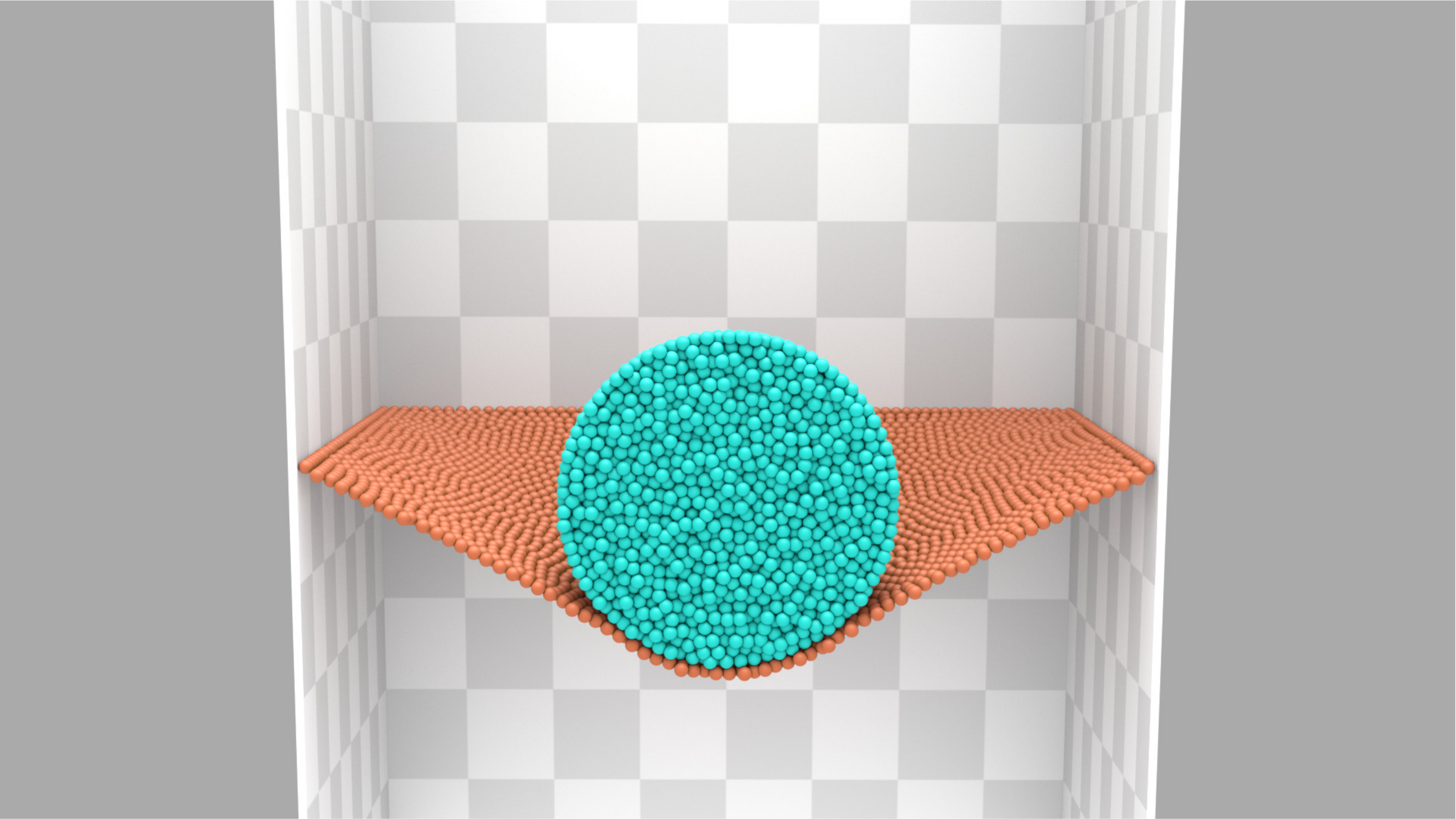}&%
\includegraphics[trim=390 0 390 150,clip,width=\figurewidth]{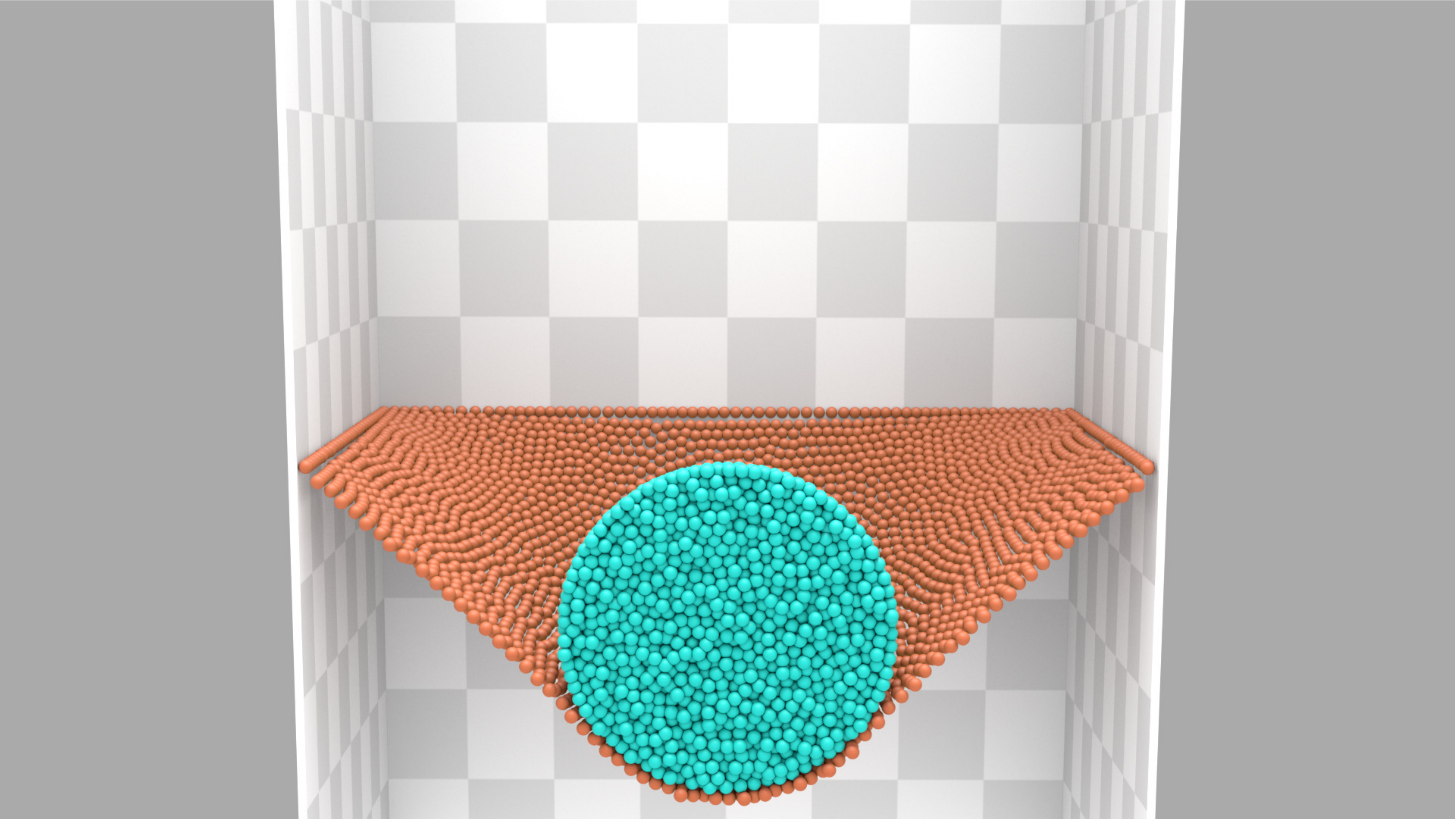} \vspace{-0.5em}\\
$\text{mass ratio: } 1/1$ &
$\text{mass ratio: } 1/10$ &
$\text{mass ratio: } 1/100$ 
		\\
	\end{tabular}
	\caption{\textbf{Different mass ratios:} cross-section views of simulations including a soft elastic cloth hit by a solid ball made up of the same or heavier particles using our merging-and-splitting method.}
	\label{fig:ms_variant_mass_ratio_and_velocity}
\end{figure}

Our merging-and-splitting method can stably handle variations in mass, since meta particles allow significant energy transfer between objects during each time step without introducing instabilities, as demonstrated in \autoref{fig:ms_variant_mass_ratio_and_velocity}.

\subsection{Solid-Fluid Coupling}

\begin{figure}[tb]
    \centering
\newcommand{\figurewidth}{0.33\linewidth}
\includegraphics[trim=400 20 300 0, clip, width=\figurewidth]{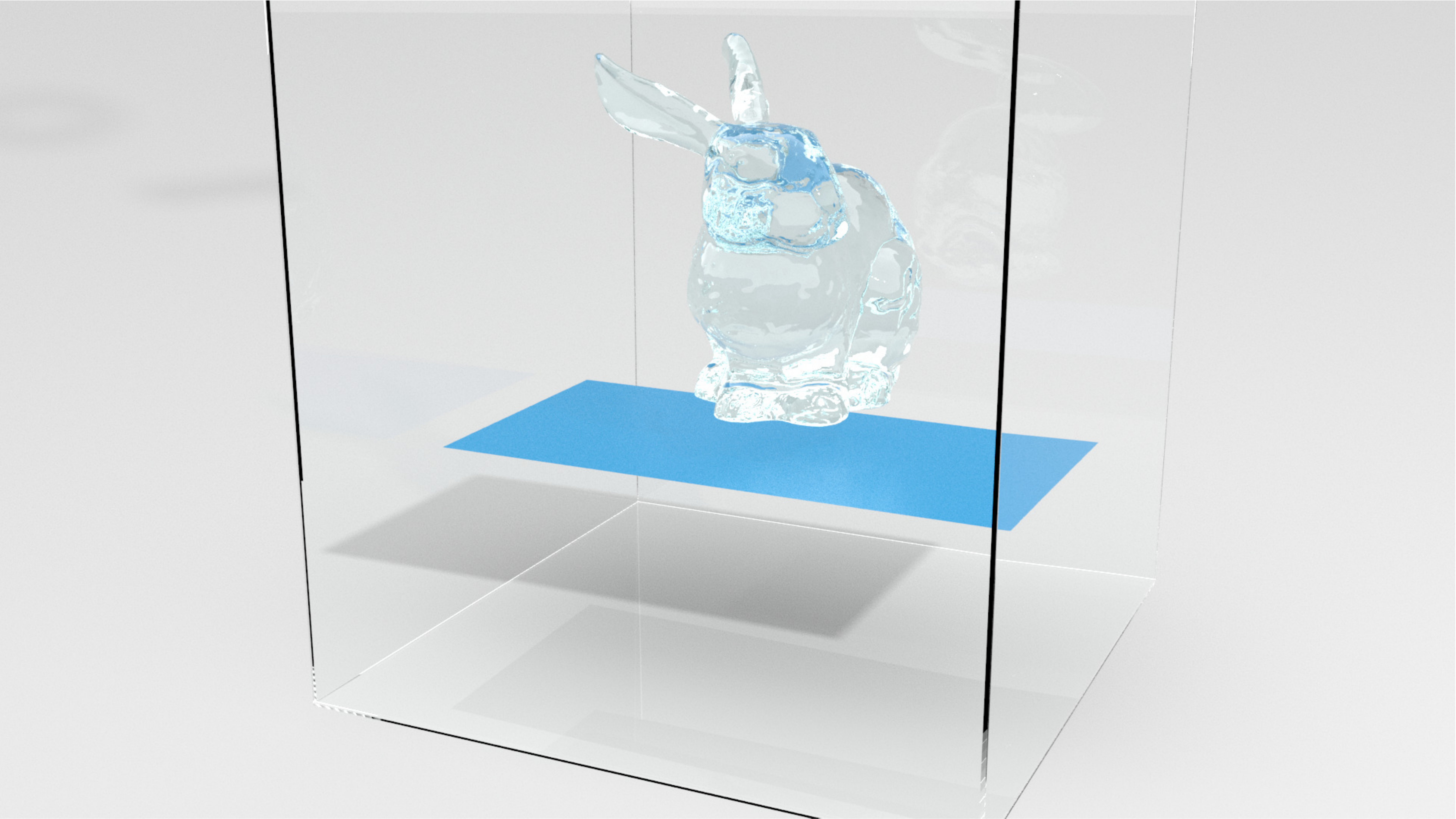}\hfill%
\includegraphics[trim=400 20 300 0, clip, width=\figurewidth]{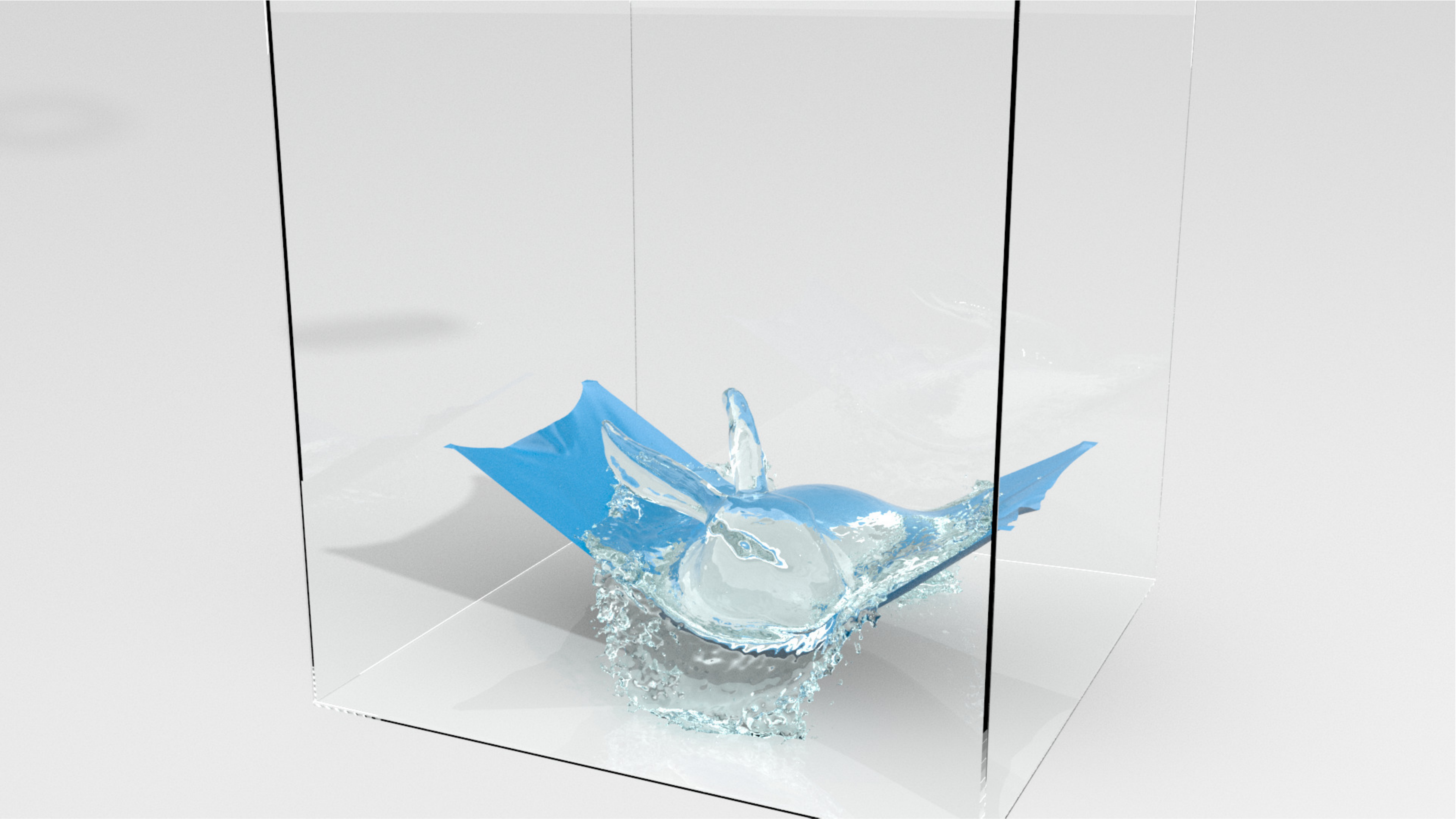}\hfill%
\includegraphics[trim=400 20 300 0, clip, width=\figurewidth]{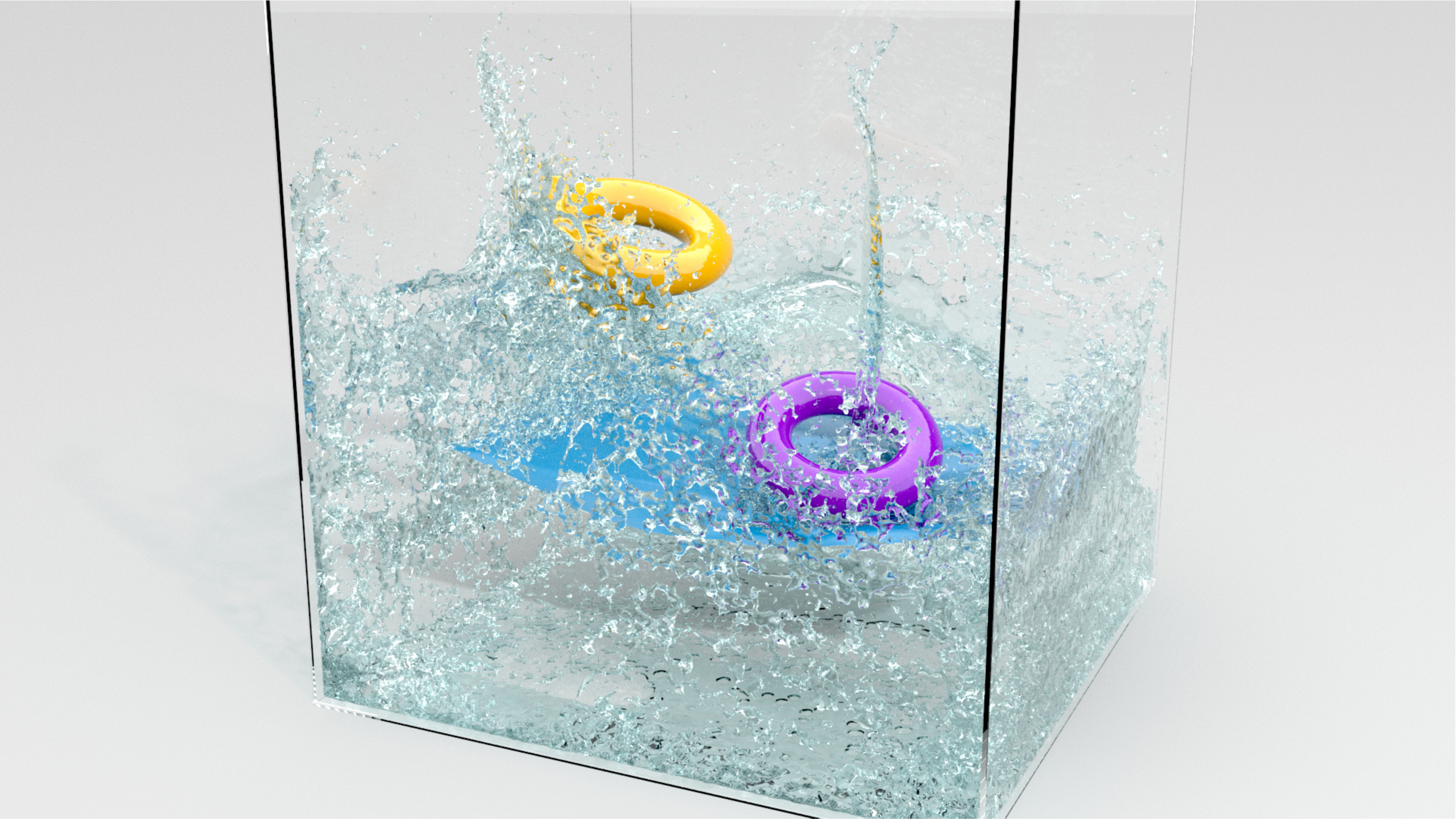}%
%\hfill%
\vspace{0.1em}\\
\includegraphics[trim=400 20 450 0, clip, width=\figurewidth]{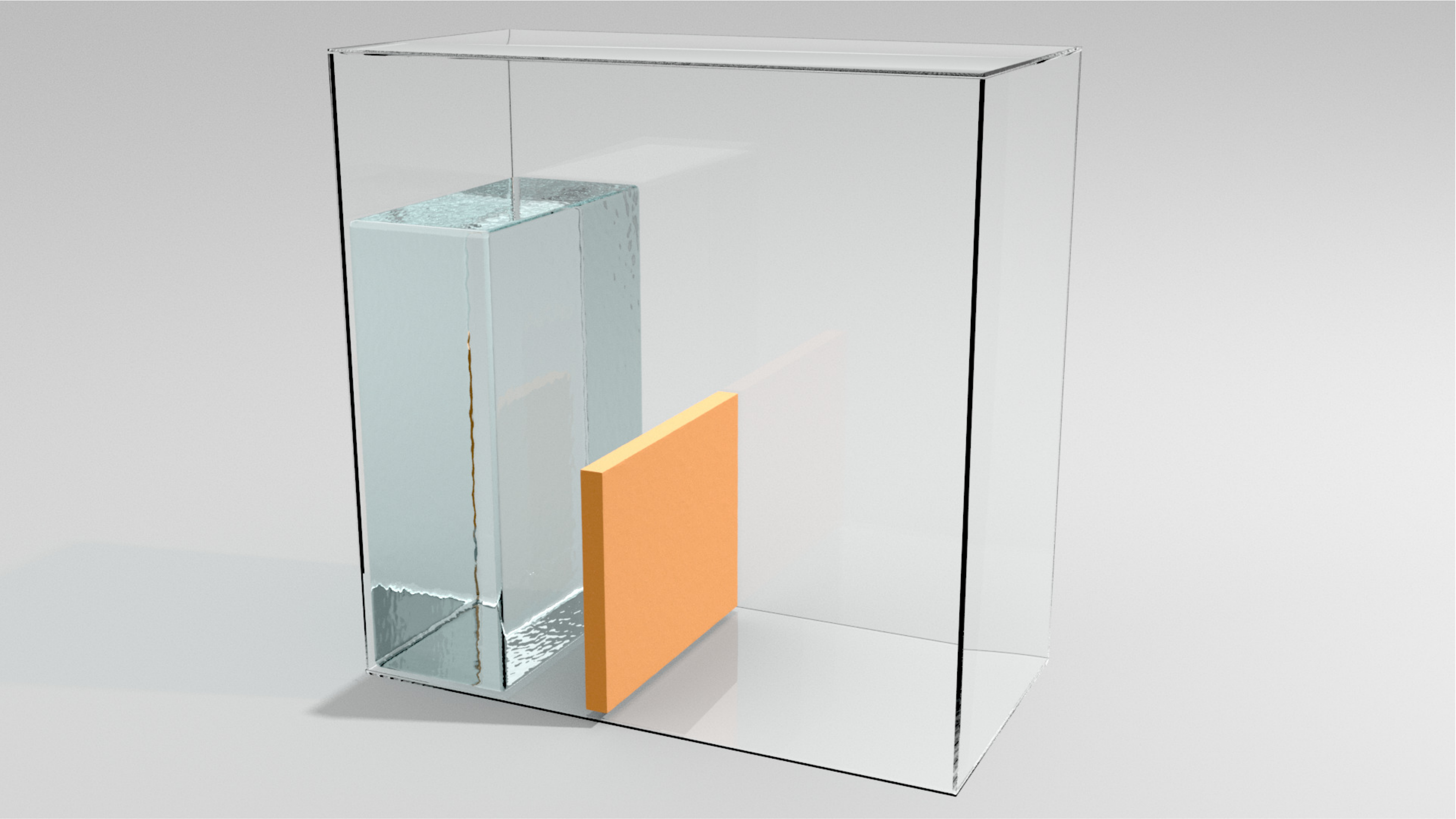}\hfill%
\includegraphics[trim=400 20 450 0, clip, width=\figurewidth]{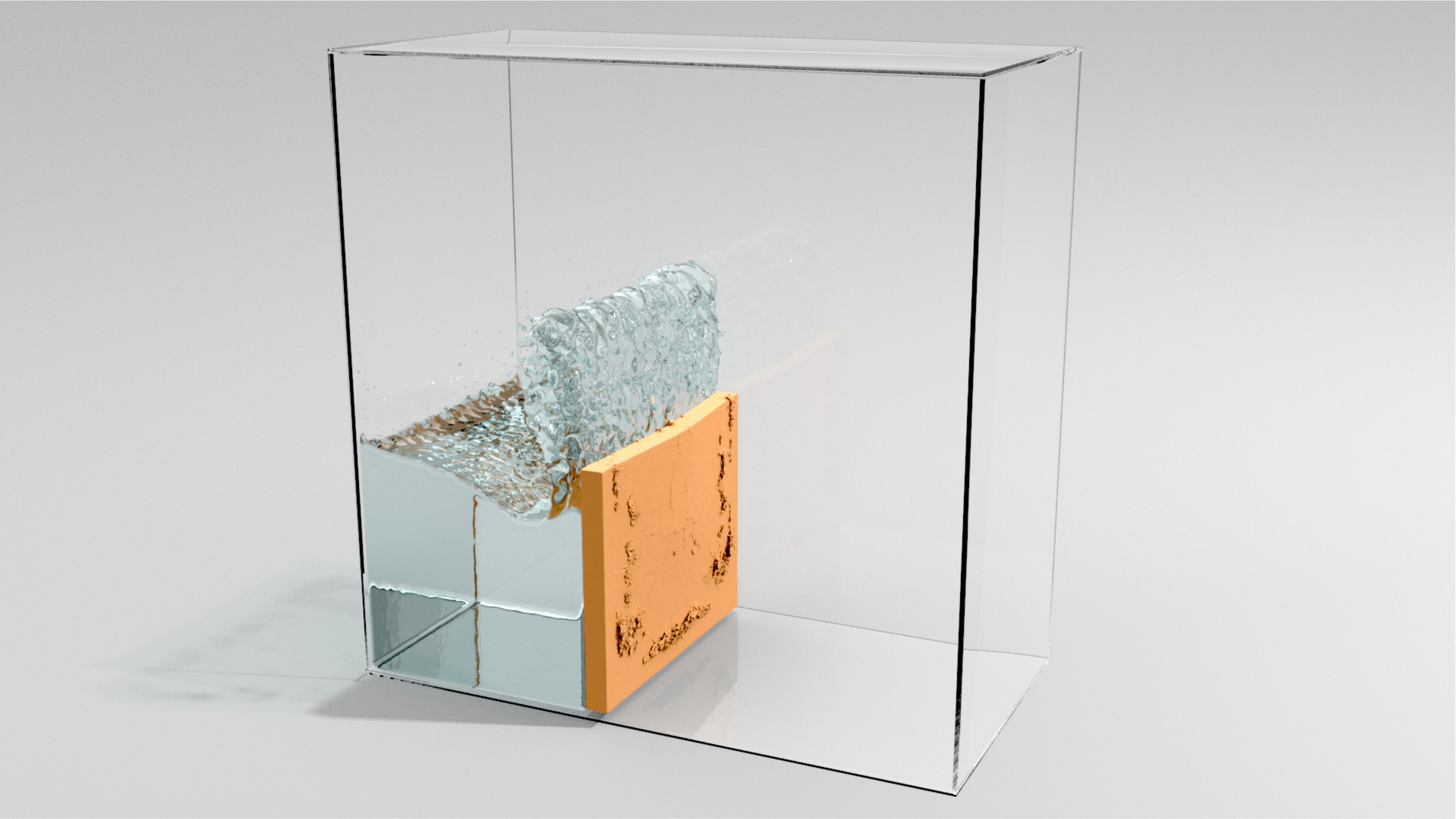}\hfill%
\includegraphics[trim=400 20 450 0, clip, width=\figurewidth]{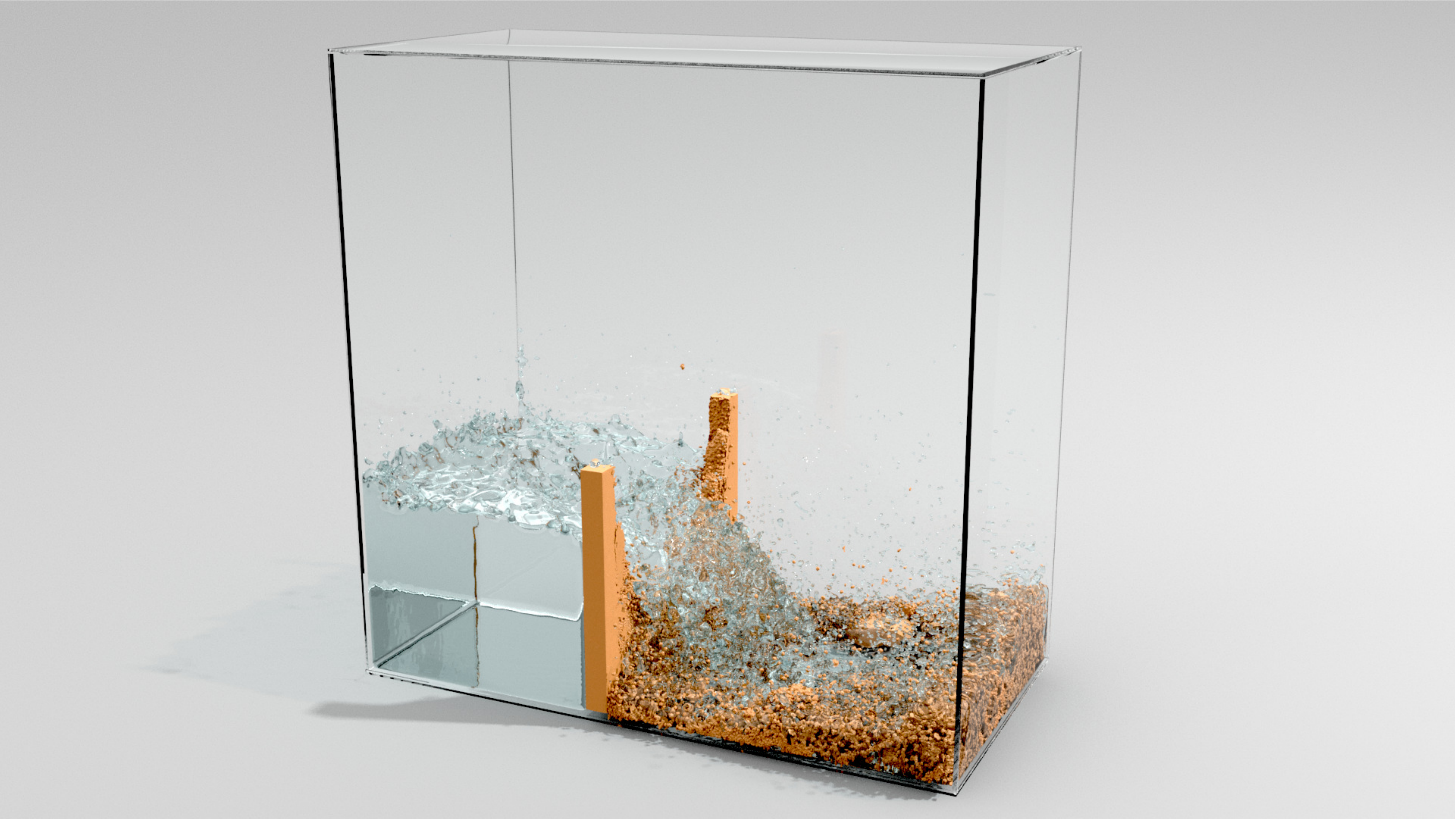}%
\vspace{0.1em}\\%
\includegraphics[trim=420 20 420 70, clip, width=\figurewidth]{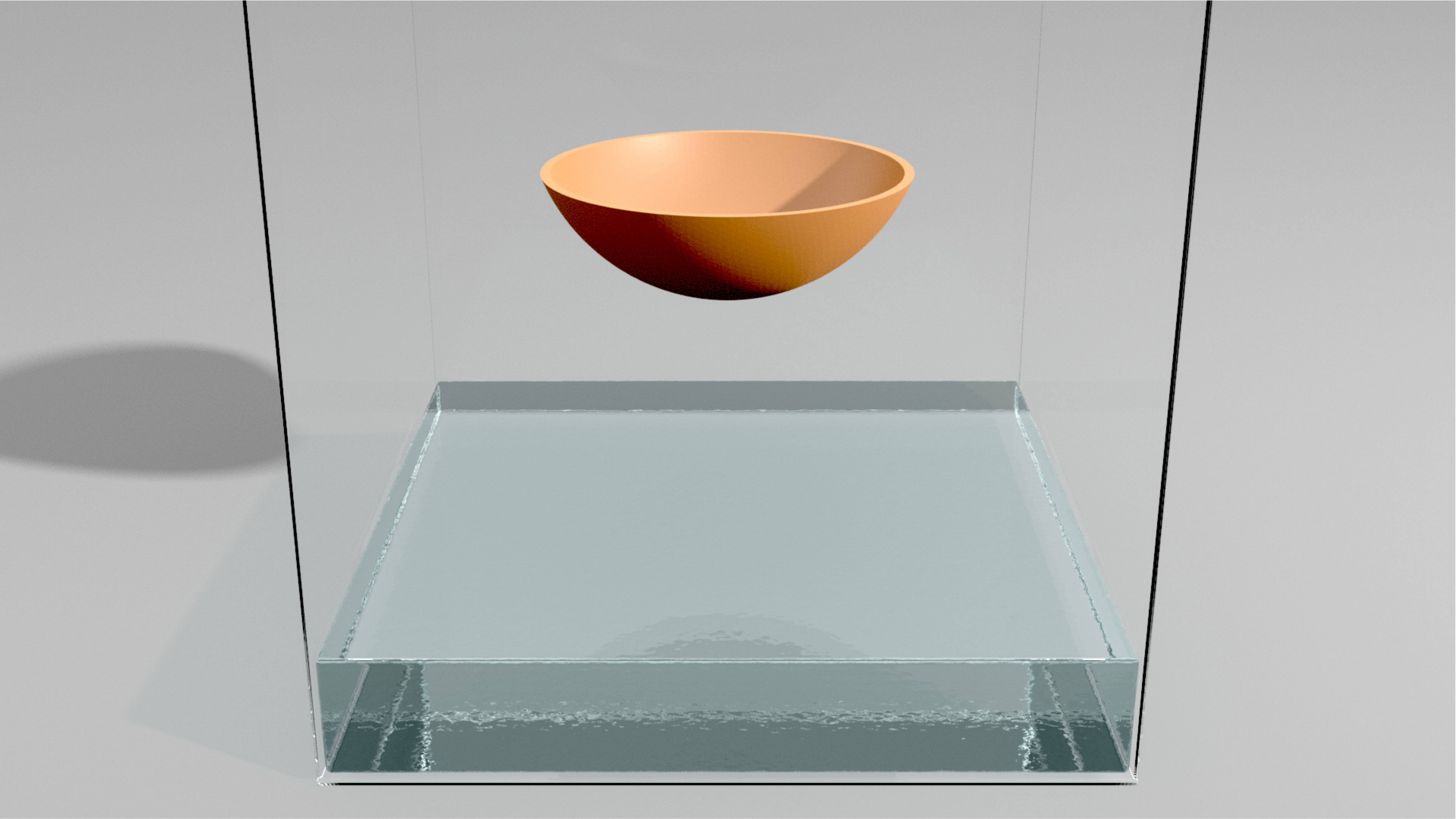}\hfill%
\includegraphics[trim=420 20 420 70, clip, width=\figurewidth]{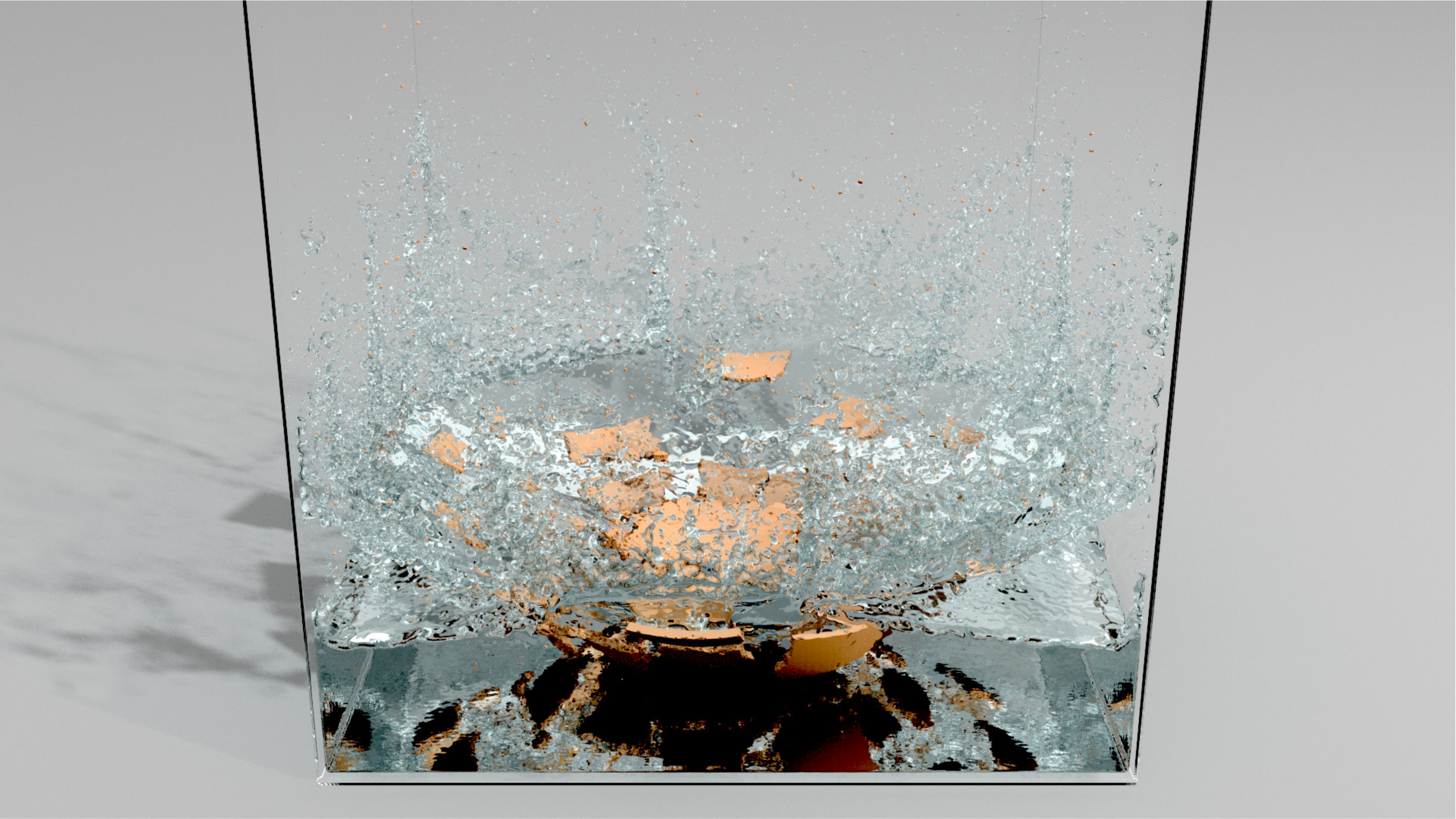}\hfill%
\includegraphics[trim=420 20 420 70, clip, width=\figurewidth]{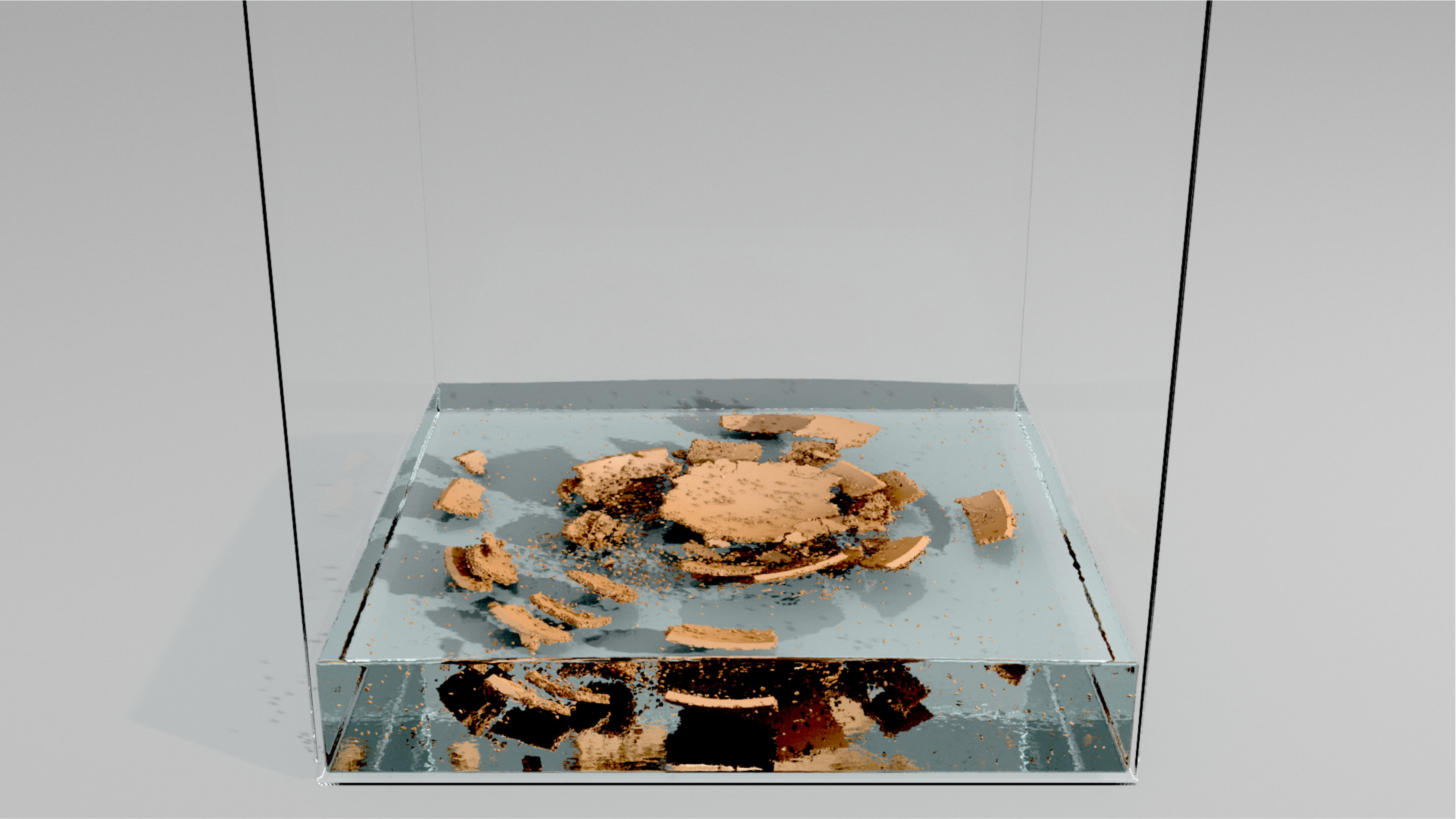}%
\\%
\caption{Example frames from our simulation tests for coupling peridynamics and
SPH using merging-and-splitting.}
  \label{fig:other_examples}
\end{figure}

Different examples of solid-fluid coupling using merging-and-splitting are shown in \autoref{fig:other_examples}. In these examples the fluid particles are simulated using SPH, the cloth is simulated using a mass-spring system, and the brittle solid objects are simulated using peridynamics. The cloth model contains only a single layer of particles. Nonetheless, no fluid particle penetrates through the cloth surface. 
The interactions using merging-and-splitting provide two-way coupling between the fluid and the cloth model as well as the cloth and the tori. The other examples, showing SPH and peridynamics coupling, present solid fracturing due to fluid interaction, enabling new forms of simulation scenarios that can be robustly handled using our merging-and-splitting approach. In the dam break example, the first impact of fluid particles lead to small fractures around the wall. These fractures form weak points that ultimately break the wall and the broken pieces are carried away by the fluid. The next example shows a bowl breaking as it hits the water surface. The reflected waves push the bottom part of the bowl up, together with fluid particles that were previously gathered on it, forming a secondary splash. 
All of these complex examples provide clear evidence of two-way coupling between the two simulation systems, including high-velocity impact situations that are traditionally challenging to handle robustly.

\begin{figure}[tb]
    \centering
\newcommand{\figurewidth}{0.5\linewidth}
\includegraphics[trim=330 0 300 100, clip, width=\figurewidth]{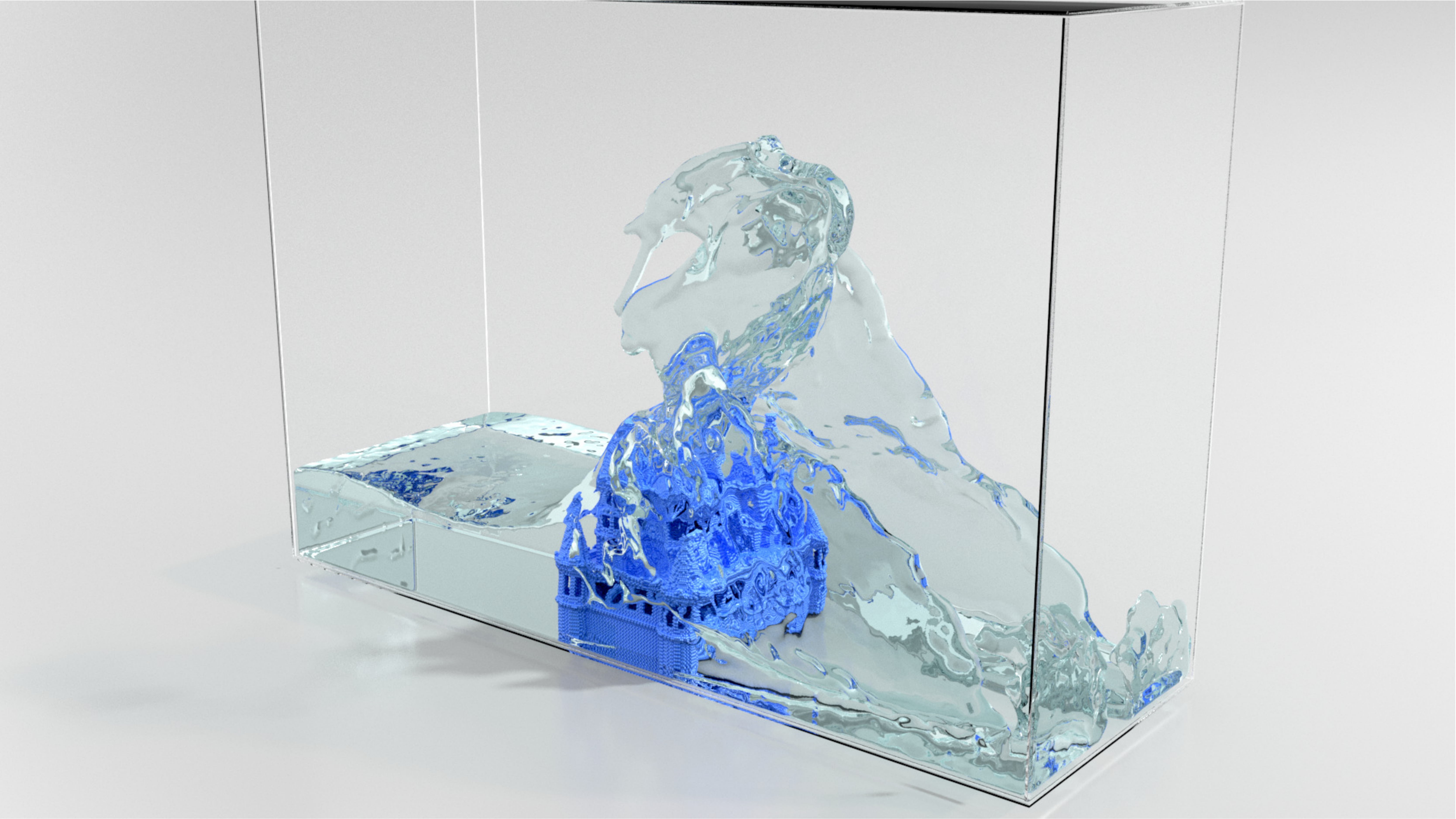}\hfill%
\includegraphics[trim=330 0 300 100, clip, width=\figurewidth]{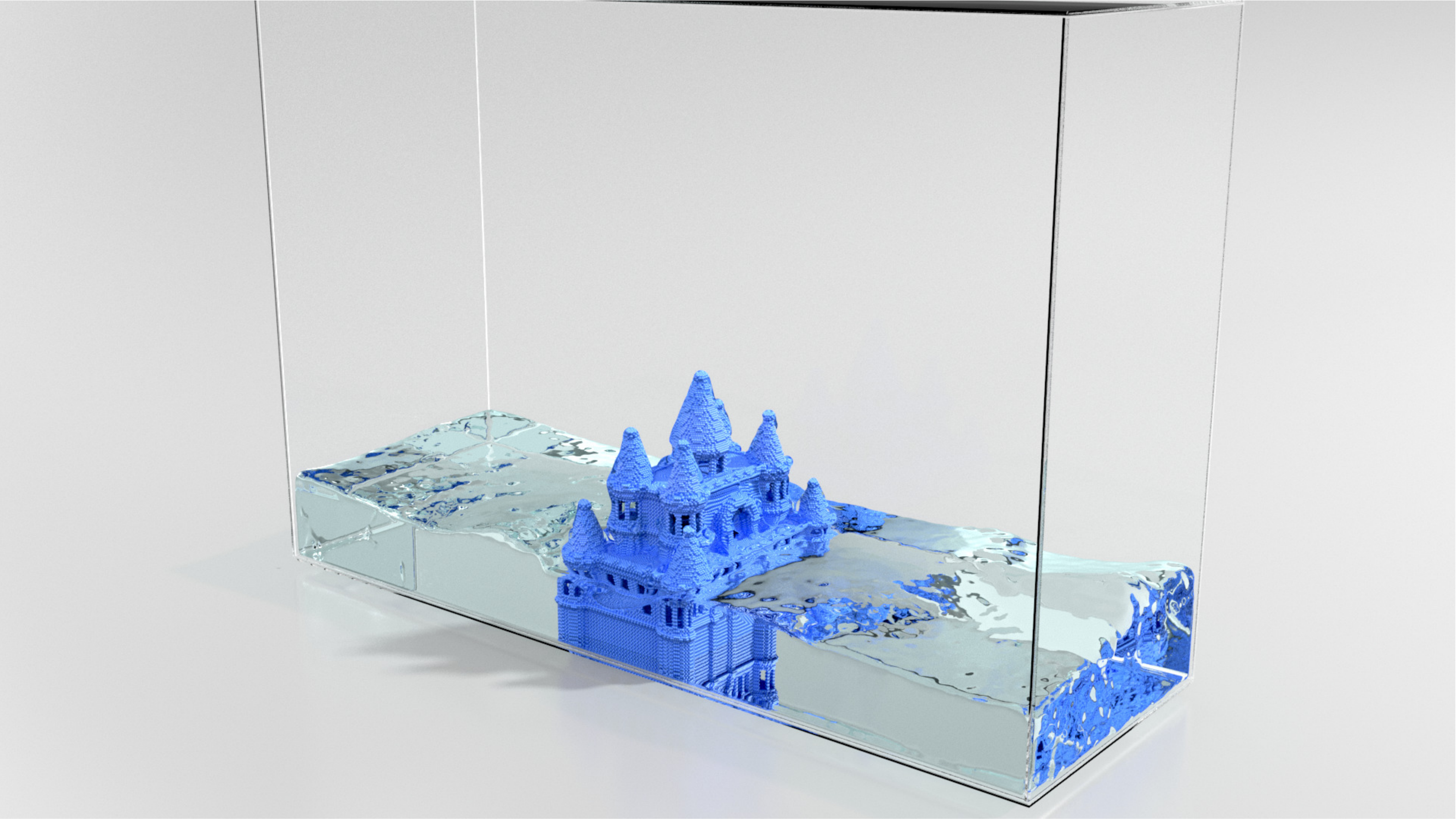}\vspace{0.4em}\\
\includegraphics[trim=330 0 300 100, clip, width=\figurewidth]{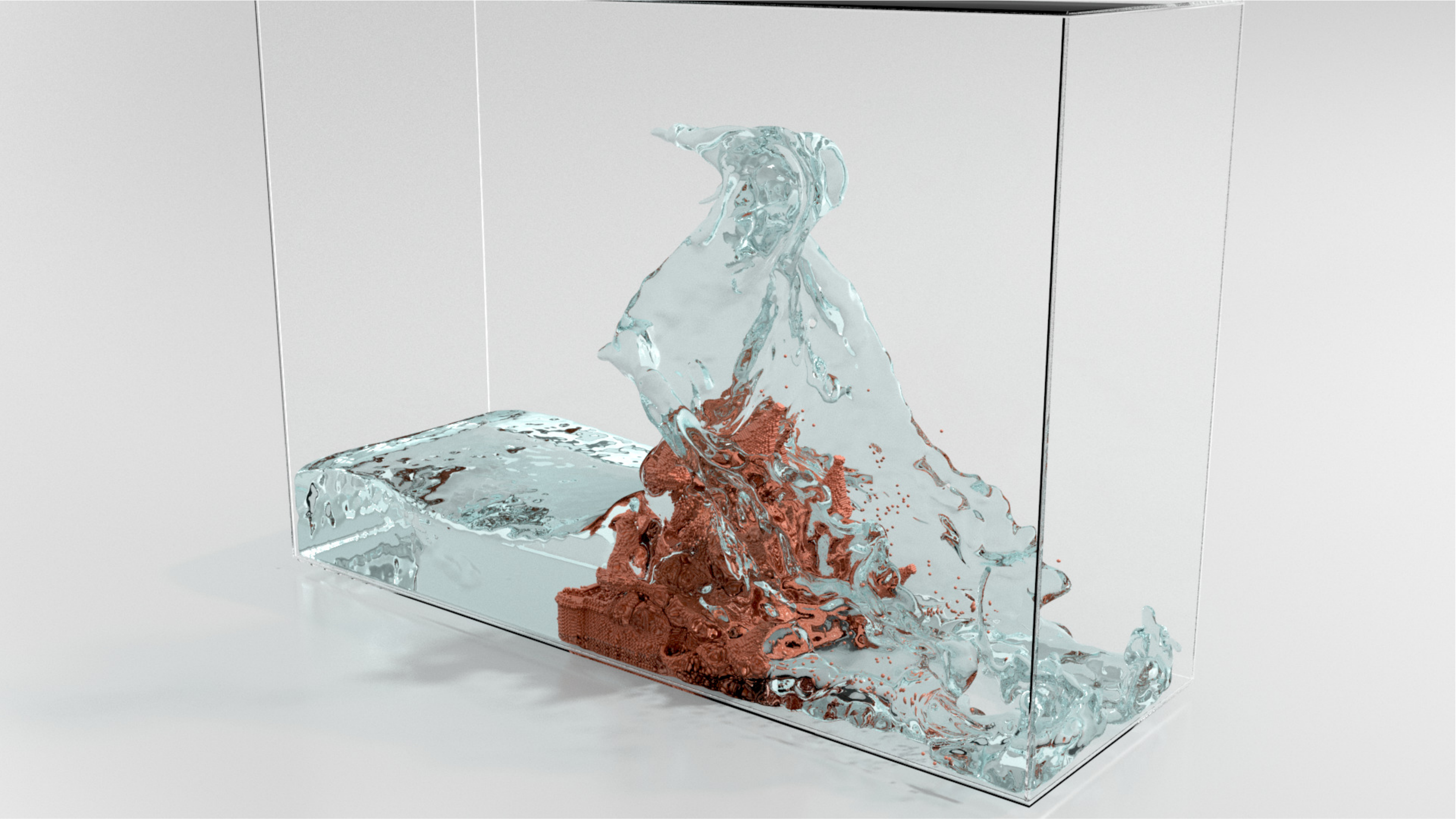}\hfill%
\includegraphics[trim=330 0 300 100, clip, width=\figurewidth]{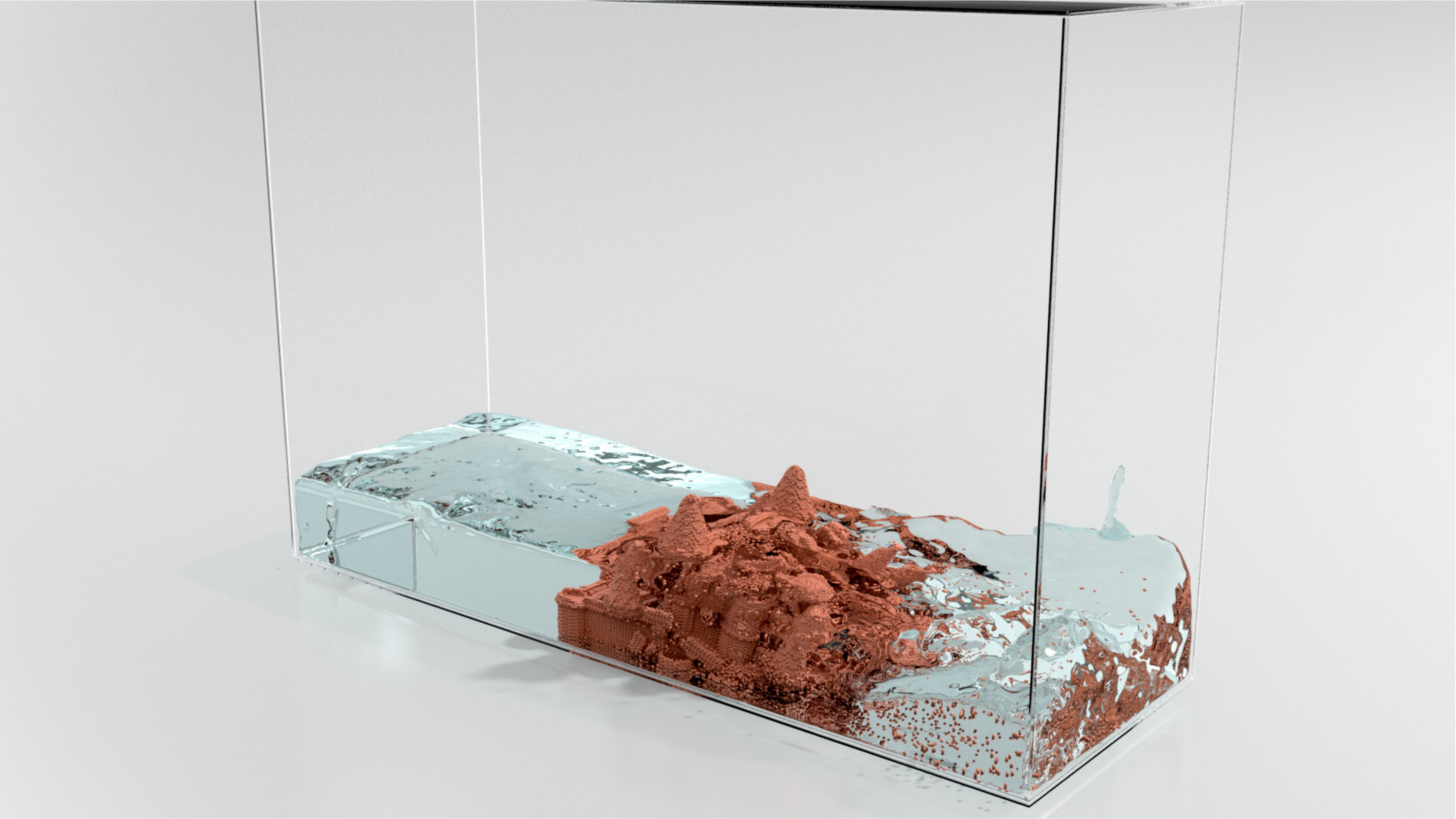}
\\
\caption{Frames from our simulation system coupling FLIP with (top)~mass-spring system and (bottom)~peridynamics.}
  \label{fig:FLIP_example}
\end{figure}

\autoref{fig:FLIP_example} shows similar coupling examples using FLIP and a mass-spring system or peridynamics, simulated using implicit integration. 
Although FLIP typically uses massless marker particles, we assigned mass to fluid particles for simulating interactions with solid particles using our merging and splitting method.
As can be seen in the figure, similar levels of complex interactions between solid and fluid simulations can be achieved using FLIP as well. 
Since FLIP uses an Eulerian pressure solver, in our implementation, coupling FLIP with another particle-based simulation system also involves marking the grid cells occupied by \emph{all} particles in both systems. This is necessary to ensure that the pressure projection step of FLIP correctly identifies which cells are occupied with particles and which cells are empty.
Other than this minor modification to FLIP, the two simulation systems handling solids and fluids run separately, exchanging information via merging and splitting alone.

\newcommand{\tblinf}{$\infty$\hspace*{1em}}
\begin{comment}
\begin{table*}[tb]
    \caption{Average computation times per time step.}\label{tab:performance}
    \centering
        \begin{tabular}{|l|rr|rrrrr|r|}
            \hline&&&&&&&&\\[-1em]
            &\multicolumn{2}{r|}{Particle Count}&Collision & \textbf{Particle}& Integration & Integration &\textbf{Particle} &Total\\
            Scene&\# Solid&\# Fluid& Detection&\textbf{Merging}&Stage-1&Stage-2&\textbf{Splitting}& Time
            \\[0.1em]
            \hline&&&&&&&& \\[-1em]
            \autoref{fig:ball_shoots_wall}                      &$247.9K$&  --   &  53 ms&  3.1 ms& 0.8 sec& 0.7 sec & 0.4 ms& 3.5 sec\\
            \autoref{fig:other_examples} (top)                  &$ 102K$&$ 845K$&  73 ms& 34.0 ms& 0.5 sec& 0.4 sec & 0.4 ms& 1.6 sec\\
            \autoref{fig:other_examples} (middle)               &$112.5K$&$ 655K$& 270 ms& 66.5 ms& 2.6 sec& 1.9 sec & 8.9 ms& 5.3 sec\\
            \autoref{fig:other_examples} (bottom)               &$121.9K$&$ 6.5M$& 392 ms&222.6 ms&12.9 sec&12.1 sec &18.2 ms&27.4 sec\\
            \autoref{fig:FLIP_example} (top)        &$  191K$&$ 1.4M$&  42 ms& 40.5 ms& 4.5 sec& 4.2 sec & 10.4 ms& 10.6 sec\\
            \autoref{fig:FLIP_example} (bottom)     &$  191K$&$ 1.4M$&  43 ms& 37.0 ms& 2.2 sec& 2.1 sec & 9.7 ms& 6.4 sec\\
            \hline
    \end{tabular}
\end{table*}
\end{comment}

\newcommand{\tblskip}{@{\hskip 0.7em}}
\newcommand{\tblfigheight}{0.1\linewidth}
\begin{table}[tb]
    \caption{Average computation times per time step.}\label{tab:performance}
    \centering
    \resizebox{\linewidth}{!}{%
     \begin{tabular}{@{\hskip 0in}l\tblskip r\tblskip r\tblskip r\tblskip r\tblskip r\tblskip r@{\hskip 0in}}
        & &
        \autoref{fig:other_examples} &
        \autoref{fig:other_examples} &
        \autoref{fig:other_examples} &
        \autoref{fig:FLIP_example} &
        \autoref{fig:FLIP_example} \\
        Scene&
        \autoref{fig:ball_shoots_wall} & top & middle & bottom & top & bottom\\
        \hline&&&&&&\\[-1em]
        Solid Particles & 247.9K & 102K & 112.5K & 121.9K & 191K & 191K\\
        Fluid Particles & -- & 845K & 655K & 6.5M & 1.4M & 1.4M\\
        \hline&&&&&&\\[-1em]
        Collision Detection & 53 ms & 73 ms & 270 ms & 392 ms & 42 ms & 43 ms\\
        \textbf{Particle Merging} & 3.1 ms & 34.0 ms & 66.5 ms & 222.6 ms & 40.5 ms & 37.0 ms\\
        Integration Stage-1 & 0.8 sec & 0.5 sec & 2.6 sec & 12.9 sec & 4.5 sec & 2.2 sec \\
        Integration Stage-2 & 0.7 sec & 0.4 sec & 1.9 sec & 12.1 sec & 4.2 sec & 2.1 sec \\
        \textbf{Particle Splitting} &  0.4 ms & 0.4 ms & 8.9 ms & 18.2 ms & 10.4 ms & 9.7 ms\\
        \hline&&&&&&\\[-1em]
        Total Time & 3.5 sec & 1.6 sec & 5.3 sec & 27.4 sec & 10.6 sec & 6.4 sec \\
    \end{tabular}
    }
\end{table}

\subsection{Performance}

The performance of our simulations largely depends on the performances of the underlying particle-based systems used. The performance results of our tests are included in the \autoref{tab:performance}. Notice that particle merging and particle splitting operations take only a negligible fraction of the computation time. Most of the computation time is spent in the integration steps. In particular, peridynamics integration, involving an implicit solver for a large number of particles, each of which is connected to hundreds of other particles via stiff springs, can be considerably slow. Yet, because our implementation uses a two-stage integration scheme, our merging-and-splitting implementation effectively doubles the computation time (by introducing an additional integration step). This additional overhead can be reduced by using the result of the first integration step as the initial guess for the second integration step, an optimization that is not included in our tests.

\begin{figure}[tb]
\setlength{\tabcolsep}{0.1em}
\renewcommand{\arraystretch}{1.5}
\newcommand{\figurewidth}{0.33\columnwidth}
\centering
\begin{tabular}{ccc}
\includegraphics[trim=390 100 520 270, clip,width=\figurewidth]{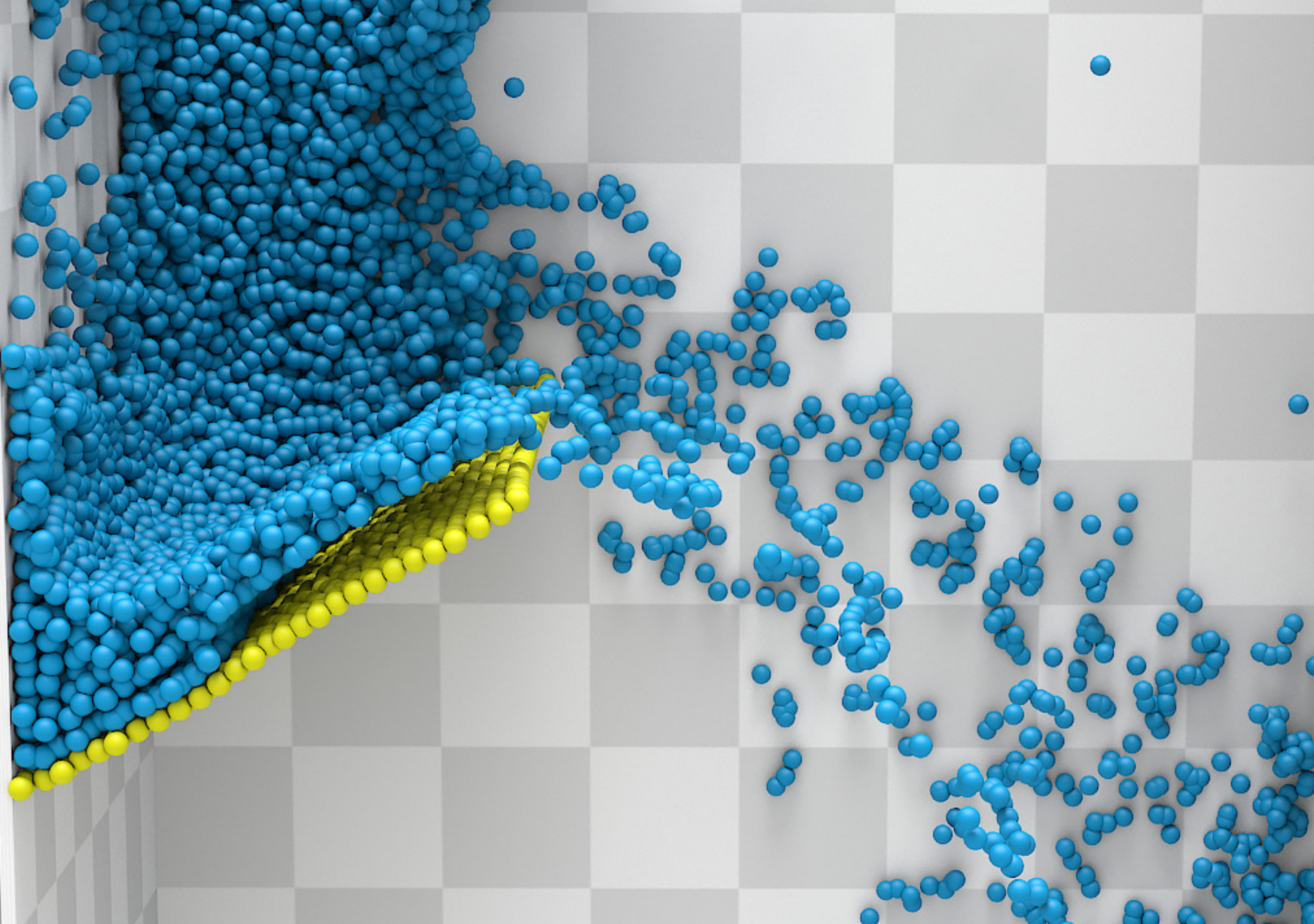}&%
\includegraphics[trim=390 100 520 270, clip,width=\figurewidth]{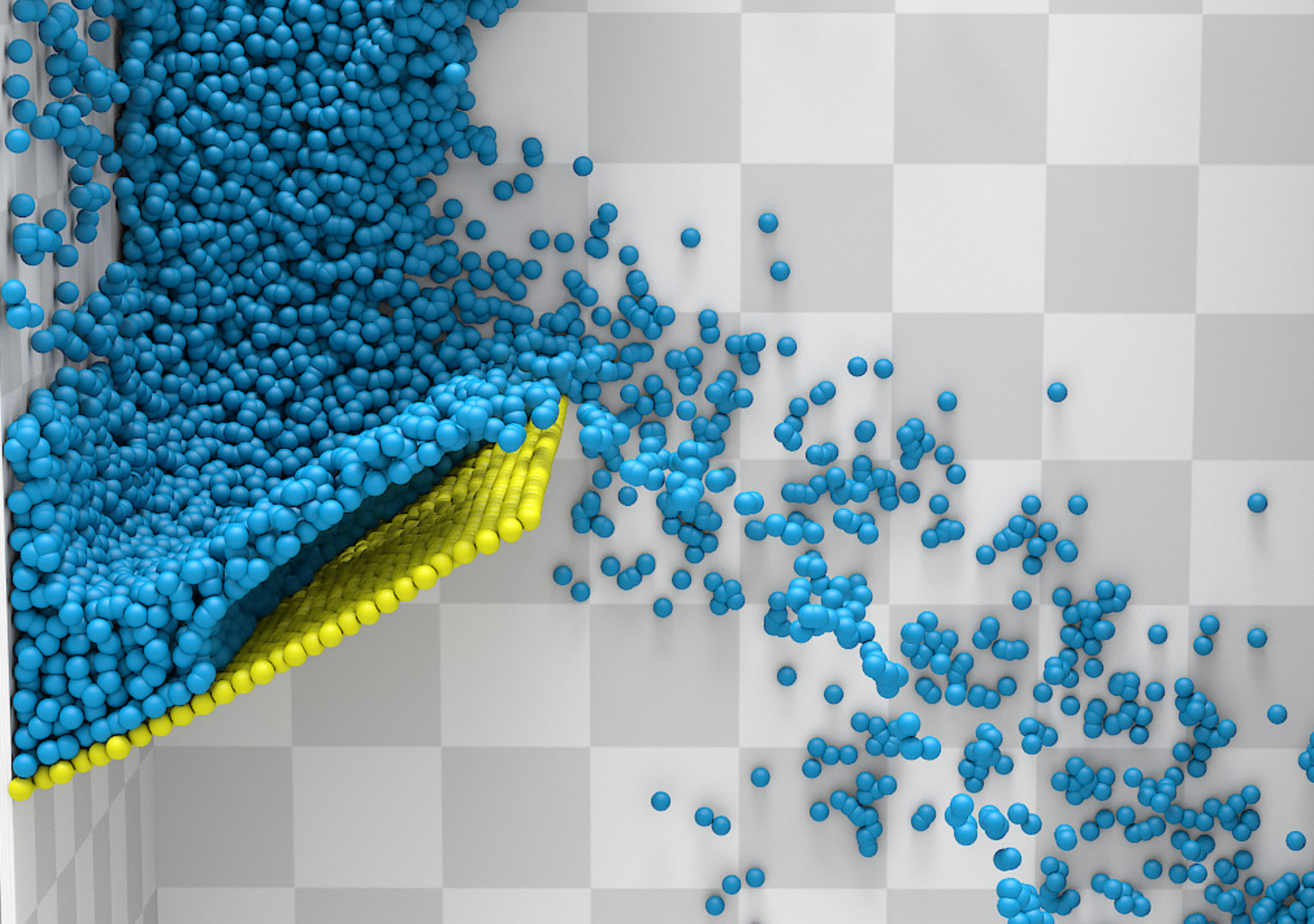}&%
\includegraphics[trim=390 100 520 270, clip,width=\figurewidth]{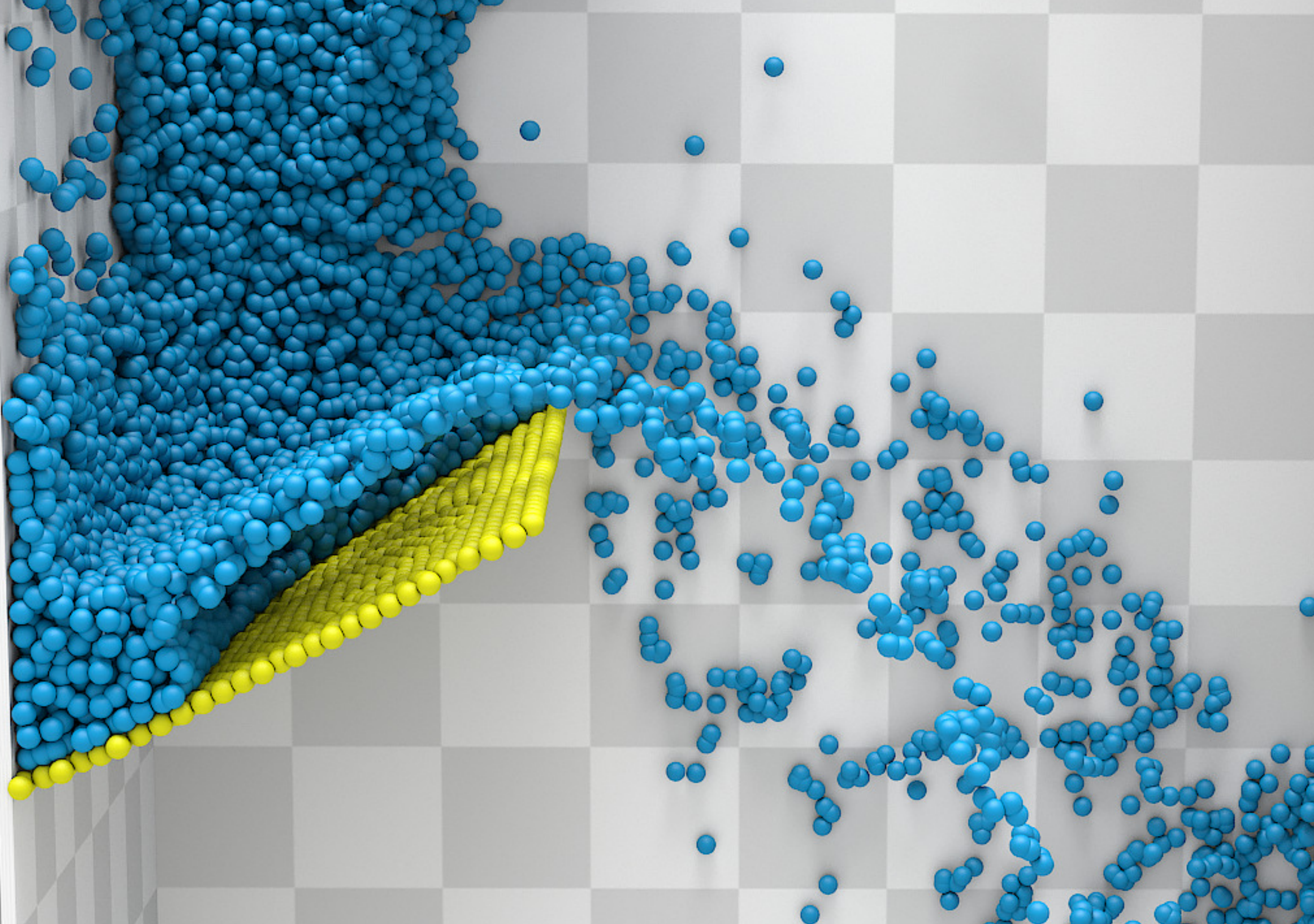}\vspace{-0.5em}\\
(a) $\alpha=0$, $\beta=0$&%
(b) $\alpha=1$, $\beta=0$&%
(c) $\alpha=1$, $\beta=1$\\%
\end{tabular}
\caption{Cross-section view of a simple solid-fluid coupling test, including a column of fluid falling onto an elastic cloth model, demonstrating the impact of different energy conservation parameters: (a)~no energy preservation with $\alpha=0$ and $\beta=0$, (b)~full energy preservation for merging using $\alpha=1$ and no energy preservation with velocity synchronization using $\beta=0$, and (c)~full energy preservation with $\alpha=1$ and $\beta=1$.}
\label{fig:beta_test}
\end{figure}

\begin{figure*}[tb]
\newcommand{\figurewidth}{0.164\linewidth} % for 4 columns
\newcommand{\figspaceA}{\hspace*{1.5em}}
\newcommand{\figspaceB}{\hspace*{1.3em}}
\setlength{\tabcolsep}{0.1em}
\renewcommand{\arraystretch}{1.5}
\centering
\begin{tabular}{c|lllll}
\includegraphics[trim=465 100 465 250,clip,width=\figurewidth]{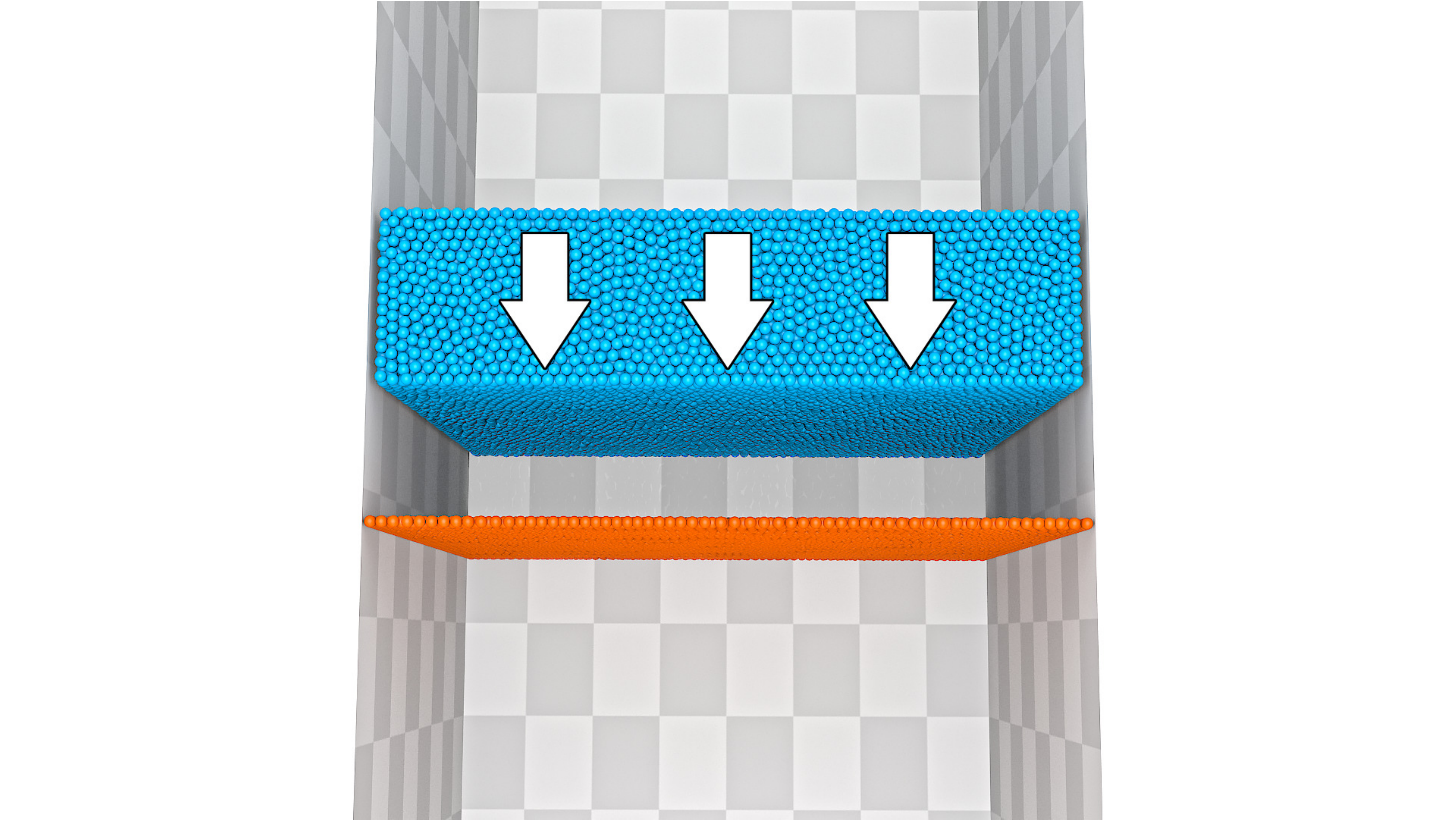}&%
\includegraphics[trim=465 100 465 250,clip,width=\figurewidth]{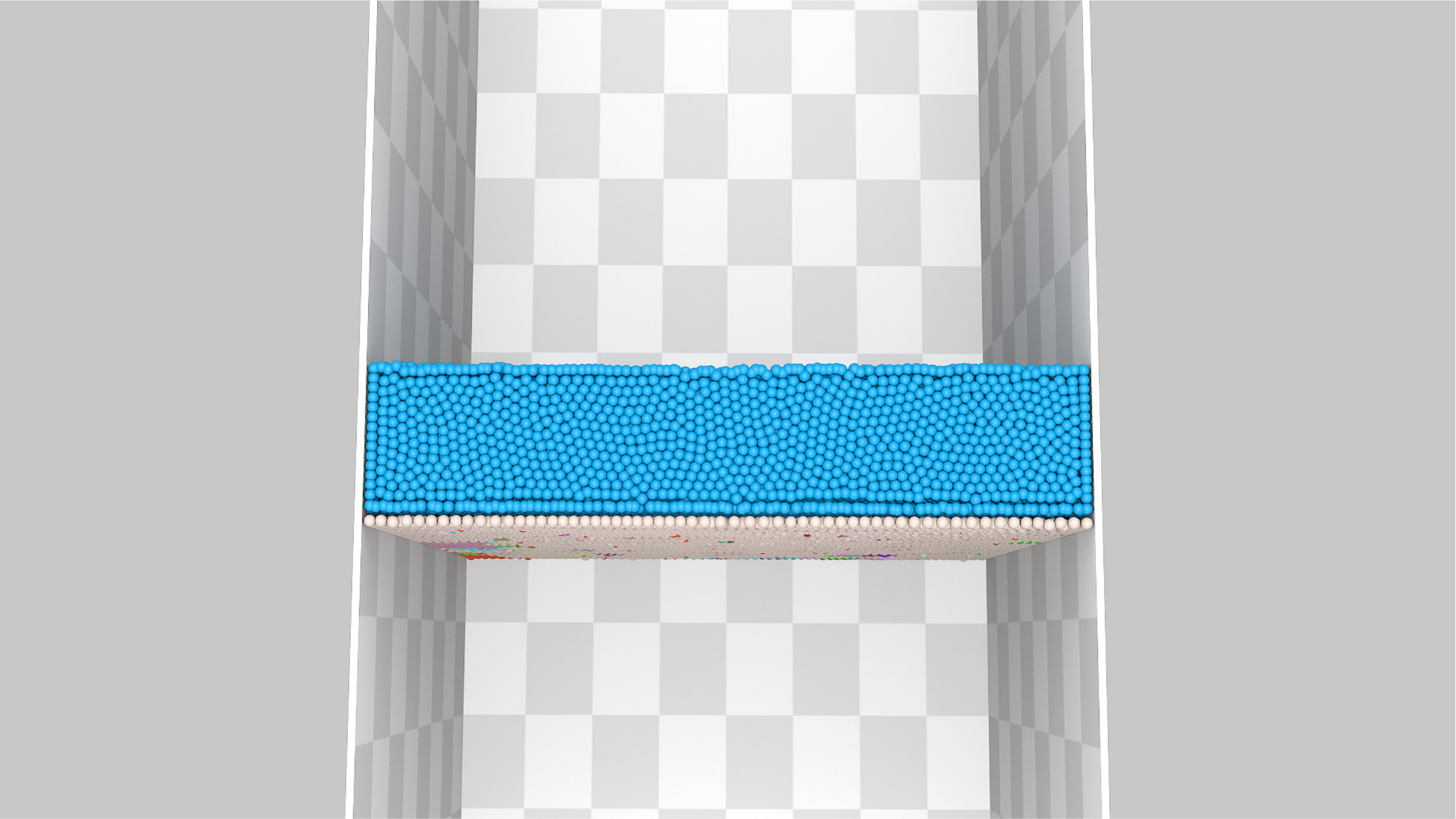}&%
\includegraphics[trim=465 100 465 250,clip,width=\figurewidth]{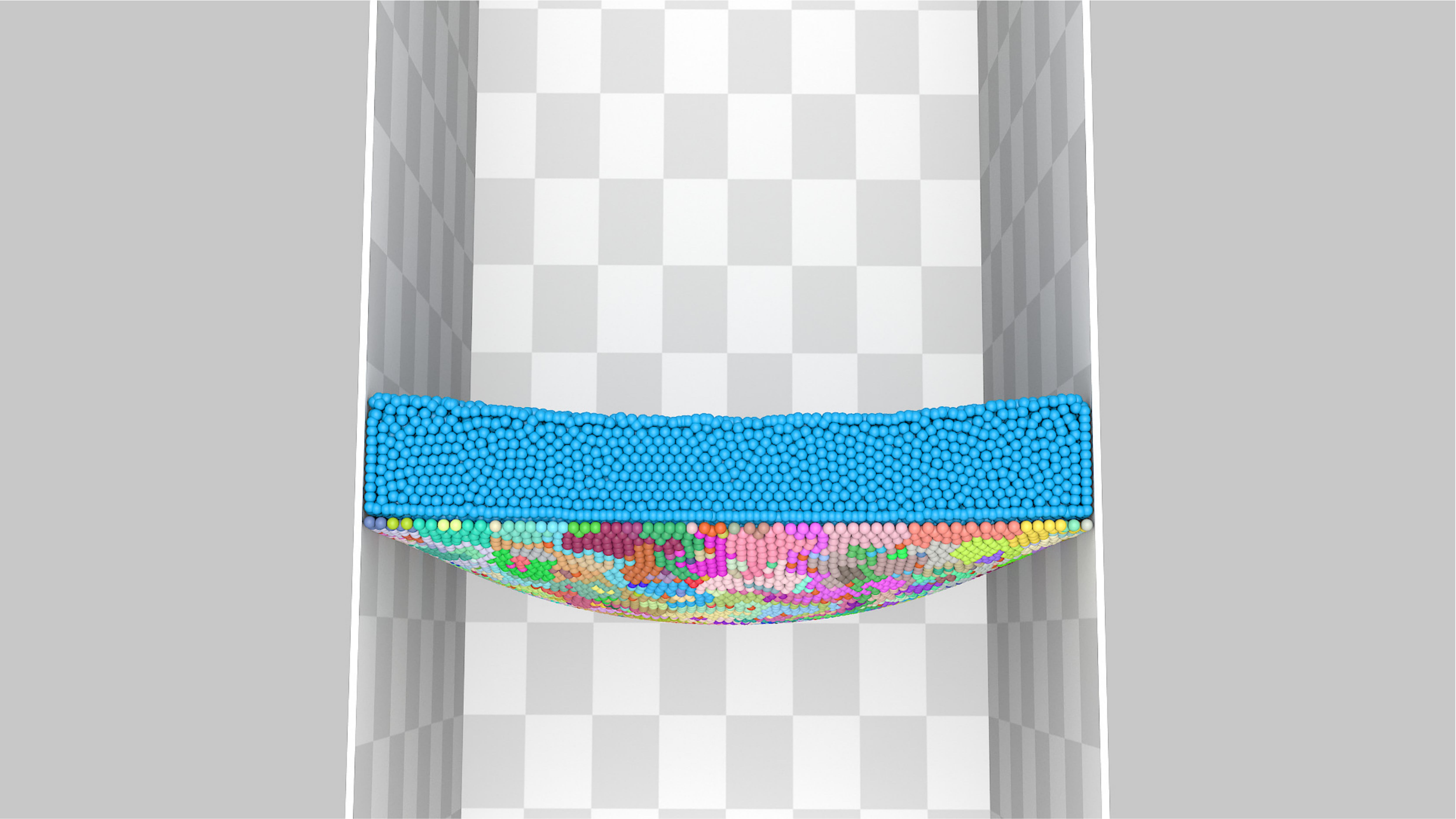}&%
\includegraphics[trim=465 100 465 250,clip,width=\figurewidth]{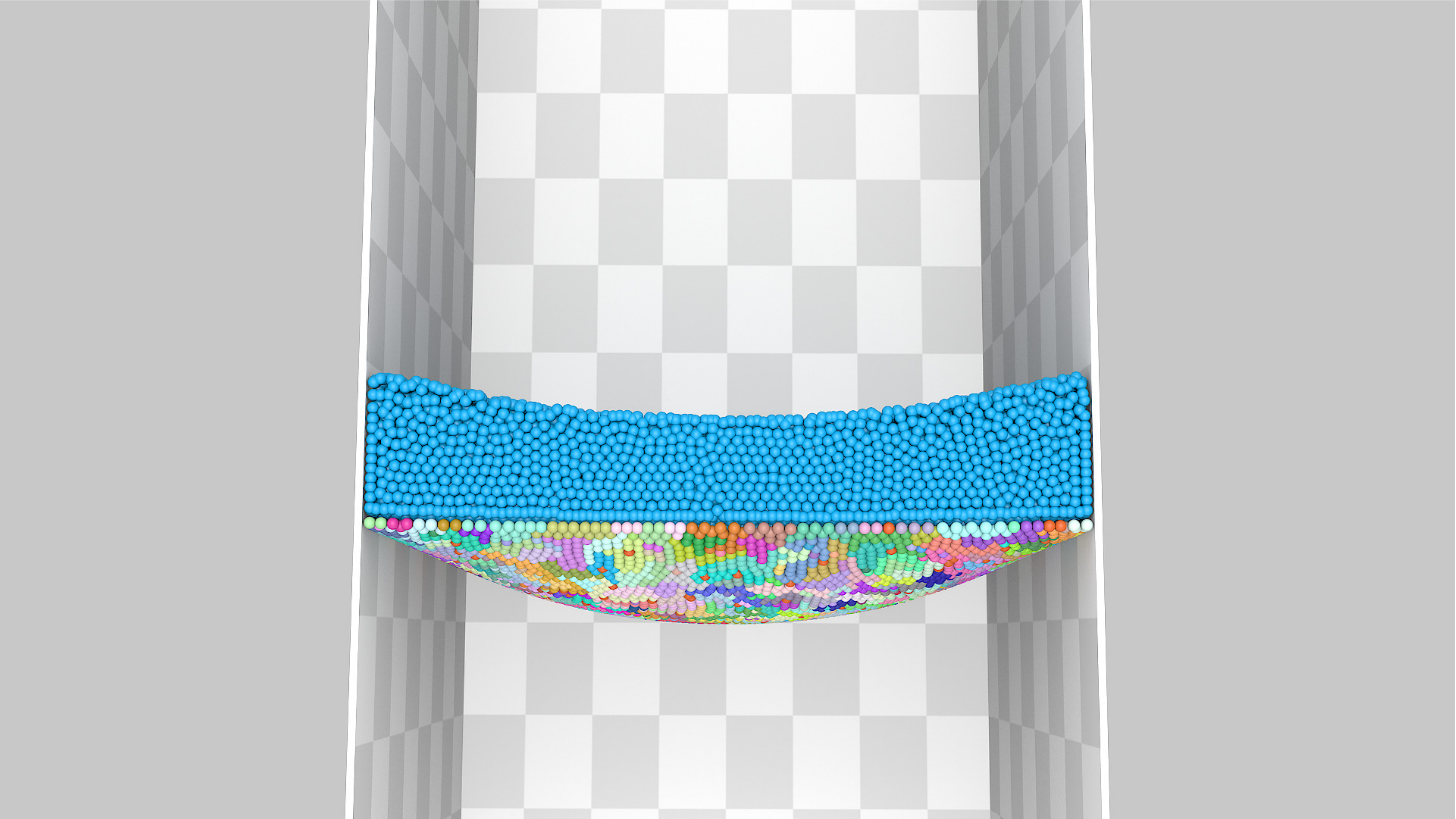}&%
\includegraphics[trim=465 100 465 250,clip,width=\figurewidth]{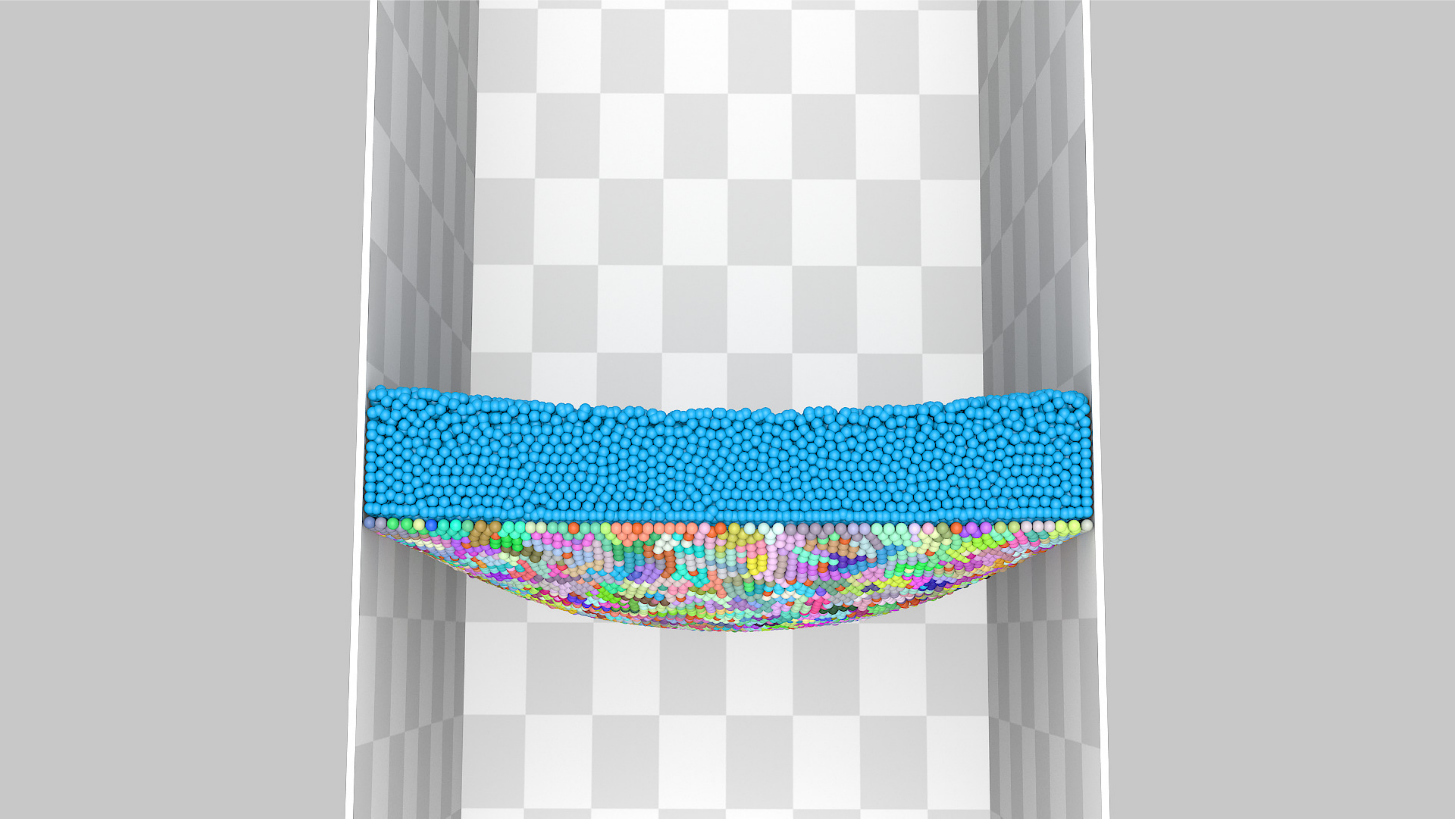}&%
\includegraphics[trim=465 100 465 250,clip,width=\figurewidth]{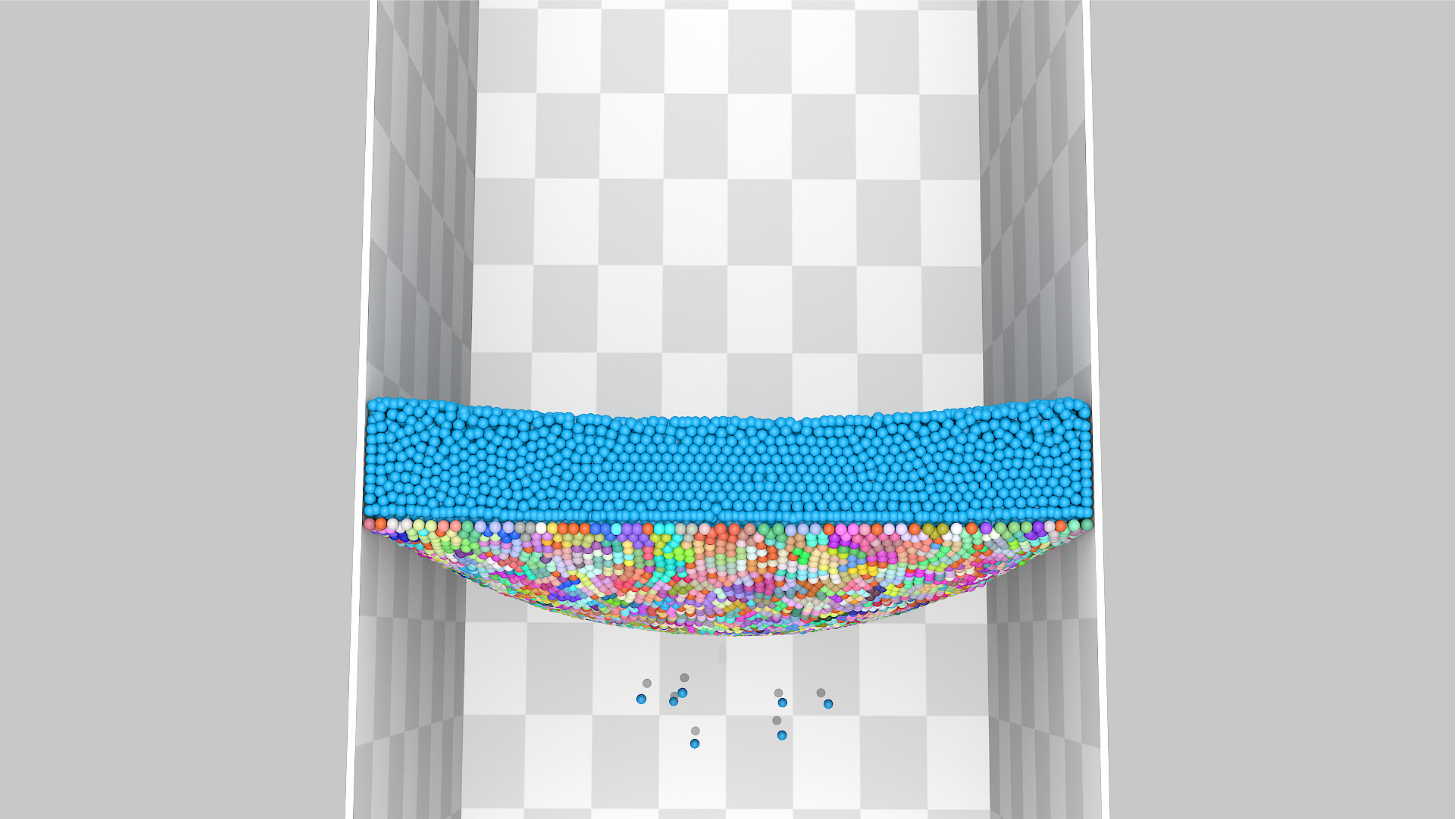}%
\vspace{-0.5em}\\
\small (a)~Initial\hspace*{2em}&
\small\figspaceA (b) $n_{\min} = \infty$, &%
\small\figspaceA (c) $n_{\min} = 32$, &%
\small\figspaceA (d) $n_{\min} = 16$, &%
\small\figspaceA (e) $n_{\min} = 8$, &%
\small\figspaceA (f) $n_{\min} = 4$,\vspace{-0.7em}\\
\small\figspaceA conditions&
\small\figspaceA\figspaceB $n_{\max} = \infty$&%
\small\figspaceA\figspaceB $n_{\max} = 64$ &%
\small\figspaceA\figspaceB $n_{\max} = 32$ &%
\small\figspaceA\figspaceB $n_{\max} = 16$ &%
\small\figspaceA\figspaceB $n_{\max} = 8$ \\
\end{tabular}
\caption{A block of fluid falling with gravity onto a soft elastic cloth: (a)~the initial conditions before the collision and (b-f)~the simulation state a few seconds after the collision, comparing different parameters for $n_{\min}$, and $n_{\max}$ that limit the number of particles in a meta-particle.
Cloth particles having the same color are merged into the same meta-particle. Notice that (b)~using an unlimited number of particles in a meta-particle with $n_{\min}=n_{\max}=\infty$ leads to rigidification and (f)~using too small limits causes leaks.}
\label{fig:comparison_merge_list_size}
\end{figure*}

\subsection{Parameters}

Our merging-and-splitting approach has only $\alpha$ and $\beta$ parameters that control the energy conservation behavior for collisions and coupling different integrators (\autoref{eq:s2}). With $\alpha=\beta=1$ all merging-and-splitting operations fully conserve energy and all energy loss is due to damping or time-step integration. Using $\alpha=\beta=0$, a portion of the energy is still conserved, but only as needed for momentum conservation, and merging-and-splitting operations dissipate the rest.
The impact of different $\alpha$ and $\beta$ parameters with merging-and-splitting is shown with a simple example in \autoref{fig:beta_test}. Notice that energy conservation in this example has a relatively minor impact on the final result, causing the fluid to jump back slightly higher after impact. 
This is typical for all simulations we have tested, but it may be possible to design counter examples where energy conservation can play a more prominent role.

Our implementation also includes $n_{\min}$ and $n_{\max}$ parameters that limit the number of particles within a meta-particle.
\autoref{fig:comparison_merge_list_size} shows impact of limiting the number of particles in a meta-particle. In this example, a block of fluid fall onto an elastic cloth, such that the initial impact covers the entire cloth surface (\autoref{fig:comparison_merge_list_size}a). Therefore, allowing an unlimited number of particles to merge into a single meta-particle effectively rigidifies the entire cloth, significantly impacting the simulation outcome (\autoref{fig:comparison_merge_list_size}b). In this case, the rigidification remains permanent, as the solid-fluid contact persists, preventing any deformation. Limiting the number of particles in a meta-particle using limit parameters $n_{\min}$ and $n_{\max}$ leads to similar results (\autoref{fig:comparison_merge_list_size}c-e). However, when the limit is too small, it prevents properly resolving some collision events, failing to stop some fluid particles to pass through the cloth model (\autoref{fig:comparison_merge_list_size}f). While this example has been carefully chosen to present the problems of using a very small limit (i.e. ${n_{\min}=4}$) or an unlimited number of particles in a meta-particle, with all our experiments (including this one) we observed similar results when using parameter values for $n_{\min}$ and $n_{\max}$ set to 8 to 64, respectively. Therefore, we conclude that our implementation is not sensitive to the values of the limit parameters for the scenes we tested.

\section{Discussion}

Conceptually, our merging-and-splitting scheme can be considered similar to impulse-based collisions. However, unlike explicit impulse-based collisions that instantaneously resolve the collisions, we keep the particles in contact for the entire duration of a time step. This is a crucial component of our method, which allows colliding objects to exchange a substantial amount of momentum, beyond what can be stably accomplished using force-based or impulse-based formulations within a time step. 
When using explicit integration, this guarantees that the particles do not penetrate further during the time step. With implicit integration, merging allows information exchange between different colliding bodies through the meta-particles while solving the implicit system. This makes merging-and-splitting particularly favorable for implicit integration.

Since meta-particles are split at the end of each time step with relative velocities of colliding particles guaranteed to be pointing away from each other, colliding particles should not be considered ``glued,'' though they maintain contact throughout a time step. Meta-particles merely facilitate neighboring particles on either side to exchange information.

Note that most of the problems we present regarding force-based collision formulations can be resolved by increasing the stiffness and reducing the time-step size or by solving for the magnitude of the force/impulse within an implicit system. 

Position correction is a common approach used in computer graphics for preventing penetrations. 
In fact, position correction (along with parameter tuning) plays a crucial role in various simulations in prior work using force-based collision models to demonstrate seemingly stable collision responses. 
However, position correction artificially injects or removes energy. This can have catastrophic results especially with fracture simulations of brittle materials using peridynamics. Furthermore, position correction may move particles into collisions with other particles. Therefore, we have entirely avoided position corrections in all our tests in this paper. Consequently, we do not compare our approach to position-based dynamics \cite{Muller:2006, Macklin:2014, Bender:2014}, which completely relies on position updates.

Merging-and-splitting using the implementation we describe in this paper can properly handle rest-in-contact situations. When a particle rests on another particle, both with no velocity, they are not merged before our first integration step. After the first integration step, however, the resulting velocity changes indicate that the particles must be merged. As a result, they preserve their relative positions during the recomputation in the second integration step.

While our merging and splitting operations are based on the principles of momentum and energy conservation, we impose no restrictions on the integrators, which are treated as black-boxes. Therefore, the velocity update provided by the integrators do not necessarily conserve energy or momentum.

While we have used similar particle sizes in most of our tests, merging-and-splitting does not inherently require the particles to have a uniform size. Particles with significantly different sizes can be handled, as long as collisions can be safely detected. Similarly, if the particle distribution is not dense enough and that the material includes large-enough holes, penetrations may occur.

The orthogonal momentum exchange in \autoref{eq:ray} can be considered a form of frictional contact. Yet, this is not a physically-based friction formulation. Instead, our formulation aims to minimize orthogonal momentum transfer between colliding particles. Thus, the lack of a physically-based friction model is a limitation of our meta-particle splitting formulation. It is important to note that properly modeling friction may require modifying time-step integration accordingly, which we deliberately avoided to provide a general coupling solution for otherwise incompatible simulation systems.

Another limitation of our formulation is that we assume that meta-particles maintain the relative positions of colliding particles throughout the time step. An interesting future direction would be exploring angular momentum conservation for meta particles and introducing rotations during the time-step integration or prior to splitting. This could be handled by treating meta-particles as rigid bodies, similar to rigid impact zones \cite{Bridson:2002}. However, this would also require modifying the particle-based simulation system accordingly, so that it can handle rigid bodies along with particles.

\section{Conclusion}
We have introduced merging-and-splitting, a new model for robustly handling collisions with particle-based simulations.
This approach also allows coupling different particle-based simulation systems using different integrators that are designed for
representing different material types and phases. We have shown that our method is effective in handling collision within
a simulation system and coupling separate simulations of different materials. We have also shown novel simulation examples involving solid fracture due to fluid interaction.

An interesting direction for future research would be testing the effectiveness of the
merging-and-splitting approach for reproducing macro-scale behavior in solid-fluid
coupling, such as computing drag and lift forces, capillary effect, as well as buoyancy. 
Also, using merging-and-splitting for coupling different fluids, such as gases and liquids, may
reveal interesting challenges and new applications for future research.
In addition, since merging-and-splitting only considers kinetic energies and linear momenta of the colliding particles, taking angular momentum into account with merging-and-splitting would be an interesting future research direction for physics-based animation.

\ifCLASSOPTIONcaptionsoff
  \newpage
\fi

% \ifCLASSOPTIONcompsoc
%   \section*{Acknowledgments}
% \else
% \fi

% We would like to thank ......

\bibliographystyle{IEEEtran}

%\addbibresource{ParticleMergingSplitting.bib}

\bibliography{ParticleMergingSplitting}

\newcommand{\biospace}{\vspace{-4em}}

\begin{IEEEbiography}[{\includegraphics[width=1in,height=1.25in,keepaspectratio]{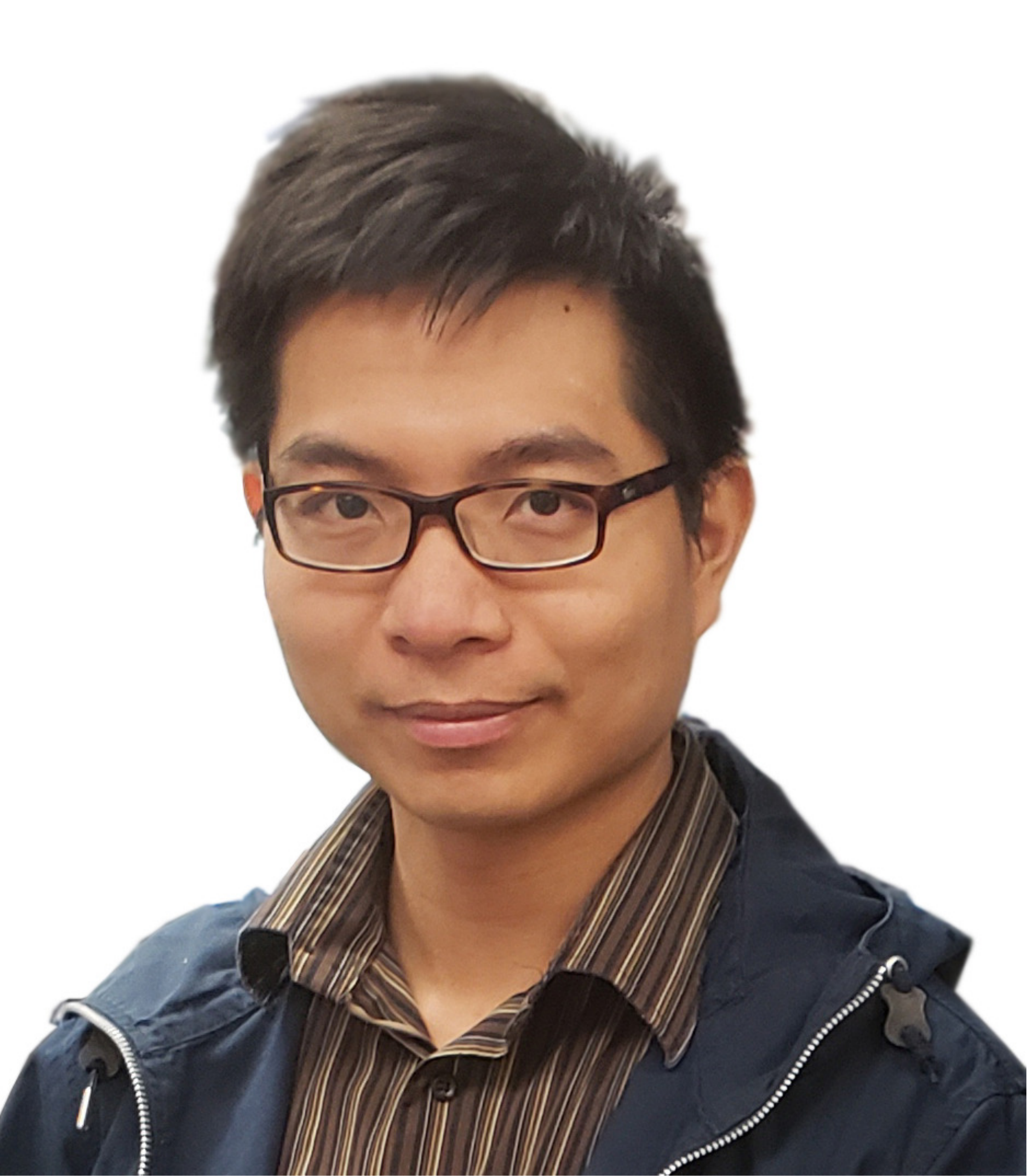}}]{Nghia Truong}
received a PhD degree in Scientific Computing from School of Computing, University of Utah in 2020, under the supervision of Prof. Cem Yuksel.
Prior to starting his PhD program, he completed a BS and MS degrees in Applied Mathematics and Computer Science from Tula State University, Russian Federation.
His research interests are in the related fields of computer graphics, machine learning/deep learning, and data science.
\biospace
\end{IEEEbiography}

\begin{IEEEbiography}[{\includegraphics[width=1in,height=1.25in,keepaspectratio]{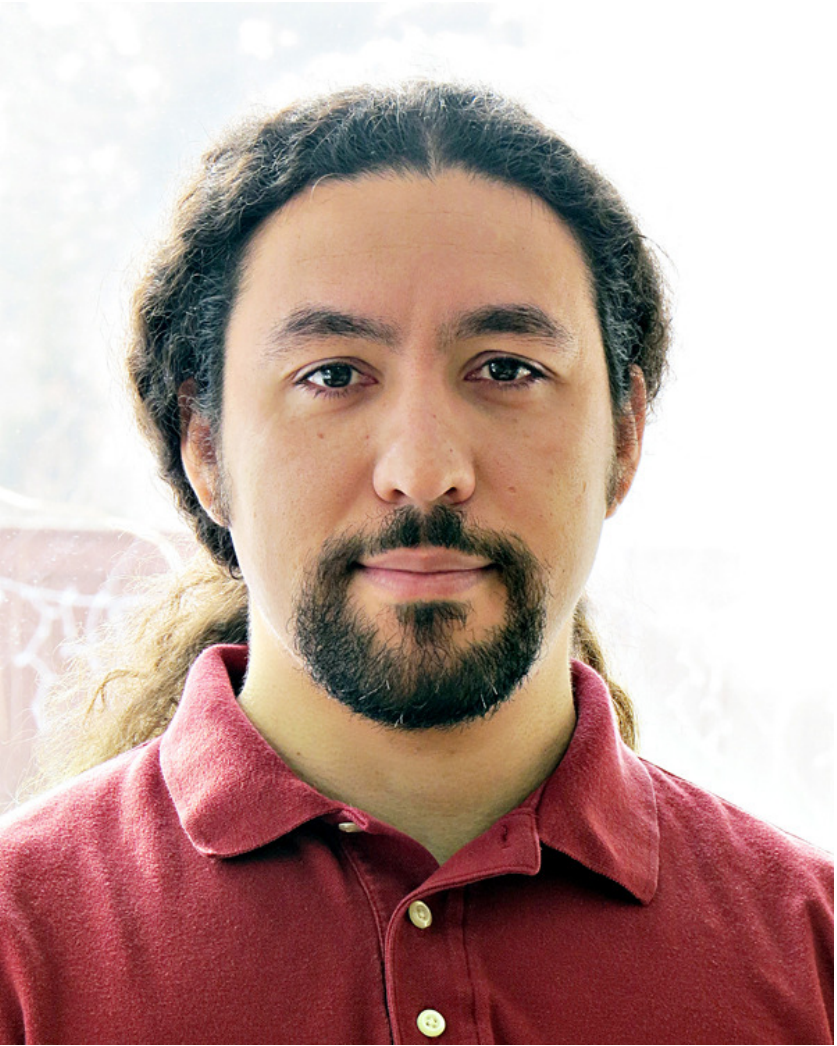}}]{Cem Yuksel}
is an associate professor in the School of Computing at the University of Utah. Previously, he was a postdoctoral fellow at Cornell University, after receiving his PhD in Computer Science from Texas A\&M University in 2010. His research interests are in computer graphics and related fields, including physically-based simulations, rendering techniques, global illumination, sampling, GPU algorithms, graphics hardware, knitted structures, and hair modeling, animation, and rendering.
\biospace
\end{IEEEbiography}

\begin{IEEEbiography}[{\includegraphics[width=1in,height=1.25in,keepaspectratio]{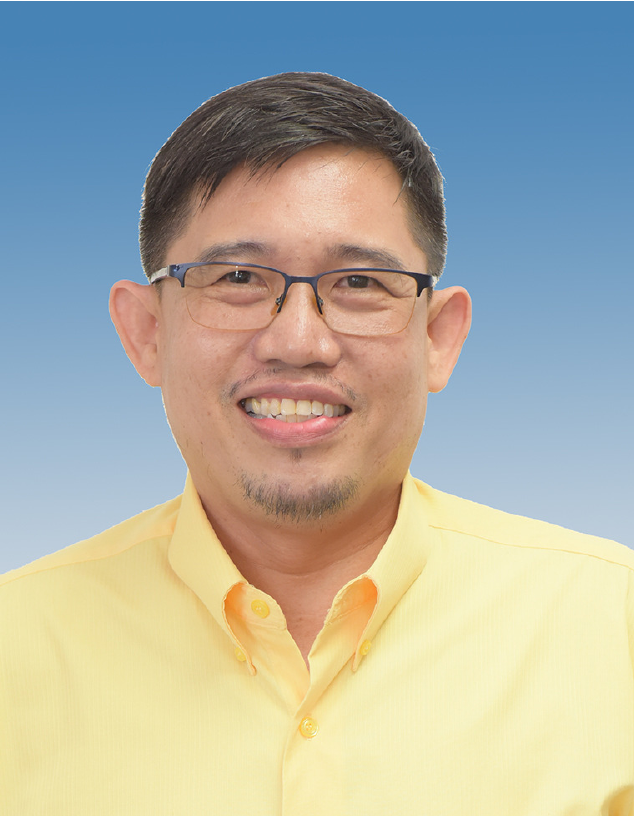}}]{Chakrit Watcharopas} is an assistant professor in the Department of Computer Science at Kasetsart University, Thailand. 
Previously, he received his PhD in Computer Science from Clemson University in 2004 and the MS degree in Computer Science from University of Southern California in 1997. He is currently the 
Deputy of the Department of Computer Science at Kasetsart and a member of the Artificial Intelligence Innovation for Healthtech (Kasetsart) where he is also the director of the Deep AI-Assisted Healthcare Group. His research interests include fracture simulation, surface extraction, phase transition, computer graphics, and deep learning for medical data and imaging. 
\biospace
\end{IEEEbiography}

\begin{IEEEbiography}[{\includegraphics[width=1in,height=1.25in,keepaspectratio]{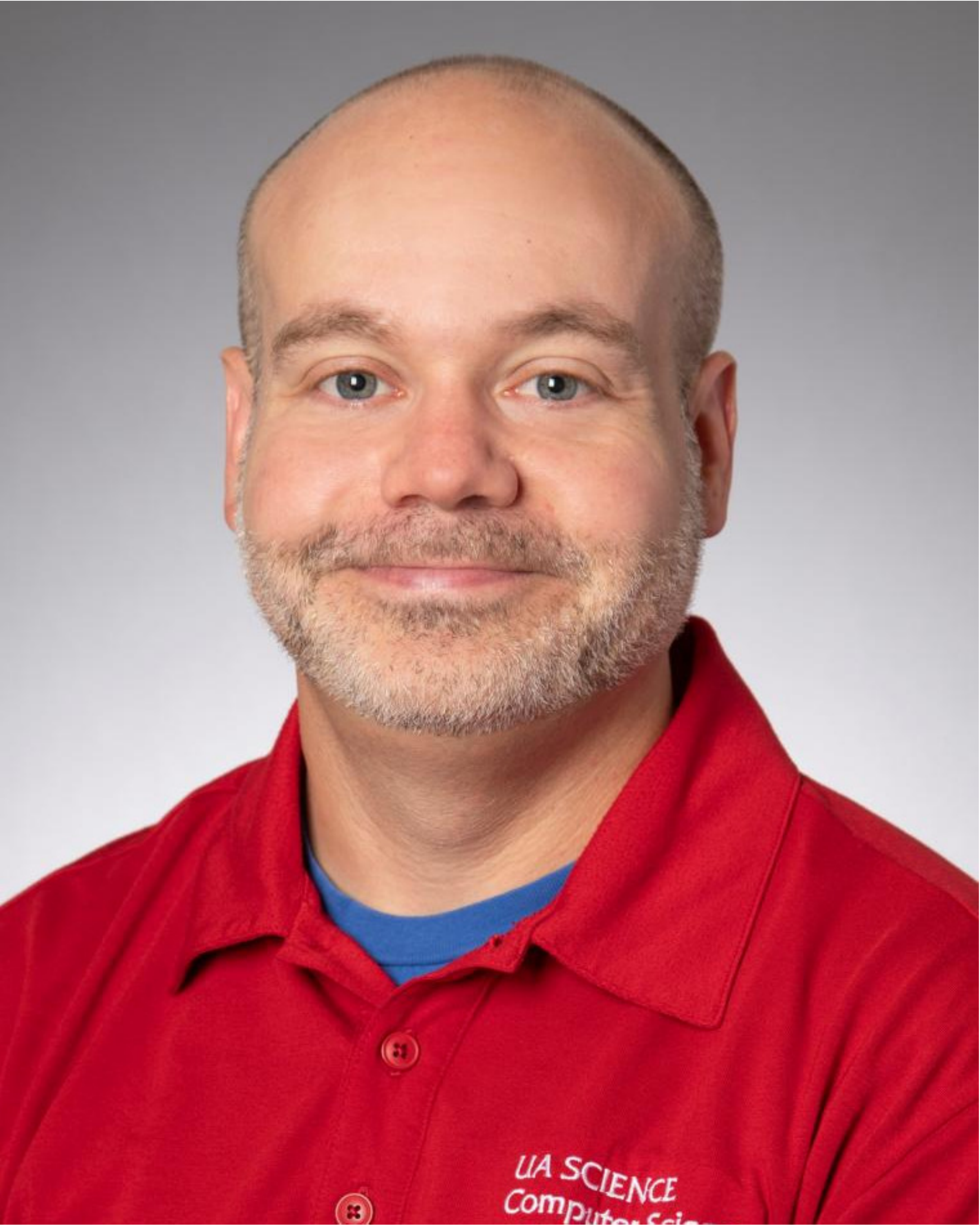}}]{Joshua A. Levine}
is an associate professor in the Department of Computer Science at University of Arizona. 
Previously, he was an assistant professor at Clemson University from 2012 to 2016, and a postdoctoral research associate at the University of Utah’s SCI Institute from 2009 to 2012. He received his PhD in Computer Science from Ohio State University (2009) after completing BS degrees in Computer Engineering and Mathematics (2003) and an MS in Computer Science (2004) from Case Western Reserve University. His research interests include visualization, geometric modeling, topological analysis, mesh generation, and 
computer 
graphics.
\biospace
\end{IEEEbiography}

\begin{IEEEbiography}[{\includegraphics[width=1in,height=1.25in,keepaspectratio,trim=0 12 0 0,clip]{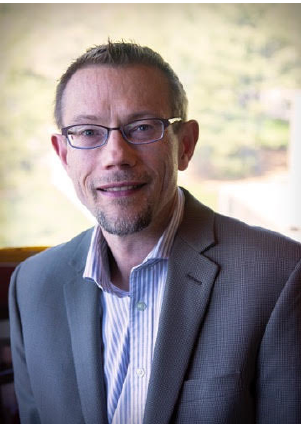}}]{Robert M. (Mike) Kirby} received the MS and PhD degrees in applied mathematics, and the MS degree in computer science from Brown University, Providence, in 1999, 2002, and 2001, respectively. He is currently the Executive Director of the Utah Informatics Initiative (UI2), the Interim Director of the Scientific Computing and Imaging (SCI) Institute, and a Professor of Computing and the Associate Director of the School of Computing, University of Utah.
His current research interests include scientific and data computing and visualization.  He is a Senior Member of the IEEE.
\end{IEEEbiography}

\end{document}